\documentclass[10pt,twocolumn,twoside]{IEEEtran}

\usepackage{graphicx}
\usepackage{cite}
\usepackage[cmex10]{amsmath}
\usepackage{amssymb}
\usepackage{subcaption}
\usepackage{graphicx}

\usepackage[per=slash,decimalsymbol=comma,loctolang={DE:ngerman},]{siunitx}
\usepackage{tikz}
\usepackage{pgfplots}
\pgfplotsset{compat=newest}
\pgfplotsset{legend style={rounded corners=2pt,nodes=right}}
\usetikzlibrary{arrows,decorations,decorations.markings,shapes}

\DeclareMathAlphabet{\mathbit}{OML}{cmr}{bx}{it}

\DeclareMathOperator{\E}{E}
\DeclareMathOperator{\T}{T}
\DeclareMathOperator{\Tr}{tr}

\DeclareMathOperator{\Diag}{diag}
\DeclareMathOperator{\Diagvec}{diagvec}

\renewcommand\vec[1]{\operatorname{vec}\left(#1\right)}

\DeclareMathOperator{\fieldR}{\mathbb{R}}

\DeclareMathOperator{\fieldN}{\mathbb{N}}
\DeclareMathOperator{\fieldB}{\mathbb{B}}

\DeclareMathOperator{\Sin}{sin}
\DeclareMathOperator{\Cos}{cos}

\DeclareMathOperator{\Ic}{I}
\DeclareMathOperator{\Qc}{Q}

\newcommand{\diag}[1]{\Diag{\left(#1\right)}}
\newcommand{\diagvec}[1]{\Diagvec{\left(#1\right)}}

\newcommand{\sign}[1]{\Sign{\left(#1\right)}}

\newcommand{\ve}[1]{\boldsymbol{#1}}
\newcommand\vect[1]{\operatorname{vec}\left(#1\right)}
\newcommand\vecT[1]{\operatorname{vec}^{\T}\left(#1\right)}

\newcommand{\exdi}[2]{\E_{#1} \left[#2\right]}

\renewcommand{\exp}[1]{\operatorname{exp}\left(#1\right)}

\newcommand{\arcsine}[1]{\sin^{-1} \left(#1\right)}

\newcommand\Sign{\operatorname{sign}}

\newcommand{\sinbr}[1]{\Sin \left(#1\right)}
\newcommand{\cosbr}[1]{\Cos \left(#1\right)}

\tikzstyle{vecArrow} = [thick, decoration={markings,mark=at position
   1 with {\arrow[semithick]{open triangle 60}}},
   double distance=1.4pt, shorten >= 5.5pt,
   preaction = {decorate},
   postaction = {draw,line width=1.4pt, white,shorten >= 4.5pt}]
\tikzstyle{innerWhite} = [semithick, white,line width=1.4pt, shorten >= 4.5pt]

\title{Glancing Through Massive Binary Radio Lenses:\\Hardware-Aware Interferometry With $1$-Bit Sensors}
\author{Manuel~S.~Stein
\thanks{The research leading to this publication was supported by the Deutsche Forschungsgemeinschaft (DFG, German Research Foundation) under the individual research grant STE 2787/1-1 (Research Fellowship Programme).}
\thanks{M. S. Stein is with the Department of Microelectronics, Delft University of Technology, The Netherlands (e-mail: M.S.Stein@tudelft.nl).}
}

\begin{document}
\newlength{\figurewidth}
\newlength{\figureheight}
\newlength{\scalefactor}
\maketitle
\begin{abstract}
Energy consumption and hardware cost of signal digitization together with the management of the resulting data volume form serious issues for high-rate measurement systems with multiple sensors. Switching to binary sensing front-ends results in a resource-efficient layout but is commonly associated with significant distortion due to the nonlinear signal acquisition. In particular, for applications that require to solve high-resolution processing tasks under extreme conditions, it is a widely held belief that low-complexity $1$-bit analog-to-digital conversion leads to unacceptable performance degradation. In the Big Science context of low-frequency radio astronomy, we propose a telescope architecture based on simplistic binary sampling, precise probabilistic modeling, and likelihood-oriented data processing. The main principles, building blocks, and advantages of such a radio telescope system, which we refer to as \emph{The Massive Binary Radio Lenses}, are sketched. The open engineering science questions which have to be answered before building a prototype are outlined. We set sail for the academic technology study by deriving a statistical algorithm for interferometric imaging from binary array measurements. The method aims at extracting the full discriminative information about the spatial power distribution embedded in a binary sensor data stream without bias. Radio measurements obtained with LOFAR are used to test the developed imaging technique and discuss visual and quantitative results. These assessments shed light on the fact that binary radio telescopes are suited for surveying the universe.
\end{abstract}
\section{Introduction}
Using a large number of small antennas and synthetic aperture methods to reconstruct the radio image produced by electromagnetic emissions of celestial objects \cite{Jansky33,Reber40} has, in recent years, become state-of-the-art in the design of low-frequency radio astronomy systems \cite{Vos09,Ellingson09,Lonsdale09,Dewdney09,Haarlem13}. In contrast to systems with a few large dishes, see, e.g., \cite{Napier83,Ananthakrishnan95}, which are costly to move and maintain, such large-scale array architectures offer faster surveying speed, higher sensitivity and resolution together with a flexible design that can continuously be updated with new chip and software technology. This holds, in particular, for all-digital architectures where each antenna features dedicated receive paths with numerical output, such that digital processing units have access to the raw measurement data of each radio sensor. The low-band antennas (LBA) of the Low-Frequency Array (LOFAR) radio telescope system \cite{Vos09}, which have been in operation in the European Union since the year 2010, represent a milestone in this technical development and mark the completion of a design paradigm shift towards all-digital system architectures in low-frequency radio astronomy. While LOFAR features approximately \num{5000} LBAs, already next-generation systems with more than \num{100000} antennas, like the low-frequency receiver of the Square Kilometre Array (SKA-LOW) are under development \cite{Dewdney09}. For such large-scale all-digital arrays, energy consumption and hardware cost of the individual receiver as well as the measurement data volume become factors that limit the viable number of sensing devices and their proliferation. Currently, also plans are being made for all-digital arrays in space, see, e.g., \cite{Rajan16} and the references therein. Here the deployment location of the radio measurement equipment imposes stringent constraints on the available power resources, instrumentation payload, and data communication rates. So, the principles for the design of future radio telescope systems, which can efficiently collect, combine, and process the observations of millions of digital sensors on Earth and in space, form open research questions.
\subsection{Approach - Resource-Efficient Binary Sensing Technology}
An option that minimizes the analog complexity of the individual receive path and the size of the resulting measurement data stream in all-digital systems, is the use of low-complexity $1$-bit analog-to-digital (A/D) conversion where the A/D converter associated with each antenna output is realized by a single symmetric comparator. In contrast to $\Sigma\Delta$-modulation \cite{Inose63,Aziz96,Daubechies03,Boufounos15}, the comparator comes without error feedback and operates with moderate oversampling \cite{SteinICASSP17}. The resulting digital sensor data features a single bit per sample and exclusively preserves information about the sign of the analog signals. Low-level processing in digital computation units, e.g., correlation of the different sensor streams, can then be performed without multiplications. Nevertheless, switching to binary sensing and data processing technology results in significant signal distortion due to the nonlinear data acquisition. It is a common belief that for challenging practical applications with high-performance requirements, like radio astronomy, $1$-bit A/D conversion leads to an unacceptable degradation of sensing performance. Fortunately, the fundamental measurement principle of interferometry is based on the time-offsets of the faint signal reception at many physical sensors. Amplitude sampling distortion at the individual receiver is then not of decisive importance for the sensitivity and spatial resolution achieved when merging the full time-offset information embedded in the measurement data to one interferometric image. Numerical signal acquisition with high A/D resolution, however, allocates most of the available resources to digitizing the uninformative amplitude of the sensor measurement noise.
\subsection{Technology Vision - Massive Binary Radio Telescopes}
We propose a radio telescope architecture that aims to engrave the maximum amount of information about the astronomical phenomena into the sensor measurement data while minimizing digitization resource dissipation. To this end, the amplitude resolution of the numerical signal acquisition process is set to a single bit. With technology specified in recent standards, binary sensing enables the use of a massive number of widely distributed all-digital sensors and storing the high-rate measurements without temporal, spectral, or spatial compression. The digital observations of all sensors can then be made accessible to supercomputers, forming a software-based radio telescope system with highly scalable collecting area and spatial proliferation. Exploiting computational power and efficient binary processing algorithms, research groups in parallel receive the temporal, spectral, and spatial information required for their specific astronomical survey. The physical radio instruments are in operation only once for acquiring and storing the sensor data for a large number of science cases.

Key to such a progressive scientific platform is the transfer of theoretic advances in hardware-aware probabilistic signal modeling and likelihood-oriented statistical data processing \cite{Stein15, SteinFauss19} to the practical design of all-digital radio telescope systems. As a firm orientation point for this challenging endeavor, the envisaged physical system is defined by sketching the architecture and components of a radio telescope which we refer to as \emph{The Massive Binary Radio Lenses}. We outline the main properties and advantages of such a measurement system together with the open engineering science questions. 
\subsection{Contribution - Binary Radio Interferometric Imaging}
To demonstrate the potential, a processing method which retrieves the interferometric image from binary sensor measurements is derived. The technique is developed along Fisher's principle of maximum-likelihood, which retrieves the full information about the probabilistic model parameters from a given dataset of sufficient size. For multivariate binary data, the required likelihood is mathematically and computationally intractable, forcing to switch to an auxiliary perspective on probabilistic data models under which statistical complexity becomes controllable \cite{Stein15}. This way, a tractable interferometric technique exploiting the spatial correlation structure of the binary array data is obtained. Leaning on Fisher's notion of information, we provide a quantitative measure characterizing the achievable interferometric sensitivity with binary telescope systems. Algorithms are tested by reconstructing all-sky images with radio measurements from LOFAR stations. To this end, calibrated $64$-bit LOFAR datasets are hard-limited. The results produced with the $1$-bit data are visually compared to the interferometric images reconstructed from the $64$-bit radio measurements. Further, we illustrate the imaging uncertainty under both approaches. These experimental and analytical results show that high-performance interferometric imaging is possible with low-complexity sensors when precise probabilistic modeling of the nonlinear and noisy signal acquisition chain is accomplished, and the measurement data is processed with likelihood-oriented statistical techniques. The performed uncertainty analysis confirms that doubling the observation time or the number of receiving devices compensates the information loss due to the binary digitization at the sensors.
\section{The Massive Binary Radio Lenses}
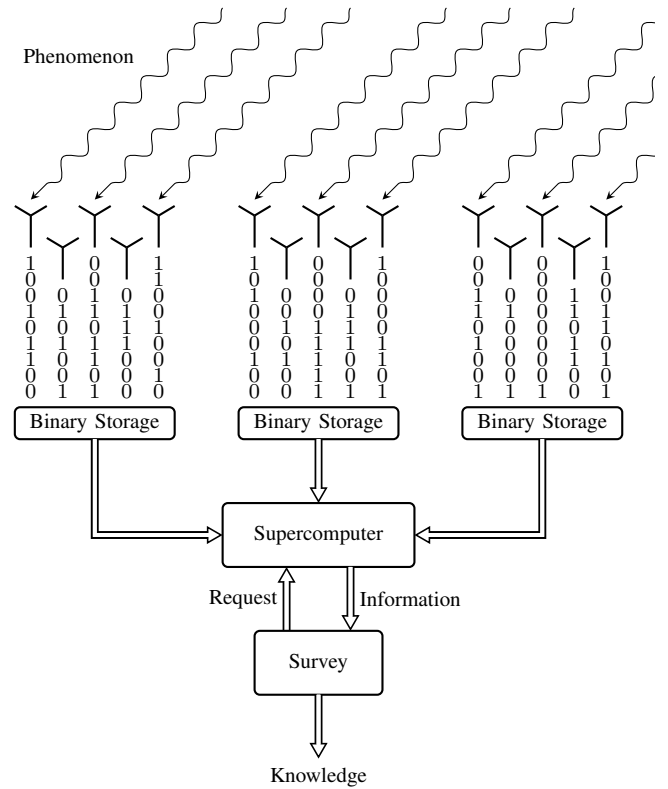
\begin{figure}[htbp!]

\def\stleftx{-3.5}
\def\strightx{3.5}


\begin{tikzpicture}[scale=0.85]

\clip(-5,-8.5) rectangle (5.25,3.75);

\draw (-3.75,3) node {\footnotesize Phenomenon};

\draw [-stealth,decorate,decoration={snake,amplitude=1mm,segment length=5mm,post length=2mm}]  (3-1+\stleftx,3.75)--(0-1+\stleftx,0.75);
\draw [-stealth,decorate,decoration={snake,amplitude=1mm,segment length=5mm,post length=2mm}]  (3+\stleftx,3.75)--(0+\stleftx,0.75);
\draw [-stealth,decorate,decoration={snake,amplitude=1mm,segment length=5mm,post length=2mm}]  (3+1+\stleftx,3.75)--(0+1+\stleftx,0.75);

\draw [-stealth,decorate,decoration={snake,amplitude=1mm,segment length=5mm,post length=2mm}]  (3-1,3.75)--(0-1,0.75);
\draw [-stealth,decorate,decoration={snake,amplitude=1mm,segment length=5mm,post length=2mm}]  (3,3.75)--(0,0.75);
\draw [-stealth,decorate,decoration={snake,amplitude=1mm,segment length=5mm,post length=2mm}]  (3+1,3.75)--(0+1,0.75);

\draw [-stealth,decorate,decoration={snake,amplitude=1mm,segment length=5mm,post length=2mm}]  (3-1+\strightx,3.75)--(0-1+\strightx,0.75);
\draw [-stealth,decorate,decoration={snake,amplitude=1mm,segment length=5mm,post length=2mm}]  (3+\strightx,3.75)--(0+\strightx,0.75);
\draw [-stealth,decorate,decoration={snake,amplitude=1mm,segment length=5mm,post length=2mm}]  (3+1+\strightx,3.75)--(0+1+\strightx,0.75);

\draw [thick,rounded corners=2pt] (-1+\stleftx,0.5) -- (0-1+\stleftx,0);
\draw [thick,rounded corners=2pt] (-1+\stleftx,0.5) -- (-0.25-1+\stleftx,0.65) ;
\draw [thick,rounded corners=2pt] (-1+\stleftx,0.5) -- (0.25-1+\stleftx,0.65) ;

\draw [thick,rounded corners=2pt] (+\stleftx,0.5) -- (0+\stleftx,0);
\draw [thick,rounded corners=2pt] (+\stleftx,0.5) -- (-0.25+\stleftx,0.65) ;
\draw [thick,rounded corners=2pt] (+\stleftx,0.5) -- (0.25+\stleftx,0.65) ;

\draw [thick,rounded corners=2pt] (+1+\stleftx,0.5) -- (0+1+\stleftx,0);
\draw [thick,rounded corners=2pt] (+1+\stleftx,0.5) -- (-0.25+1+\stleftx,0.65) ;
\draw [thick,rounded corners=2pt] (+1+\stleftx,0.5) -- (0.25+1+\stleftx,0.65) ;

\draw [thick,rounded corners=2pt] (-0.5+\stleftx,0.5-0.5) -- (0-0.5+\stleftx,0-0.5);
\draw [thick,rounded corners=2pt] (-0.5+\stleftx,0.5-0.5) -- (-0.25-0.5+\stleftx,0.65-0.5) ;
\draw [thick,rounded corners=2pt] (-0.5+\stleftx,0.5-0.5) -- (0.25-0.5+\stleftx,0.65-0.5) ;

\draw [thick,rounded corners=2pt] (-0.5+1+\stleftx,0.5-0.5) -- (0-0.5+1+\stleftx,0-0.5);
\draw [thick,rounded corners=2pt] (-0.5+1+\stleftx,0.5-0.5) -- (-0.25-0.5+1+\stleftx,0.65-0.5) ;
\draw [thick,rounded corners=2pt] (-0.5+1+\stleftx,0.5-0.5) -- (0.25-0.5+1+\stleftx,0.65-0.5) ;

\draw (-1+\stleftx,-0.25) node {\footnotesize $1$};
\draw (-1+\stleftx,-0.25-0.25) node {\footnotesize $0$};
\draw (-1+\stleftx,-0.25-0.5) node {\footnotesize $0$};
\draw (-1+\stleftx,-0.25-0.75) node {\footnotesize $1$};
\draw (-1+\stleftx,-0.25-1) node {\footnotesize $0$};
\draw (-1+\stleftx,-0.25-1.25) node {\footnotesize $1$};
\draw (-1+\stleftx,-0.25-1.5) node {\footnotesize $1$};
\draw (-1+\stleftx,-0.25-1.75) node {\footnotesize $0$};
\draw (-1+\stleftx,-0.25-2) node {\footnotesize $0$};

\draw (\stleftx,-0.25) node {\footnotesize $0$};
\draw (\stleftx,-0.25-0.25) node {\footnotesize $0$};
\draw (\stleftx,-0.25-0.5) node {\footnotesize $1$};
\draw (\stleftx,-0.25-0.75) node {\footnotesize $1$};
\draw (\stleftx,-0.25-1) node {\footnotesize $0$};
\draw (\stleftx,-0.25-1.25) node {\footnotesize $1$};
\draw (\stleftx,-0.25-1.5) node {\footnotesize $1$};
\draw (\stleftx,-0.25-1.75) node {\footnotesize $0$};
\draw (\stleftx,-0.25-2) node {\footnotesize $1$};

\draw (1+\stleftx,-0.25) node {\footnotesize $1$};
\draw (1+\stleftx,-0.25-0.25) node {\footnotesize $1$};
\draw (1+\stleftx,-0.25-0.5) node {\footnotesize $0$};
\draw (1+\stleftx,-0.25-0.75) node {\footnotesize $0$};
\draw (1+\stleftx,-0.25-1) node {\footnotesize $1$};
\draw (1+\stleftx,-0.25-1.25) node {\footnotesize $0$};
\draw (1+\stleftx,-0.25-1.5) node {\footnotesize $0$};
\draw (1+\stleftx,-0.25-1.75) node {\footnotesize $1$};
\draw (1+\stleftx,-0.25-2) node {\footnotesize $0$};

\draw (-0.5+\stleftx,-0.25-0.5) node {\footnotesize $0$};
\draw (-0.5+\stleftx,-0.25-0.75) node {\footnotesize $1$};
\draw (-0.5+\stleftx,-0.25-1) node {\footnotesize $0$};
\draw (-0.5+\stleftx,-0.25-1.25) node {\footnotesize $1$};
\draw (-0.5+\stleftx,-0.25-1.5) node {\footnotesize $0$};
\draw (-0.5+\stleftx,-0.25-1.75) node {\footnotesize $0$};
\draw (-0.5+\stleftx,-0.25-2) node {\footnotesize $1$};

\draw (+0.5+\stleftx,-0.25-0.5) node {\footnotesize $0$};
\draw (+0.5+\stleftx,-0.25-0.75) node {\footnotesize $1$};
\draw (+0.5+\stleftx,-0.25-1) node {\footnotesize $1$};
\draw (+0.5+\stleftx,-0.25-1.25) node {\footnotesize $1$};
\draw (+0.5+\stleftx,-0.25-1.5) node {\footnotesize $0$};
\draw (+0.5+\stleftx,-0.25-1.75) node {\footnotesize $0$};
\draw (+0.5+\stleftx,-0.25-2) node {\footnotesize $0$};

\draw [thick,rounded corners=2pt] (-1.25+\stleftx,-3) rectangle (1.25+\stleftx,-2.5);
\draw (\stleftx,-2.75) node {\footnotesize Binary Storage};

\draw [thick,rounded corners=2pt] (0-1,0.5) -- (0-1,0);
\draw [thick,rounded corners=2pt] (0-1,0.5) -- (-0.25-1,0.65) ;
\draw [thick,rounded corners=2pt] (0-1,0.5) -- (0.25-1,0.65) ;

\draw [thick,rounded corners=2pt] (0,0.5) -- (0,0);
\draw [thick,rounded corners=2pt] (0,0.5) -- (-0.25,0.65) ;
\draw [thick,rounded corners=2pt] (0,0.5) -- (0.25,0.65) ;

\draw [thick,rounded corners=2pt] (1,0.5) -- (0+1,0);
\draw [thick,rounded corners=2pt] (1,0.5) -- (-0.25+1,0.65) ;
\draw [thick,rounded corners=2pt] (1,0.5) -- (0.25+1,0.65) ;

\draw [thick,rounded corners=2pt] (-0.5,0.5-0.5) -- (0-0.5,0-0.5);
\draw [thick,rounded corners=2pt] (-0.5,0.5-0.5) -- (-0.25-0.5,0.65-0.5) ;
\draw [thick,rounded corners=2pt] (-0.5,0.5-0.5) -- (0.25-0.5,0.65-0.5) ;

\draw [thick,rounded corners=2pt] (-0.5+1,0.5-0.5) -- (0-0.5+1,0-0.5);
\draw [thick,rounded corners=2pt] (-0.5+1,0.5-0.5) -- (-0.25-0.5+1,0.65-0.5) ;
\draw [thick,rounded corners=2pt] (-0.5+1,0.5-0.5) -- (0.25-0.5+1,0.65-0.5) ;

\draw (-1,-0.25) node {\footnotesize $1$};
\draw (-1,-0.25-0.25) node {\footnotesize $0$};
\draw (-1,-0.25-0.5) node {\footnotesize $1$};
\draw (-1,-0.25-0.75) node {\footnotesize $0$};
\draw (-1,-0.25-1) node {\footnotesize $0$};
\draw (-1,-0.25-1.25) node {\footnotesize $0$};
\draw (-1,-0.25-1.5) node {\footnotesize $1$};
\draw (-1,-0.25-1.75) node {\footnotesize $0$};
\draw (-1,-0.25-2) node {\footnotesize $0$};

\draw (0,-0.25) node {\footnotesize $0$};
\draw (0,-0.25-0.25) node {\footnotesize $0$};
\draw (0,-0.25-0.5) node {\footnotesize $0$};
\draw (0,-0.25-0.75) node {\footnotesize $0$};
\draw (0,-0.25-1) node {\footnotesize $1$};
\draw (0,-0.25-1.25) node {\footnotesize $1$};
\draw (0,-0.25-1.5) node {\footnotesize $1$};
\draw (0,-0.25-1.75) node {\footnotesize $1$};
\draw (0,-0.25-2) node {\footnotesize $1$};

\draw (1,-0.25) node {\footnotesize $1$};
\draw (1,-0.25-0.25) node {\footnotesize $0$};
\draw (1,-0.25-0.5) node {\footnotesize $0$};
\draw (1,-0.25-0.75) node {\footnotesize $0$};
\draw (1,-0.25-1) node {\footnotesize $0$};
\draw (1,-0.25-1.25) node {\footnotesize $1$};
\draw (1,-0.25-1.5) node {\footnotesize $1$};
\draw (1,-0.25-1.75) node {\footnotesize $0$};
\draw (1,-0.25-2) node {\footnotesize $1$};

\draw (-0.5,-0.25-0.5) node {\footnotesize $0$};
\draw (-0.5,-0.25-0.75) node {\footnotesize $0$};
\draw (-0.5,-0.25-1) node {\footnotesize $1$};
\draw (-0.5,-0.25-1.25) node {\footnotesize $0$};
\draw (-0.5,-0.25-1.5) node {\footnotesize $1$};
\draw (-0.5,-0.25-1.75) node {\footnotesize $0$};
\draw (-0.5,-0.25-2) node {\footnotesize $0$};

\draw (+0.5,-0.25-0.5) node {\footnotesize $0$};
\draw (+0.5,-0.25-0.75) node {\footnotesize $1$};
\draw (+0.5,-0.25-1) node {\footnotesize $1$};
\draw (+0.5,-0.25-1.25) node {\footnotesize $1$};
\draw (+0.5,-0.25-1.5) node {\footnotesize $0$};
\draw (+0.5,-0.25-1.75) node {\footnotesize $0$};
\draw (+0.5,-0.25-2) node {\footnotesize $1$};

\draw [thick,rounded corners=2pt] (-1.25,-3) rectangle (1.25,-2.5);
\draw (0,-2.75) node {\footnotesize Binary Storage};

\draw [thick,rounded corners=2pt] (-1+\strightx,0.5) -- (0-1+\strightx,0);
\draw [thick,rounded corners=2pt] (-1+\strightx,0.5) -- (-0.25-1+\strightx,0.65) ;
\draw [thick,rounded corners=2pt] (-1+\strightx,0.5) -- (0.25-1+\strightx,0.65) ;

\draw [thick,rounded corners=2pt] (\strightx,0.5) -- (0+\strightx,0);
\draw [thick,rounded corners=2pt] (\strightx,0.5) -- (-0.25+\strightx,0.65) ;
\draw [thick,rounded corners=2pt] (\strightx,0.5) -- (0.25+\strightx,0.65) ;

\draw [thick,rounded corners=2pt] (1+\strightx,0.5) -- (0+1+\strightx,0);
\draw [thick,rounded corners=2pt] (1+\strightx,0.5) -- (-0.25+1+\strightx,0.65) ;
\draw [thick,rounded corners=2pt] (1+\strightx,0.5) -- (0.25+1+\strightx,0.65) ;

\draw [thick,rounded corners=2pt] (-0.5+\strightx,0.5-0.5) -- (0-0.5+\strightx,0-0.5);
\draw [thick,rounded corners=2pt] (-0.5+\strightx,0.5-0.5) -- (-0.25-0.5+\strightx,0.65-0.5) ;
\draw [thick,rounded corners=2pt] (-0.5+\strightx,0.5-0.5) -- (0.25-0.5+\strightx,0.65-0.5) ;

\draw [thick,rounded corners=2pt] (-0.5+1+\strightx,0.5-0.5) -- (0-0.5+1+\strightx,0-0.5);
\draw [thick,rounded corners=2pt] (-0.5+1+\strightx,0.5-0.5) -- (-0.25-0.5+1+\strightx,0.65-0.5) ;
\draw [thick,rounded corners=2pt] (-0.5+1+\strightx,0.5-0.5) -- (0.25-0.5+1+\strightx,0.65-0.5) ;

\draw (-1+\strightx,-0.25) node {\footnotesize $0$};
\draw (-1+\strightx,-0.25-0.25) node {\footnotesize $0$};
\draw (-1+\strightx,-0.25-0.5) node {\footnotesize $1$};
\draw (-1+\strightx,-0.25-0.75) node {\footnotesize $1$};
\draw (-1+\strightx,-0.25-1) node {\footnotesize $0$};
\draw (-1+\strightx,-0.25-1.25) node {\footnotesize $1$};
\draw (-1+\strightx,-0.25-1.5) node {\footnotesize $0$};
\draw (-1+\strightx,-0.25-1.75) node {\footnotesize $0$};
\draw (-1+\strightx,-0.25-2) node {\footnotesize $1$};

\draw (\strightx,-0.25) node {\footnotesize $0$};
\draw (\strightx,-0.25-0.25) node {\footnotesize $0$};
\draw (\strightx,-0.25-0.5) node {\footnotesize $0$};
\draw (\strightx,-0.25-0.75) node {\footnotesize $0$};
\draw (\strightx,-0.25-1) node {\footnotesize $0$};
\draw (\strightx,-0.25-1.25) node {\footnotesize $0$};
\draw (\strightx,-0.25-1.5) node {\footnotesize $0$};
\draw (\strightx,-0.25-1.75) node {\footnotesize $0$};
\draw (\strightx,-0.25-2) node {\footnotesize $1$};

\draw (1+\strightx,-0.25) node {\footnotesize $1$};
\draw (1+\strightx,-0.25-0.25) node {\footnotesize $0$};
\draw (1+\strightx,-0.25-0.5) node {\footnotesize $0$};
\draw (1+\strightx,-0.25-0.75) node {\footnotesize $1$};
\draw (1+\strightx,-0.25-1) node {\footnotesize $1$};
\draw (1+\strightx,-0.25-1.25) node {\footnotesize $0$};
\draw (1+\strightx,-0.25-1.5) node {\footnotesize $1$};
\draw (1+\strightx,-0.25-1.75) node {\footnotesize $0$};
\draw (1+\strightx,-0.25-2) node {\footnotesize $1$};

\draw (-0.5+\strightx,-0.25-0.5) node {\footnotesize $0$};
\draw (-0.5+\strightx,-0.25-0.75) node {\footnotesize $1$};
\draw (-0.5+\strightx,-0.25-1) node {\footnotesize $0$};
\draw (-0.5+\strightx,-0.25-1.25) node {\footnotesize $0$};
\draw (-0.5+\strightx,-0.25-1.5) node {\footnotesize $0$};
\draw (-0.5+\strightx,-0.25-1.75) node {\footnotesize $0$};
\draw (-0.5+\strightx,-0.25-2) node {\footnotesize $1$};

\draw (0.5+\strightx,-0.25-0.5) node {\footnotesize $1$};
\draw (0.5+\strightx,-0.25-0.75) node {\footnotesize $1$};
\draw (0.5+\strightx,-0.25-1) node {\footnotesize $0$};
\draw (0.5+\strightx,-0.25-1.25) node {\footnotesize $1$};
\draw (0.5+\strightx,-0.25-1.5) node {\footnotesize $1$};
\draw (0.5+\strightx,-0.25-1.75) node {\footnotesize $0$};
\draw (0.5+\strightx,-0.25-2) node {\footnotesize $0$};

\draw [thick,rounded corners=2pt] (-1.25+\strightx,-3) rectangle (1.25+\strightx,-2.5);
\draw (\strightx,-2.75) node {\footnotesize Binary Storage};


\draw[vecArrow] (0+\stleftx,-3) |- (-1.5,-4.5);
\draw[innerWhite] (0+\stleftx,-3) |- (-1.5,-4.5);

\draw[vecArrow] (0,-3) to (0,-4);
\draw[innerWhite] (0,-3) to (0,-4);

\draw[vecArrow] (0+\strightx,-3) |- (1.5,-4.5);
\draw[innerWhite] (0+\strightx,-3) |- (1.5,-4.5);

\draw [thick,rounded corners=2pt] (-1.5,-5) rectangle (1.5,-4);
\draw (0,-4.5) node {\footnotesize Supercomputer};

\draw[vecArrow] (0-0.5,-6) to (0-0.5,-5);
\draw[innerWhite] (0-0.5,-6) to (0-0.5,-5);
\draw (-0.5,-5.5) node[left] {\footnotesize Request};

\draw[vecArrow] (0+0.5,-5) to (0+0.5,-6);
\draw[innerWhite] (0+0.5,-5) to (0+0.5,-6);
\draw (+0.5,-5.5) node[right] {\footnotesize Information};

\draw [thick,rounded corners=2pt] (-1,-7) rectangle (1,-6);
\draw (0,-6.5) node {\footnotesize Survey};

\draw[vecArrow] (0,-7) to (0,-8);
\draw[innerWhite] (0,-7) to (0,-8);
\draw (0,-8) node[below] {\footnotesize Knowledge};

\end{tikzpicture}
\caption{Binary Radio Telescope System ($M_{\text{S}}=3, M_{\text{A}}=5$)}
\label{BRTS}
\end{figure}

\emph{The Massive Binary Radio Lenses} consist of $M_{\text{S}}\in\fieldN$ telescope stations with $M_{\text{A}}\in\fieldN$ simple radio antennas, where the total amount of radio sensors $M=M_{\text{S}} M_{\text{A}}$ and their proliferation are as large as possible. The antennas perform synchronized digital signal acquisition with a spectral bandwidth of $B_{\text{Y}}\in\fieldR, B_{\text{Y}}>0,$ and store the binary version of the unprocessed radio measurements. The data recorded is archived and made available to supercomputers. Here the information required for diverse astronomical surveys is extracted using the binary data streams of all sensors. Fig. \ref{BRTS} illustrates the general idea via an exemplary binary radio telescope.

A massive version of such a sensing architecture follows from basic considerations which we outline in the following together with the main building blocks and advantages. This conceptual section is concluded by naming the open research questions surrounding the envisioned telescope technology.
\subsection{Fundamental Technical Considerations}
\paragraph{Maximizing Sensitivity and Spatial Resolution}
The measurement sensitivity and spatial resolution of all-digital radio telescopes are determined by the collecting area and the spread of the sensors. A modern system architect aligns the entire design according to such performance laws. The focus is on maximizing the flow of information from the physical sensing interface to the final numerical output algorithms.
\paragraph{Adaptation to the Fastest Technology Process}
Technology progress takes place at different speeds. For certain technologies it is more likely that one might soon reach feasibility limits than for others. A modern scientific system design relies as far as possible on those technologies with the fastest evolution and most significant room for further improvement. For decades, we have seen an exponential increase in digital computing power \cite{Schaller97}. Concerns that this quick development, often referred to as Moore's law, could soon end \cite{Powell08} are contrasted by the development of new information processing concepts inspired by quantum physics \cite{Steane98} or living cells \cite{Cavin12}. In particular, for digital storage, technical limits seem far from being exhausted, see, e.g., \cite{Patel17}. The situation is different when it comes to converting analog sensor signals into numerical measurement data. 
\begin{figure}[!htbp]
\centering
    \begin{subfigure}[t]{0.24\textwidth}
    \centering
        \begin{tikzpicture}[scale=0.5]
	\begin{semilogyaxis}[
		xlabel={Effective Resolution [bit]},
		ylabel={A/D Energy Dissipation [pJ]},
		ymax=100000000,
		ymin=0.01,
		xmax=21.5,
		xmin=2]

	\addplot[only marks, mark=*] table[x index=0, y index=1]{Figures/adc_scatter1.txt};
	\addplot[color=black,dashed] table[x index=0, y index=1]{Figures/adc_scatter1_ref.txt};
	\node[below] at (19,70000) {\scriptsize 2x per bit};

	\end{semilogyaxis}
\end{tikzpicture}
\caption{Years 1997-2004}
\label{adc:efficiency:early}
\end{subfigure}%
    ~ 
\begin{subfigure}[t]{0.25\textwidth}
\centering
\begin{tikzpicture}[scale=0.5]
	\begin{semilogyaxis}[
		xlabel={Effective Resolution [bit]},
		ymax=100000000,
		ymin=0.01,
		xmax=21.5,
		xmin=2]

	\addplot[only marks, mark=*] table[x index=0, y index=1]{Figures/adc_scatter2.txt};
	\addplot[color=black, dashed] table[x index=0, y index=1]{Figures/adc_scatter2_ref.txt};
	\node[below] at (19,50000) {\scriptsize 4x per bit};

	\end{semilogyaxis}
\end{tikzpicture}
\caption{Years 2011-2018}
\label{adc:efficiency:late}
\end{subfigure}%
\caption{Energy Efficiency - A/D Technology (Source: \cite{MurmannSurvey})}
\label{adc:efficiency}
\end{figure}
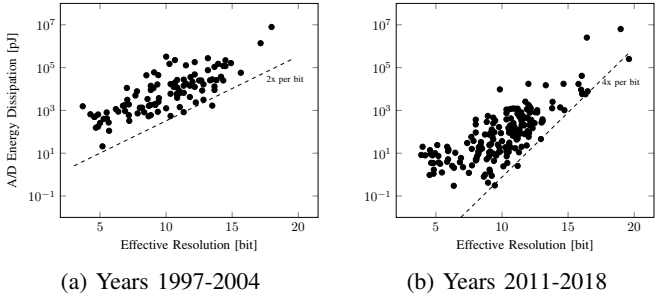
Fig. \ref{adc:efficiency:early} shows the energy dissipation per Nyquist sample as function of the effective A/D resolution for different converter designs published in the years 1997-2004 at the ISSCC \& VLSI symposia \cite{MurmannSurvey}. In Fig. \ref{adc:efficiency:late} the same figure of merit for the A/D architectures presented in the years 2011-2018 is depicted. It can be observed that the progress for A/D technology featuring high amplitude resolution has stagnated in the last twenty years. Improvements have been achieved for low amplitude resolution, but the speed of this development is much slower than Moore's estimate of four orders of magnitude technology improvement over twenty years. Fig. \ref{adc:efficiency} also shows that adding one bit to the effective A/D resolution at least doubles the digitization complexity. Therefore, from today's perspective, the components of a modern radio telescope system should be predominantly realized through digital processing and storage technologies. For the conversion of the physical sensor signals into numerical measurement data, the A/D resolution is to be minimized to adjust the development speed of analog front-end components to the evolution of digital processing modules.

\paragraph{Precise Physical and Hardware-aware Modeling}
Radio astronomy aims to discover unknown physical phenomena in the universe. Each survey has its focus on specific quantities. We summarize them here in the vector $\ve{\theta}_{\text{phy}}\in\fieldR^{D_{\text{phy}}}$. The observations are influenced by the propagation medium or radio interference which are physical effects outside the primary focus of the scientific survey. These are summarized here by $\ve{\theta}_{\text{med}}\in\fieldR^{D_{\text{med}}}$. Also the electromagnetic sensing instruments exhibit unknown properties influencing the measurements. Such technical parameters are denoted here as $\ve{\theta}_{\text{inst}}\in\fieldR^{D_{\text{inst}}}$.

By theoretic investigations and practical experiments, it is possible to obtain certain prior knowledge about $\ve{\theta}_{\text{med}}$ and $\ve{\theta}_{\text{inst}}$, which can be modeled by known distribution functions 
\begin{align}\label{prior:med:inst}
\ve{\theta}_{\text{med}}\sim p_{\ve{\theta}_{\text{med}}}(\ve{\theta}_{\text{med}})\quad\text{and }\quad\ve{\theta}_{\text{inst}}\sim p_{\ve{\theta}_{\text{inst}}}(\ve{\theta}_{\text{inst}}). 
\end{align}
Since $\ve{\theta}_{\text{phy}}$ represents physical quantities to be investigated in an unbiased manner, the consideration of prior knowledge
\begin{align}\label{prior:phy}
\ve{\theta}_{\text{phy}}\sim p_{\ve{\theta}_{\text{phy}}}(\ve{\theta}_{\text{phy}})
\end{align}
appears to be scientifically debatable. Of undisputed importance is to characterize the precise probabilistic relationship 
\begin{align}\label{definition:observation:model}
\ve{Y} \sim p_{\ve{Y}}(\ve{Y};\ve{\theta})
\end{align}
between the numerical measurements $\ve{Y}\in\ve{\mathcal{Y}}$ and all system parameters $\ve{\theta}\in\ve{\Theta},\ve{\Theta} \subset \fieldR^{D}, D=D_{\text{phy}}+D_{\text{inst}}+D_{\text{med}}$, 
\begin{align}\label{all:parameters}
\ve{\theta}=\begin{bmatrix} \ve{\theta}^{\T}_{\text{phy}} & \ve{\theta}^{\T}_{\text{med}} &\ve{\theta}^{\T}_{\text{inst}} \end{bmatrix}^{\T}.
\end{align}
Such a probabilistic model can be derived through physical and technical reasoning along with experimental verification, or it can be learned automatically utilizing large calibrated datasets. Since radio astronomy aims to observe unknown phenomena, a data-driven approach is challenging to implement due to the lack of labeled measurements. Consequently, probabilistic radio telescope technology is centered around the precise characterization of the model \eqref{definition:observation:model} such that the system architects strive to keep the sensing instruments as simple as possible.
\paragraph{Aiming for Consistent Data Processing}
Since fundamental scientific conclusions are to be drawn from astronomical surveys, it is critical to ensure consistency. Consistency implies that the knowledge $\ve{\hat{\theta}}(\ve{Y})\in\fieldR^{D}$ extracted from the measurements $\ve{Y}$ converges in probability towards reality $\ve{\theta}_t\in\fieldR^{D}$ as the amount of informative sensor data increases. It is advantageous if the estimates $\ve{\hat{\theta}}(\ve{Y})$ follow a gentle distribution of known structure $\ve{\hat{\theta}}\sim p_{\ve{\hat{\theta}}} (\ve{\hat{\theta}})$ centered at reality $\ve{\theta}_t$. Given sufficiently large data $\ve{Y}$, an algorithmic method featuring such characteristics, without incorporating prior knowledge like \eqref{prior:phy}, is the maximum-likelihood estimator (MLE)
\begin{align}\label{definition:mle}
\ve{\hat{\theta}}(\ve{Y})=\arg \max_{\ve{\theta}\in \ve{\Theta}} \ln p_{\ve{Y}}(\ve{Y};\ve{\theta}),
\end{align}
with the root of the score function
\begin{align}\label{ideal:score}
\ve{s}_{\ve{Y}}(\ve{Y};\ve{\theta}) = \bigg(\frac{ \partial \ln p_{\ve{Y}}(\ve{Y};\ve{\theta})  }{\partial \ve{\theta}}\bigg)^{\T}\in\fieldR^{D}
\end{align}
as its solution\footnote{For simplicity, we here assume that the score features a single root.}. In the large sample regime, the MLE follows a Gaussian distribution with mean $\ve{\theta}_t$ and covariance matrix
\begin{align}\label{definition:error:mle}
\exdi{\ve{Y};\ve{\theta}_t}{ \big(\ve{\hat{\theta}}(\ve{Y})-\ve{\theta}_t\big) \big(\ve{\hat{\theta}}(\ve{Y})-\ve{\theta}_t\big)^{\rm{T}} } =  \ve{F}_{\ve{Y}}^{-1}(\ve{\theta}_t),
\end{align}
where $\ve{F}_{\ve{Y}}(\ve{\theta})\in\fieldR^{D \times D}$ is Fisher's information matrix 
\begin{align}\label{definition:fisher:matrix}
&\ve{F}_{\ve{Y}}\big(\ve{\theta}\big)= \exdi{\ve{Y};\ve{\theta}}{\ve{s}_{\ve{Y}}(\ve{Y};\ve{\theta}) \ve{s}^{\T}_{\ve{Y}}(\ve{Y};\ve{\theta})}
\end{align}
of the probabilistic system model \eqref{definition:observation:model}. Classical estimation-theoretic arguments of Cram\'er and Rao ensure that \eqref{definition:error:mle} forms the performance limit for unbiased parameter estimation, i.e., the error covariance matrix of a digital processing unit which produces knowledge $\ve{\hat{\theta}}(\ve{Y})$ centered at reality $\ve{\theta}_t$ will always dominate the inverse of Fisher's information matrix \eqref{definition:fisher:matrix}. 

With a large number of parameters $D$, the probabilistic model \eqref{definition:observation:model} might present itself as non-identifiable such that $\ve{\theta}$ cannot be extracted unambiguously through \eqref{definition:mle}. In such a case, one might be tempted to incorporate regularizing assumptions through \eqref{prior:phy}. Scientific telescope technology, however, will try to avoid to deviate from the principle of unbiasedness and first concentrate on changing the observation instrument in such a way that the desired $D_{\text{phy}}$ survey parameters $\ve{\theta}_{\text{phy}}$ can be estimated without potentially biasing prior. An approach to push the probabilistic structure of the radio telescope system \eqref{definition:observation:model} towards identifiability is to reduce the complexity of the sensing front-ends and to invest the saved resources in a larger number of antennas and the extension of their proliferation.
\subsection{Station - Building Blocks}
Based on these considerations, the telescope stations of \emph{The Massive Binary Radio Lenses} are equipped with the following:
\paragraph{Low-Complexity Sensing Front-ends}
The $M_{\text{A}}$ station antennas are designed to receive with a one-sided spectral bandwidth $B_{\text{Y}}$. Each sensor features a separate analog output. A/D converters are realized by symmetric comparators which operate with coordinated clock and sampling rate $f_{\text{Y}}\in\fieldR, f_{\text{Y}}>0$. Alternatively, the measurements $\ve{Y}$ acquired with conventional high-resolution radio front-ends are hard-limited 
\begin{align}\label{binarization}
\ve{Z}=\sign{\ve{Y}}.
\end{align}
The probabilistic response of the antenna elements, the analog preprocessing chain and the signal binarization process \eqref{binarization}
\begin{align}\label{binary:prob:model}
\ve{Z} \sim p_{\ve{Z}}(\ve{Z};\ve{\theta})
\end{align}
is available as function of all unknown system parameters \eqref{all:parameters}.
\paragraph{Precise Synchronization and Positioning}
The $M_{\text{S}}$ stations are equipped with a precise temporal synchronization and spatial positioning mechanism to a joint coordinate system. This is achieved using GNSS receivers and atomic clocks. Alternatively, one binarizes the analog outputs of GNSS sensors to form additional synchronized data streams from which the station position in time and space can be extracted later. The relative positioning of the sensors within each station is precisely established during the installation of the equipment or through transmitters close to the stations.
\paragraph{Large Binary Data Storage}
Each telescope station is equipped with fast and large binary memory. For an assessment on the local data rates and storage size, let us assume a telescope station with $256$ dual-polarized low-frequency antennas with $M_{\text{A}}=512$ analog outputs. The observation bandwidth is $B_{\text{Y}}=\SI{250}{\mega\hertz}$ while temporal sampling takes place at Nyquist rate, i.e., $f_{\text{Y}}=2 B_{\text{Y}}$. Thus, the radio telescope station produces a binary measurement stream with a rate of
\begin{align}\label{rate:station:example}
2 B_{\text{Y}} M_{\text{A}} \text{ bit} =\SI{256}{\giga\bit\per\second}.
\end{align}
If $\SI{1024}{\tera\byte}$ can be stored locally, measurements of a duration
\begin{align}
\frac{ \SI{8.192e15}{\bit} }{ 2 B_{\text{Y}} M_{\text{A}}\SI{}{\bit} } = \SI{32000}{\second} \approx \SI{8}{\hour}\,\SI{ 53}{\min}
\end{align}
can be recorded. Such more than one third of the uncompressed radio sky, as seen by each of the $M_{\text{A}}$ radio sensors, can be made available to the supercomputers. With a link capacity of $\SI{100}{\giga\bit\per\second}$, the downlink transmission from the station to a supercomputer takes less than a day. Note that, for short to medium-range network transmissions, such data rates are specified in technology standards like InfiniBand or Ethernet.
\subsection{Supercomputer - Processing Modules}
The supercomputers of \emph{The Massive Binary Radio Lenses} have access to the distributed binary storage through data networks or wireless communication links. The storage-based architecture enables to dynamically adjust the data transmission rates while removing real-time processing requirements. At the supercomputers the following digital data processing tasks are performed with layered and iterative approaches:
\paragraph{Low-Level Processing of Binary Signals}
The binary measurements are preprocessed according to the probabilistic notion of sufficient statistics. We will see that, under a tractable representation of \eqref{binary:prob:model}, for interferometric imaging, this implies pairwise correlation among all $M$ binary sensor streams. These low-level computations (binary logic and counting) can be performed efficiently at massive scale on optimized hardware.
\paragraph{Local and Global Instrument Calibration}
Calibration for station-specific parameters $\ve{\theta}_{\text{inst}}$ like local offsets from the synchronization frame or attenuations of individual receive paths can be performed with the binary measurements from single telescope stations and potential prior knowledge \eqref{prior:med:inst}. Calibration for global effects, like changes of the propagation medium parameters $\ve{\theta}_{\text{med}}$, can be performed by using the binary data from a proper subset of radio stations. Note that distributed calibration for local effects helps to reduce the processing overhead at the supercomputers. Distributed radio transmitters can be used to improve the knowledge on \eqref{prior:med:inst}.

\paragraph{Local and Global RFI Identification}
Radio frequency interference (RFI) on local and global level is identified from the binary data streams together with the available priors \eqref{prior:med:inst}. The temporal, spectral, and spatial parameters in $\ve{\theta}_{\text{med}}$ characterizing the RFI are precisely determined. Local RFI identification can be performed at the telescope stations to reduce the processing overhead at the supercomputers. Note that certain radio telescope stations can be located close to common sources of RFI to increase the discriminative information on the parameters $\ve{\theta}_{\text{med}}$ in the binary data stream.

\paragraph{Request-specific High-Performance Data Processing}
The temporal, spectral, and spatial parameters $\ve{\theta}_{\text{phy}}$ of the physical phenomena of interest are extracted from the binary measurements of all telescope stations. Locally calculated low-accuracy solutions from the stations help to initialize algorithms at the supercomputers. The extraction of $\ve{\theta}_{\text{phy}}$ is performed such that potential knowledge \eqref{prior:med:inst} about the instrument and the propagation medium is taken into account. As a result, variations in the instrument hardware and the propagation medium are in probabilistic radio astronomy natural parts of \eqref{definition:observation:model}. In the classical deterministic perspective on radio astronomy, RFI is considered disruptive such that measurement instruments are installed in remote radio quiet zones or contaminated measurements are discarded. In probabilistic radio telescope systems, each signal sample carries some information about the parameters $\ve{\theta}_{\text{phy}}$. Due to consistency, there are no processing artifacts caused by the measurement instruments, propagation effects, or RFI. The main concerns in hardware-aware probabilistic telescope system design are that, under limited resources, the physical-numerical model \eqref{definition:observation:model} of the signal acquisition apparatus is precise, identifiable, and highly informative for serving all potential science cases.
\subsection{Scientific, Technical, and Institutional Advantages}
Massive binary radio telescope architectures have several advantages on scientific, technical, and institutional level:

\paragraph{Low Investment/Operation Costs and Scaleability}
Due to the simple radio equipment and the focus on digital storage and data processing technologies, the initial investment costs are low. The number of telescope stations can be extended successively while digital infrastructure can be provided on demand by third parties. The operating costs for the telescope are small as energy dissipation at the front-ends is minimum and the digital components quickly gain efficiency through their fast technological progress. Besides, the radio instruments only need to be operated once to acquire the measurements for a large variety of astronomical surveys.

\paragraph{Fast and Flexible Astronomical Observations}
Due to the software and storage-based architecture, scientists have access to a highly sensitive and flexible instrument and can conduct complex surveys without waiting. Scientific progress is rather limited by mental and algorithmic capacity than by research funds. The short operating time of the observation instruments for a large number of science cases frees surveying resources, for example, for event-triggered measurement campaigns. The review overhead  for applications requesting individual observation time on the instrument is eliminated. 

\paragraph{Progressive and Open Science}
Due to the last aspect, unconventional research approaches of young scientists can be implemented without anticipated assessment by superior colleagues or delay due to institutional preferences. The stored radio measurement data streams enable end-to-end reproducibility and in-depth validation with alternative algorithmic approaches. For verification purposes, the layout of the individual telescope stations can be deliberately designed differently while arbitrary parts of the binary sensor data stream can be flexibly excluded from the actual investigation.

\paragraph{Technology Transfer and Interdisciplinary Dialog}
The proposed system architecture is in line with recent advances in binary sensing and data processing, which have mass-market applications, see, e.g., \cite{Madsen00,Ribeiro06,Dabeer06,Ivra07,Mezghani07,Kamilov12,Jacques13,SteinTheiler15,Mo15,Jacobsson15,Mollen17,Li17,SteinTabrikian18}. The research institutions involved in the development, implementation, and operation of a binary radio telescope find themselves at the forefront of this innovation. It is, therefore, attractive for them to educate and employ engineering talents to drive the technology transfer from scientific systems to consumer markets. Revenue from intellectual property licensing and technology consulting can be used to promote high-risk high-reward research in radio astronomy, sensor system engineering, and statistical data processing. Further, patent rights held by public institutions constitute a control mechanism for the responsible use of advanced sensor and data processing technologies. 

Establishing a functional massive binary telescope system represents a grand challenge. It can only be mastered in an open and solution-oriented dialogue taking place in protected research environments characterized by flat and encouraging hierarchies which are committed to ethical leadership in research and education. The discourse forces interdisciplinary cooperation between the complementary areas of physics, mathematics, computer science, and engineering science. Common understanding between these fields and a joint lighthouse project lead to innovative unified scientific approaches and advanced educational strategies while strengthening the work culture within the involved research institutions.
\subsection{Open Engineering Science Questions}
Binary telescope systems raise scientific and technical questions. It will be impossible to answer all of them by one research team. Everyone is invited to collaborate and contribute to an open and respectful discussion which is centered around scientific arguments and rests on the mutual interest to advance science and technology. To organize the discourse and ensure efficient allocation of research resources, we disclose the aspects which have our attention during the early studies:
\paragraph{High-Performance Binary Data Processing}
The main question is how to extract the spatial, spectral, and temporal information $\ve{\theta}_{\text{phy}}$ needed for the astronomical survey from the binary radio measurements. It can be doubted that binary measurement data still contains any information about $\ve{\theta}_{\text{phy}}$. In the second part of this article, we achieve a first step in the opposite direction. We show that the spatial power distribution can be recovered from the highly nonlinear $1$-bit radio measurements. To this end, we formulate the hardware-aware probabilistic model \eqref{binary:prob:model} under simplifying assumptions and outline a tractable estimation algorithm which consistently extracts the interferometric information from binary radio telescope data. That such a systematic approach captures well the effects relevant in practice, is corroborated by image reconstruction results obtained with $1$-bit LOFAR data.
\paragraph{Radio Frequency Identification and Mitigation}
Terrestrial RFI forms a challenge when observing weak electromagnetic emission from celestial objects. This is equally true for binary sensing architectures with a highly nonlinear and probabilistic response. The possibility to use a significantly larger number of radio sensors through the simplified equipment and having at hand a precise hardware-aware probabilistic model \eqref{definition:observation:model}, however, enables advanced spatial and nonlinear processing. The interferometric capabilities of binary arrays under RFI form open questions which we will address.
\paragraph{Cost and Performance Models}
Binary radio telescopes achieve digital signal acquisition at low complexity and generate sensor data streams with dense discriminative information which can be stored and processed efficiently. For determining the final design of a physical binary radio telescope, it is crucial to derive and verify models for the final output performance and the resource costs associated with different system levels. Through such analytic models, it becomes possible to mathematically optimize the telescope architecture with respect to the number of antennas, spectral bandwidths, sampling rates, and digital processing strategies.
\section{Binary Radio Interferometric Imaging}
Next, a first proof of concept for the technology vision of massive binary radio telescopes is provided. To this end, we derive an interferometric imaging technique which is capable of extracting the spatial power distribution from $1$-bit sensor measurements. We perform hardware-aware probabilistic modeling of the signal acquisition and the binary digitization process. Based on that, an iterative solution for the interferometric image reconstruction is derived by consistently approximating the MLE for a multivariate binary data model. For the derived algorithmic procedure, we give the achieved interferometric performance level \eqref{definition:error:mle} in a quantitative form.

\subsection{Analog Signal Acquisition Model}
A calibrated radio telescope with $M$ analog sensor signals  $\breve{y}_{m}(t)\in\fieldR,t\in\fieldR$ is considered. The $m$th output consist of a superposition of $D_{\text{S}}\in\fieldN$ wide-band radio sources $\breve{x}_{d}(t)\in\fieldR$ 
\begin{align}\label{def:analog:antenna:signal}
\breve{y}_{m}(t)=\sum_{d=1}^{D_{\text{S}}} \sqrt{\theta_d}\breve{x}_{d}(t-\tau_{d,m})+\breve{\eta}_{m}(t),
\end{align}
with individual signal strength parameter $\theta_d\in\fieldR, \theta_d>0$ and time-offset $\tau_{d,m}\in\fieldR$. The output of each sensor is distorted by independent additive measurement noise $\breve{\eta}_{m}(t) \in\fieldR$.
Each of the sources consists of two orthogonal components
\begin{align}\label{def:analog:source:signal}
\breve{x}_{d}(t)=\breve{x}_{\text{I},d}(t)\cosbr{{\omega}_d t} -\breve{x}_{\text{Q},d}(t)\sinbr{{\omega}_d t},
\end{align}
$\breve{x}_{\Ic,d}(t),\breve{x}_{\Qc,d}(t)\in\fieldR$, centered at frequency ${\omega}_d \in\fieldR$. Both signal components are considered to be zero-mean wide-sense stationary Gaussian processes. Here each of the $M$ received signals \eqref{def:analog:antenna:signal} is band-pass filtered with non-overlapping versions of the prototype $h(t; B, \omega) \in\fieldR$ to obtain $K$ channels (radio frequency binning), where the $k$th one is
\begin{align}\label{def:analog:bandpass}
{y}_{m}^{(k)}(t)&=\breve{y}_{m}(t) * h(t;B_{\text{Y}}, \omega^{(k)}),
\end{align}
${y}_{m}^{(k)}(t)\in\fieldR$, with one-sided bandwidth $B_{\text{Y}}$ and center frequency $\omega^{(k)} \in\fieldR$. Note, that $\breve{u}(t)$ stands for an analog wide-band signal while $u(t)$ denotes its narrow-band version after filtering. Each channel is divided into two real-valued signals
\begin{align}\label{def:demodulation}
{y}_{\Ic,m}^{(k)}(t)&=\phantom{-}{2} \cosbr{\omega^{(k)} t} \cdot {y}_{m}^{(k)}(t),\notag\\
{y}_{\Qc,m}^{(k)}(t)&=-{2} \sinbr{\omega^{(k)} t} \cdot {y}_{m}^{(k)}(t),
\end{align}
${y}_{\Ic,m}^{(k)}(t), {y}_{\Qc,m}^{(k)}(t)\in\fieldR$, by multiplying with sinusoids of frequency $\omega^{(k)}$ (demodulation). In the following, we omit the channel superscript $k$ and only consider the two outputs for one of the $K$ radio frequency bands. With narrow-band source components $x_{\Ic,d}(t),x_{\Qc,d}(t)\in\fieldR$ and a sufficiently small proliferation of the $M$ radio sensors, one can approximate
\begin{align}\label{narrow:band:assumption}
x_{\Ic/\Qc,d}(t-\tau_{d,m})&\approx x_{\Ic/\Qc,d}(t),\quad \forall d,m.
\end{align}
Therefore, after removing modulation products by low-pass filtering, the real-valued in-phase sensor output
\begin{align}\label{i:antenna:outputs}
&y_{\Ic,m}(t)=\eta_{\Ic,m}(t)\notag\\
&+\sum_{d=1}^{D} \sqrt{\theta_d} \Big(x_{\Ic,d}(t)\cosbr{\omega\tau_{d,m}} + x_{\Qc,d}(t)\sinbr{\omega\tau_{d,m}}\Big)
\end{align}
and the quadrature sensor output
\begin{align}\label{q:antenna:outputs}
&y_{\Qc,m}(t)=\eta_{\Qc,m}(t)\notag\\
&+\sum_{d=1}^{D} \sqrt{\theta_d} \Big(x_{\Qc,d}(t)\cosbr{\omega\tau_{d,m}}-x_{\Ic,d}(t)\sinbr{\omega \tau_{d,m}}\Big)
\end{align}
are obtained. In the literature covering radio front-ends with high-resolution A/D conversion, the real-valued analog outputs \eqref{i:antenna:outputs} and \eqref{q:antenna:outputs} are usually denoted by one complex-valued output. Note, that we use a real-valued notation to keep the framework applicable to different front-end architectures with low-resolution A/D conversion as outlined in \cite[Sec. VI]{SteinFauss19}.
\subsection{Data Model for Ideal All-Digital Radio Telescopes}
With $\infty$-bit A/D converters, a sample of all antennas
\begin{align}\label{digital:array:snapshot}
\ve{y}=\begin{bmatrix} y_{\Ic,1} &\ldots &y_{\Ic,M} &y_{\Qc,1} &\ldots &y_{\Qc,M} \end{bmatrix}^{\T}\in\fieldR^{2M}
\end{align}
has the structure
\begin{align}\label{digital:array:structure}
\ve{y}=\sum_{d=1}^{D_{\text{S}}} \sqrt{\theta_d} \ve{A}(\ve{\tau}_d)\ve{x}_d +\ve{\eta},
\end{align}
with the array steering matrix
\begin{align}\label{steering:matrix}
\ve{A}(\ve{\tau}_d)=\begin{bmatrix}\ve{A}^{\T}_\text{I}(\ve{\tau}_d) &\ve{A}^{\T}_\text{Q}(\ve{\tau}_d) \end{bmatrix}^{\T}\in\fieldR^{2M\times 2},
\end{align}
featuring the sub-matrices
\begin{align*}
\ve{A}_\text{I}(\ve{\tau}_d)=\begin{bmatrix}
\cosbr{\omega\tau_{d,1}} &\sinbr{\omega\tau_{d,1}}\\ 
\vdots &\vdots\\ 
\cosbr{\omega\tau_{d,M}} &\sinbr{\omega\tau_{d,M}}
\end{bmatrix}\in\fieldR^{M \times 2},
\end{align*}
and
\begin{align*}
\ve{A}_\text{Q}(\ve{\tau}_d)=\begin{bmatrix}
-\sinbr{\omega \tau_{d,1}} &\cosbr{\omega\tau_{d,1}}\\ 
\vdots &\vdots\\ 
-\sinbr{\omega \tau_{d,M}} &\cosbr{\omega\tau_{d,M}}
\end{bmatrix}\in\fieldR^{M \times 2},
\end{align*}
while the $d$th source's delay profile is denoted as
\begin{align}\label{delay:profile}
\ve{\tau}_d=\begin{bmatrix}\tau_{d,1} &\ldots &\tau_{d,M}\end{bmatrix}^{\T}\in\fieldR^{M},
\end{align}
the $d$th source signal as
\begin{align}\label{source:component:vector}
\ve{x}_d=\begin{bmatrix} x_{\Ic,d} & x_{\Qc,d}\end{bmatrix}^{\T}\in\fieldR^{2},
\end{align}
and the sensor noise samples as
\begin{align}
\ve{\eta}=\begin{bmatrix} \eta_{\Ic,1} &\ldots & \eta_{\Ic,M} & \eta_{\Qc,1} &\ldots &\eta_{\Qc,M}\end{bmatrix}^{\T}\in\fieldR^{2M}.
\end{align}
Assuming that the two Gaussian signals in \eqref{source:component:vector} are jointly independent and of unit variance, mutual antenna coupling can be neglected, ideal filters are employed and sampling is conducted at Nyquist rate ($f_{\text{Y}}=2B_{\text{Y}}$), the covariance matrix 
\begin{align}\label{covariance:unquantized}
\ve{R}_{\ve{y}}(\ve{\theta})&=\exdi{\ve{y};\ve{\theta}}{\ve{y}\ve{y}^{\T}},
\end{align}
$\ve{R}_{\ve{y}}(\ve{\theta})\in\fieldR^{2M\times 2M}$, of the digital sensor array signal \eqref{digital:array:structure} is
\begin{align}\label{definition:received:covariance}
\ve{R}_{\ve{y}}(\ve{\theta})&=\ve{R}_{\ve{x}}(\ve{\theta})+\ve{R}_{\ve{\eta}}(\ve{\theta})\notag\\
&=\sum_{d=1}^{D_{\text{S}}} \theta_d \ve{A}(\ve{\tau}_d)\ve{A}^{\T}(\ve{\tau}_d) + \theta_0\ve{I},
\end{align}
where $\theta_0\in\fieldR,\theta_0>0,$ denotes the power level of the spatially white noise. Assuming that the delay profile \eqref{delay:profile} for each source is known, the $D=D_{\text{S}}+1$ model parameters are
\begin{align}
\ve{\theta}=\begin{bmatrix} \theta_0 &\theta_1 &\ldots &\theta_{D_{\text{S}}}\end{bmatrix}^{\T}\in\fieldR^{D}.
\end{align}
For notational convenience, we also introduce a vectorized version of the asymptotic receive covariance matrix \eqref{definition:received:covariance}
\begin{align}\label{vectorized:covariance}
\ve{r}_{\ve{y}}(\ve{\theta})=\vec{\ve{R}_{\ve{y}}(\ve{\theta})}=\ve{M}\ve{\theta},
\end{align}
$\ve{r}_{\ve{y}}(\ve{\theta})\in\fieldR^{4M^2}$, with the steering matrix 
\begin{align}
\ve{M}&=\begin{bmatrix} \vec{\ve{I}} &\ve{a}(\ve{\tau}_1) &\ldots &\ve{a}(\ve{\tau}_{D_{\text{S}}}) \end{bmatrix}\in\fieldR^{4M^2 \times D},
\end{align}
containing the array steering vectors
\begin{align}
\ve{a}(\ve{\tau}_d)=\vec{\ve{A}(\ve{\tau}_d)\ve{A}^{\T}(\ve{\tau}_d)}.
\end{align}
During signal acquisition $N\in\fieldN$ samples \eqref{digital:array:snapshot} are collected
\begin{align}\label{data:matrix}
\ve{Y}=\begin{bmatrix} \ve{y}_1 &\ve{y}_2 &\ldots &\ve{y}_N\end{bmatrix} \in \fieldR^{2M \times N},
\end{align}
which independently follow the Gaussian distribution
\begin{align}\label{multivariate:gauss:covariance:parameter}
p_{\ve{y}}(\ve{y};\ve{\theta})\sim\frac{\exp{-\frac{1}{2} \ve{y}^{\T} \ve{R}_{\ve{y}}^{-1}(\ve{\theta}) \ve{y}}}{ \sqrt{(2\pi)^{2M} \det{(\ve{R}_{\ve{y}}(\ve{\theta}))}} }.
\end{align}
\subsection{Statistical Interferometric Imaging with $\infty$-bit Data}
With the data \eqref{data:matrix} and the model \eqref{multivariate:gauss:covariance:parameter}, the MLE is
\begin{align}\label{definition:mle:gaussian}
\ve{\hat{\theta}}(\ve{Y}) &= \arg \max_{\ve{\theta}\in\ve{\Theta}} \,\, \ln p_{\ve{Y}}(\ve{Y};\ve{\theta})\\
&= \arg \max_{\ve{\theta}\in\ve{\Theta}} \sum_{n=1}^{N} \ln p_{\ve{y}}(\ve{y}_n;\ve{\theta})\notag\\
&= \arg \min_{\ve{\theta}\in\ve{\Theta}} \,\,  \ln \big(\det(\ve{R}_{\ve{y}}(\ve{\theta}))\big) + \Tr{\big(\ve{\hat{R}}_{\ve{y}}(\ve{Y})\ve{R}_{\ve{y}}^{-1}(\ve{\theta}) \big)},\notag
\end{align}
with the empirical receive covariance matrix
\begin{align}\label{definition:empirical:cov:y}
\ve{\hat{R}_{\ve{y}}}(\ve{Y}) = \frac{1}{N} \sum_{n=1}^{N} \ve{y}_n\ve{y}_n^{\T},
\end{align}
$\ve{\hat{R}_{\ve{y}}}(\ve{Y})\in\fieldR^{2M\times 2M}$. The solution of the optimization problem \eqref{definition:mle:gaussian}, is the root of the score of the Gaussian model \eqref{multivariate:gauss:covariance:parameter}, which can be found by Fisher's scoring method \cite{Mak93}. With this approach, in the $i$th of $I\in\fieldN$ iterations, one calculates
\begin{align}\label{update:scoring:rule}
\ve{\hat{\theta}}^{(i)}&=\ve{\hat{\theta}}^{(i-1)}+\Delta \ve{\hat{\theta}}(\ve{Y};\ve{\hat{\theta}}^{(i-1)}),
\end{align}
with the update term
\begin{align}\label{update:scoring:many:bits}
\Delta \ve{\hat{\theta}}(\ve{Y};\ve{\theta})&=\ve{F}_{\ve{Y}}^{-1}\big(\ve{\theta}\big) \ve{s}_{\ve{Y}}(\ve{Y};\ve{\theta}) 
\end{align}
being characterized by the Fisher information matrix 
\begin{align}\label{def:fisher:matrix}
\ve{F}_{\ve{Y}}\big(\ve{\theta}\big)&=N \exdi{\ve{y};\ve{\theta}}{ \ve{s}_{\ve{y}}(\ve{y};\ve{\theta}) \ve{s}^{\T}_{\ve{y}}(\ve{y};\ve{\theta}) }\notag\\
&=N\ve{F}_{\ve{y}}\big(\ve{\theta}\big),
\end{align}
where the single-sample score is
\begin{align}
\ve{s}_{\ve{y}}(\ve{y};\ve{\theta})=\bigg( \frac{ \partial \ln p_{\ve{y}}(\ve{y};\ve{\theta})  }{\partial \ve{\theta}} \bigg)^{\T},
\end{align}
and, due to independency, the $N$-sample score writes
\begin{align}
\ve{s}_{\ve{Y}}(\ve{Y};\ve{\theta}) &= \sum_{n=1}^{N}  \ve{s}_{\ve{y}}(\ve{y}_n;\ve{\theta}).
\end{align}
The results of \eqref{update:scoring:rule} converge quickly to the solution of the MLE \eqref{definition:mle:gaussian} while back-projection after each iteration \eqref{update:scoring:rule}
\begin{align}\label{back:projection}
\ve{\hat{\theta}}^{(i)}=\operatorname{max}(\theta_\Delta,\ve{\hat{\theta}}^{(i)}),
\end{align}
with $\theta_\Delta>0$, ensures that the estimates satisfy $\ve{\hat{\theta}}^{(i)}\in \ve{\Theta}$.

The Fisher information matrix \eqref{definition:fisher:matrix} of the multivariate Gaussian model \eqref{multivariate:gauss:covariance:parameter} with the covariance structure \eqref{definition:received:covariance} is
\begin{align}\label{fisher:many:bits}
\ve{F}_{\ve{y}}(\ve{\theta})&=\frac{1}{2}\ve{M}^{\T} \big(\ve{R}^{-1}_{\ve{y}}(\ve{\theta}) \otimes \ve{R}^{-1}_{\ve{y}}(\ve{\theta}) \big)\ve{M}
\end{align}
and the score function with $N$ samples
\begin{align}\label{score:many:bits}
\ve{s}_{\ve{Y}}(\ve{Y};\ve{\theta})=\frac{N}{2} \ve{M}^{\T} \big(\ve{R}^{-1}_{\ve{y}}(\ve{\theta}) \otimes \ve{R}^{-1}_{\ve{y}}(\ve{\theta}) \big)  \big(\ve{\hat{r}}_{\ve{y}}(\ve{Y}) - \ve{M}\ve{\theta}\big),
\end{align}
where we used 
\begin{align}
\frac{\partial \ln \det(\ve{R}_{\ve{y}}(\ve{\theta}))}{\partial \ve{\theta} }&= \ve{\theta}^{\T} \ve{M}^{\T} \big(\ve{R}^{-1}_{\ve{y}}(\ve{\theta}) \otimes \ve{R}^{-1}_{\ve{y}}(\ve{\theta}) \big) \ve{M},\\
\frac{\partial \Tr{\big(\ve{\hat{R}}_{\ve{y}}(\ve{Y})\ve{R}_{\ve{y}}^{-1}(\ve{\theta}) \big)}}{\partial \ve{\theta} }&=- \ve{\hat{r}}_{\ve{y}}^{\T}(\ve{Y}) \big(\ve{R}^{-1}_{\ve{y}}(\ve{\theta}) \otimes \ve{R}^{-1}_{\ve{y}}(\ve{\theta}) \big)\ve{M},
\end{align}
under the definition
\begin{align}
\ve{\hat{r}}_{\ve{y}}(\ve{Y})=\vec{\ve{\hat{R}}_{\ve{y}}(\ve{Y})}.
\end{align}
Therefore, substituting \eqref{def:fisher:matrix} with \eqref{fisher:many:bits} and using it together with \eqref{score:many:bits} in the update expression \eqref{update:scoring:many:bits}, results in
\begin{align}\label{scoring:update:substitution}
\Delta \ve{\hat{\theta}}(\ve{Y};\ve{\theta})&= \big(\ve{M}^{\T} \big(\ve{R}^{-1}_{\ve{y}}(\ve{\theta}) \otimes \ve{R}^{-1}_{\ve{y}}(\ve{\theta}) \big)\ve{M}\big)^{-1}\\
&\phantom{=}\cdot\ve{M}^{\T} \big(\ve{R}^{-1}_{\ve{y}}(\ve{\theta}) \otimes \ve{R}^{-1}_{\ve{y}}(\ve{\theta}) \big)\big(\ve{\hat{r}}_{\ve{y}}(\ve{Y}) - \ve{M}\ve{\theta}\big).\notag
\end{align}
\subsection{Data Model for Binary Radio Telescopes}
In the following, we consider the hard-limited version of the radio array measurements \eqref{digital:array:snapshot}
\begin{align}\label{hard:limiter}
\ve{z} = \sign{\ve{y}},
\end{align}
where $\sign{\cdot}$ is the element-wise signum function
\begin{align}
\left[\sign{\ve{y}}\right]_i=
\begin{cases}
+1& \text{if } [\ve{y}]_i \geq 0,\\
-1 & \text{if } [\ve{y}]_i < 0.
\end{cases}
\end{align}
The resulting dataset 
\begin{align}\label{one:bit:dataset}
\ve{Z}=\begin{bmatrix} \ve{z}_1 &\ve{z}_2 &\ldots &\ve{z}_N\end{bmatrix} \in \fieldB^{2M \times N}
\end{align}
mimics the measurements of a radio telescope which employs low-complexity $1$-bit analog-to-digital conversion. As the variance of the individual inputs to the quantizer \eqref{hard:limiter} can not be resolved from its outputs, one can reduce the parameter vector
\begin{align}
\ve{\theta}=\begin{bmatrix} \theta_1 &\ldots &\theta_{D_\text{S}}\end{bmatrix}^{\T}\in\fieldR^{D_\text{S}},
\end{align}
and set the sensor noise variance to a fixed value, e.g., $\theta_0=1$. This removes the problem of having to estimate the noise level. Consequently, the receive covariance \eqref{covariance:unquantized} takes the structure
\begin{align}\label{covariance:unquantized:reduced}
\ve{R}_{\ve{y}}(\ve{\theta})=\sum_{d=1}^{D_\text{S}} \theta_d \ve{A}(\ve{\tau}_d)\ve{A}^{\T}(\ve{\tau}_d) + \ve{I}.
\end{align}
For convenience, we define a receive correlation matrix
\begin{align}\label{definition:receive:corr}
\ve{\Sigma}_{\ve{y}}(\ve{\theta})&=\diag{\ve{R}_{\ve{y}}(\ve{\theta})}^{-\frac{1}{2}}\ve{R}_{\ve{y}}(\ve{\theta})\diag{\ve{R}_{\ve{y}}(\ve{\theta})}^{-\frac{1}{2}}\notag\\
&=  \frac{1}{\sum_{d=1}^{D_\text{S}} \theta_d + 1} \Bigg( \sum_{d=1}^{D_\text{S}} \theta_d \ve{A}(\ve{\tau}_d)\ve{A}^{\T}(\ve{\tau}_d) + \ve{I} \Bigg),
\end{align}
$\ve{\Sigma}_{\ve{y}}(\ve{\theta})\in\fieldR^{2M \times 2M }$, where, for a square matrix $\ve{U}$, $\diag{\ve{U}}$ 
returns the matrix with off-diagonal values switched to zero.
\subsection{Statistical Interferometric Imaging with $1$-bit Data}
For performing interferometric imaging with the $1$-bit sensor data \eqref{one:bit:dataset} and scoring \eqref{update:scoring:rule}, the Fisher information matrix 
\begin{align}\label{definition:fisher:matrix:quantized}
&\ve{F}_{\ve{z}}\big(\ve{\theta}\big)= \exdi{\ve{z}}{ \ve{s}_{\ve{z}}(\ve{z};\ve{\theta}) \ve{s}^{\T}_{\ve{z}}(\ve{z};\ve{\theta}) }
\end{align}
and the score function 
\begin{align}\label{binary:score}
\ve{s}_{\ve{z}}(\ve{z};\ve{\theta}) = \bigg(\frac{ \partial \ln p_{\ve{z}}(\ve{z};\ve{\theta})  }{\partial \ve{\theta}}\bigg)^{\T}
\end{align}
of the multivariate binary model $p_{\ve{z}}(\ve{z};\ve{\theta})$, characterizing the output data of the hard-limiter \eqref{hard:limiter}, are required. Providing exact formulations is computationally intractable as the multivariate binary distribution, in general, exhibits a number of sufficient statistics growing as $\mathcal{O}(2^{2M})$ \cite{Dai13}. Additionally, the missing analytic characterization for the orthant probabilities of multivariate Gaussian distributions with more than four variables renders the multivariate binary likelihood $p_{\ve{z}}(\ve{z};\ve{\theta})$ mathematically inaccessible. A conceptual observation that turns out helpful, is that distributions of binary signals belong to the exponential family. Like multivariate Gaussian data \eqref{multivariate:gauss:covariance:parameter}, their probabilistic characterization admits a factorization
\begin{align}\label{definition:exponential:family}
{p}_{\ve{z}}(\ve{z};\ve{\theta})=\exp{\ve{\beta}^{\T}(\ve{\theta}) \ve{\phi}(\ve{z}) - \lambda(\ve{\theta})+\nu(\ve{z})},
\end{align}
where $\ve{\beta}(\ve{\theta})\colon \fieldR^{D} \to \fieldR^{C}$ are the statistical weights\footnote{In the literature one finds the term ``natural parameters''. Trying to distinguish between components of the probabilistic data model and the physical system parameters $\ve{\theta}$, we here use the terminology ``statistical weights''.}, $\ve{\phi}(\ve{z})\colon \fieldB^{2M} \to\fieldR^{C}$ the sufficient statistics, $\lambda(\ve{\theta})\colon \fieldR^D \to \fieldR$ the log-normalizer and $\nu(\ve{z})\colon \fieldB^{2M} \to\fieldR$ the carrier measure. This can be exploited by formulating an auxiliary model of the form \eqref{definition:exponential:family} for which the number $\tilde{C}$ and form of the statistics 
\begin{align}\label{definition:aux:stat}
\ve{\tilde{\phi}}(\ve{z}): \fieldB^{2M} \to \fieldR^{\tilde{C}}
\end{align}
are user-defined. On these specific statistics the auxiliary model $\tilde{p}_{\ve{z}}(\ve{z};\ve{\theta})$ features the same mean and covariance matrix
\begin{align}\label{definition:mean:transformed}
\ve{\mu}_{\ve{\tilde{\phi}}}(\ve{\theta})&=\exdi{\ve{\tilde{z}};\ve{\theta}}{  \ve{\tilde{\phi}}(\ve{z}) },\\
\label{definition:covariance:transformed}
\ve{R}_{\ve{\tilde{\phi}}}(\ve{\theta}) &=\exdi{\ve{\tilde{z}};\ve{\theta}}{\big(\ve{\tilde{\phi}}(\ve{z})-\ve{\mu}_{\ve{\tilde{\phi}}}(\ve{\theta})\big) \big(\ve{\tilde{\phi}}(\ve{z})-\ve{\mu}_{\ve{\tilde{\phi}}}(\ve{\theta})\big)^{\T} },
\end{align}
$\ve{\mu}_{\ve{\tilde{\phi}}}(\ve{\theta}) \in\fieldR^{\tilde{C}}, \ve{R}_{\ve{\tilde{\phi}}}(\ve{\theta})\in\fieldR^{\tilde{C} \times \tilde{C}}$, like under the exact model $p_{\ve{z}}(\ve{z};\ve{\theta})$. Such an auxiliary likelihood, admits the score \cite{Stein15}
\begin{align}\label{conservative:score}
\ve{\tilde{s}}_{\ve{z}}(\ve{z};\ve{\theta}) = \bigg( \frac{\partial \ve{\mu}_{\ve{\tilde{\phi}}}(\ve{\theta})}{ \partial \ve{\theta}} \bigg)^{\T} \ve{R}_{\ve{\tilde{\phi}}}^{-1}(\ve{\theta})\big(\ve{\tilde{\phi}}(\ve{z}) - \ve{\mu}_{\ve{\tilde{\phi}}}(\ve{\theta})\big)
\end{align}
and the Fisher information matrix
\begin{align}\label{pessimistic:fisher:matrix}
\ve{\tilde{F}}_{\ve{z}}(\ve{\theta}) = \bigg(\frac{\partial \ve{\mu}_{\ve{\tilde{\phi}}}(\ve{\theta})}{ \partial \ve{\theta}} \bigg)^{\T} \ve{R}_{\ve{\tilde{\phi}}}^{-1}(\ve{\theta}) \frac{\partial \ve{\mu}_{\ve{\tilde{\phi}}}(\ve{\theta})}{ \partial \ve{\theta}}.
\end{align}
Therefore, through a suitable choice of \eqref{definition:aux:stat}, it becomes possible to control statistical complexity and perform iterations
\begin{align}\label{scoring:rule:quant}
\ve{\hat{\theta}}^{(i)}&=\ve{\hat{\theta}}^{(i-1)}+\Delta \ve{\hat{\theta}}(\ve{Z};\ve{\hat{\theta}}^{(i-1)}),
\end{align}
each followed by back-projection \eqref{back:projection}, with the update term
\begin{align}\label{scoring:update:quant}
\Delta \ve{\hat{\theta}}(\ve{Z};\ve{\theta})&= \ve{\tilde{F}}^{-1}_{\ve{Z}}(\ve{\theta}) \ve{\tilde{s}}_{\ve{Z}}(\ve{Z};\ve{\theta})\\
&= \Bigg( \bigg(\frac{\partial \ve{\mu}_{\ve{\tilde{\phi}}}(\ve{\theta})}{ \partial \ve{\theta}} \bigg)^{\T} \ve{R}_{\ve{\tilde{\phi}}}^{-1}(\ve{\theta}) \frac{\partial \ve{\mu}_{\ve{\tilde{\phi}}}(\ve{\theta})}{ \partial \ve{\theta}} \Bigg)^{-1}\notag\\
&\phantom{=}\cdot \bigg( \frac{\partial \ve{\mu}_{\ve{\tilde{\phi}}}(\ve{\theta})}{ \partial \ve{\theta}} \bigg)^{\T} \ve{R}_{\ve{\tilde{\phi}}}^{-1}(\ve{\theta})\big(\ve{\hat{\mu}}_{\ve{\tilde{\phi}}}(\ve{Z}) - \ve{\mu}_{\ve{\tilde{\phi}}}(\ve{\theta})\big),\notag
\end{align}
where the empirical mean of \eqref{definition:aux:stat} is defined by
\begin{align}\label{empirical:auxiliary:statistics}
\ve{\hat{\mu}}_{\ve{\tilde{\phi}}}(\ve{Z})=\frac{1}{N}\sum_{k=1}^{N}\ve{\tilde{\phi}}(\ve{z}_n),
\end{align}
$\ve{\hat{\mu}}_{\ve{\tilde{\phi}}}(\ve{Z})\in\fieldR^{\tilde{C}}$. The iterations \eqref{scoring:rule:quant} converge to a possible solution of the maximization problem \eqref{definition:mle} defined on the basis of the auxiliary likelihood $\tilde{p}_{\ve{Z}}(\ve{Z};\ve{\theta})$. If this interferometric imaging solution is the global maximum of the auxiliary likelihood it is guaranteed to be consistent and for a sufficiently large number of independent array snapshots $N$ reaches \cite{Stein15}
\begin{align}\label{definition:error:mle:quant}
\exdi{\ve{Z};\ve{\theta}_t}{ \big(\ve{\hat{\theta}}(\ve{Z})-\ve{\theta}_t\big) \big(\ve{\hat{\theta}}(\ve{Z})-\ve{\theta}_t\big)^{\rm{T}} } &=  \ve{\tilde{F}}_{\ve{Z}}^{-1}(\ve{\theta}_t)\notag\\
&=\frac{1}{N} \ve{\tilde{F}}_{\ve{z}}^{-1}(\ve{\theta}_t).
\end{align}
Due to the conservative nature of the information matrix \eqref{pessimistic:fisher:matrix}
\begin{align}
\ve{F}_{\ve{z}}(\ve{\theta})\succeq \ve{\tilde{F}}_{\ve{z}}(\ve{\theta}),\quad \forall \ve{\theta},
\end{align}
equation \eqref{definition:error:mle:quant} can not overestimate the interferometric capabilities with the exact binary observation data model ${p}_{\ve{Z}}(\ve{Z};\ve{\theta})$ while the expression \eqref{pessimistic:fisher:matrix} enables to quantitatively determine the performance achieved with an auxiliary MLE algorithm.

For statistical array processing with hard-limited zero-mean Gaussian sensor signals, the particular choice \cite{SteinWSA16}
\begin{align}\label{aux:stat:quant}
\ve{\tilde{\phi}}(\ve{z})&=\ve{\Phi}\vect{\ve{z} \ve{z}^{\T}},
\end{align}
with $\ve{\Phi}\in\{ 0, 1\}^{\tilde{C} \times 4M^2}$ being an elimination matrix canceling duplicate and constant statistics on the outer product $\ve{z} \ve{z}^{\T}$, leads to analytically tractable expressions for \eqref{definition:mean:transformed} and \eqref{definition:covariance:transformed} while complexity ($\tilde{C}=2M^2-M$) grows quadratic with the number of antennas. The empirical statistics are then given by
\begin{align}\label{empirical:auxiliary:statistics:quant}
\ve{\hat{\mu}}_{\ve{\tilde{\phi}}}(\ve{Z})=\frac{1}{N}\ve{\Phi}\sum_{k=1}^{N}\vect{\ve{z}_n \ve{z}_n^{\T}}
\end{align}
and imply to pairwise correlate all binary sensor outputs with each other. Defining the covariance of the quantized data
\begin{align}\label{covariance:quantized}
\ve{R}_{\ve{z}}(\ve{\theta})&=\exdi{\ve{z}}{ \ve{z}\ve{z}^{\T} },
\end{align} 
$\ve{R}_{\ve{z}}(\ve{\theta})\in\fieldR^{2M \times 2M}$, and using the arc-sine law \cite[p. 284]{Thomas69}
\begin{align}\label{arcsin:law}
\ve{R}_{\ve{z}}(\ve{\theta})&=\frac{2}{\pi} \arcsine{ \ve{\Sigma}_{\ve{y}}(\ve{\theta}) },
\end{align} 
the mean of the statistics \eqref{aux:stat:quant} is
\begin{align}\label{mean:transformed:quant}
\ve{\mu}_{\ve{\tilde{\phi}}}(\ve{\theta})
&= \ve{\Phi}\vect{ \ve{R}_{\ve{z}}(\ve{\theta})}\notag\\
&= \frac{2}{\pi} \ve{\Phi}\vect{ \arcsine{ \ve{\Sigma}_{\ve{y}}(\ve{\theta}) } }.
\end{align}
Further, its derivative
\begin{align}
\frac{\partial \ve{\mu}_{\ve{\tilde{\phi}}}(\ve{\theta})}{\partial \ve{\theta}} = \ve{\Phi}\frac{\partial \vect{ \ve{R}_{\ve{z}}(\ve{\theta}) }}{\partial \ve{\theta}}
\end{align}
is found by considering that the $d$th column is
\begin{align}
\left[ \frac{\partial \ve{\mu}_{\ve{\tilde{\phi}}}(\ve{\theta})}{\partial \ve{\theta}} \right]_{d} &= \ve{\Phi}\frac{\partial \ve{\mu}_{\ve{\tilde{\phi}}}(\ve{\theta})}{\partial \theta_d}=\ve{\Phi} \vect{ \frac{\partial \ve{R}_{\ve{z}}(\ve{\theta}) }{\partial \theta_d}},
\end{align}
with the matrix entries
\begin{align}\label{quantizer:covariance:derivative:entries}
&\left[\frac{\partial \ve{R}_{\ve{z}}(\ve{\theta}) }{\partial \theta_{d} }\right]_{ij}=\frac{2}{\pi} \frac{ \left[\frac{\partial \ve{\Sigma}_{\ve{y}}(\ve{\theta}) }{\partial \theta_{d} }\right]_{ij} }{ \sqrt{ 1 - \left[ \ve{\Sigma}_{\ve{y}}(\ve{\theta})  \right]^2_{ij}} }, \quad \forall i,j: i\neq j,
\end{align}
while the diagonal entries vanish, i.e.,
\begin{align}\label{quantizer:covariance:derivative:entries:diagonal}
\left[\frac{\partial \ve{R}_{\ve{z}}(\ve{\theta}) }{\partial \theta_{d}  }\right]_{ij}&=0, \quad \forall i,j: i= j.
\end{align}
With the receive correlation matrix \eqref{definition:receive:corr}, we obtain
\begin{align}
\left[\frac{\partial \ve{\Sigma}_{\ve{y}}(\ve{\theta}) }{\partial \theta_{d} }\right]_{ij}&=\frac{ \left[ \ve{A}(\ve{\tau}_d)\ve{A}^{\T}(\ve{\tau}_d) \right]_{ij}}{\sum_{d'=1}^{D_\text{S}} \theta_{d'} + 1} \notag\\
&- \frac{ \left[ \sum_{d'=1}^{D_\text{S}} \theta_{d'} \ve{A}(\ve{\tau}_{d'})\ve{A}^{\T}(\ve{\tau}_{d'}) +\ve{I}\right]_{ij}}{(\sum_{d'=1}^{D_\text{S}} \theta_{d'} + 1)^2}.
\end{align}
The numerical evaluation of the covariance matrix \eqref{definition:covariance:transformed} with the auxiliary statistics \eqref{aux:stat:quant} is achieved \cite{SteinWSA16} by using the arc-sine law \eqref{arcsin:law} and the quadrivariate orthant probabilities \cite{Sinn11}.

\section{Radio Image Reconstruction with LOFAR Data}
To evaluate and visualize the potential of the derived binary interferometric method under real-word conditions, we use data from the LOFAR radio telescope. The goal is to produce an all-sky radio image with a height and width of $\num{100}$ pixels which translates to $D_\text{S}=7668$ sources\footnote{Note that not all pixels constitute physically feasible steering directions.}. With the considered array configurations such a setting leads to a non-identifiable model \eqref{definition:observation:model}. Therefore, we solve the interferometric image reconstruction by using a smaller number of signal sources and generating the final all-sky image by linear interpolation.

\subsection{LOFAR Dataset 1 - CS$302$ Station}
\begin{figure}[!ht]
\centering
	\setlength{\figurewidth}{5.5cm}
	\setlength{\figureheight}{5cm}
%
%
\begin{tikzpicture}[scale=0.75]

\begin{axis}[%
width=\figurewidth,
height=\figureheight,
at={(1.011in,0.642in)},
scale only axis,
xmin=-50,
xmax=40,
xlabel style={font=\color{white!15!black}},
xlabel={Sensor x-Position [m]},
ymin=-50,
ymax=40,
ylabel style={font=\color{white!15!black}},
ylabel={Sensor y-Position [m]},
axis background/.style={fill=white},
axis x line*=bottom,
axis y line*=left,
legend style={legend cell align=left, align=left, draw=white!15!black}
]
\addplot[only marks, mark=asterisk, mark options={}, mark size=1.5000pt, draw=black] table[row sep=crcr]{%
x	y\\
14.6975100083695	-10.3112119530929\\
15.2324280320102	-13.3384962655803\\
1.37741620221628	-20.345190531294\\
-13.6077132094751	-14.4105453313673\\
-17.2812940137403	-11.1227082507728\\
-18.3045829399152	0.971327523619987\\
-12.4294714674814	12.8865970984015\\
-7.39950543997608	19.0579771148748\\
6.83681946539114	16.7231038616779\\
14.4164643726426	13.6659314857176\\
18.1919180339377	4.45837410891769\\
24.8398623594574	-0.328176289205203\\
21.4102810243986	-12.8819952686502\\
12.4431424330835	-22.123496434284\\
6.72580832682837	-24.7048198832129\\
-1.47610970731153	-26.6866275410678\\
-16.4696506247377	-17.4015989097884\\
-21.6665203208869	-11.3509080631208\\
-20.3494069234606	14.7715453190226\\
-14.2514501439615	21.805490165724\\
-2.86053345892769	24.6163292633981\\
13.2238276454007	23.568112453304\\
24.9930847632429	10.2892950808388\\
34.640587683886	0.399676662455009\\
28.4843372299849	-13.1724861115445\\
19.3574681191534	-24.588884199835\\
13.303883505308	-29.8302393255355\\
-5.55291599073894	-31.0939317880264\\
-22.720168619898	-25.5199135637687\\
-30.2130245147472	-15.7749496983766\\
-29.2977268361448	-6.4513449353637\\
-33.3389194423975	7.63121882148774\\
-23.8627666936831	24.1178584533735\\
-16.6337905937038	29.4218290537137\\
-1.5569086211578	34.6478219912408\\
6.14337034428121	31.745817870644\\
24.5524139076066	22.4818286501937\\
26.1853938130046	15.9668560921377\\
32.2381309898725	17.636297299468\\
35.8771738257321	-21.4007165985872\\
15.4445635485435	-36.9111948538705\\
-4.02828590145351	-41.7859081402734\\
-18.8856028372713	-35.6214309367523\\
-41.0074356969125	3.64376118925477\\
-28.8637727324209	22.439329005392\\
-20.0852797104074	36.2568945414694\\
22.8028631625156	31.3686435992984\\
};

\end{axis}
\end{tikzpicture}%
	\caption{LOFAR - CS$302$ Station ($M=47$)}
	\label{fig:LOFAR:cs302:pos}
\end{figure}
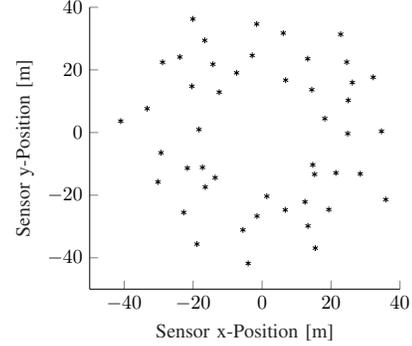

The first dataset was obtained with the LOFAR station CS$302$ and $M=47$ active single-polarized LBA antennas. Fig. \ref{fig:LOFAR:cs302:pos} depicts the placement of the radio sensors on the plane. The time-delay profile \eqref{delay:profile} associated with sources from different directions can be deduced from the positions of the antennas, such that the sensor locations define the steering matrices \eqref{steering:matrix}.
The LOFAR sensor data streams were obtained by sampling the wide-band signal \eqref{def:analog:antenna:signal} at each element with a $12$-bit A/D converter and using demodulation, filtering, and decimation to implement \eqref{def:analog:bandpass} and \eqref{def:demodulation} in the digital domain. Additionally, calibration operations were applied such that the radio measurements have $64$-bit floating-point precision. The considered data can be interpreted as the digital sensor outputs \eqref{digital:array:snapshot} of an ideal narrow-band radio telescope with a two-sided receive bandwidth of $2B_\text{Y} = \SI{763}{\hertz}$, centered at  approximately $\SI{44.48}{\mega\hertz}$, and zero-IF front-ends where at each single-polarized antenna two A/D converters with $64$-bit resolution sample the in-phase \eqref{i:antenna:outputs} and quadrature outputs \eqref{q:antenna:outputs} at Nyquist rate. The data for each radio sensor consists of two real-valued channels with $N = \num{58073}$ samples which corresponds to an observation duration of $T=\SI{76}{\second}$. For back-projection \eqref{back:projection}, we set $\theta_\Delta=\SI{-80}{\decibel}$ while using the current noise estimate $\hat{\theta}_0^{(i)}$ as the power level reference.

\begin{figure}
\centering
    \begin{subfigure}[t]{0.24\textwidth}
    \centering
	\setlength{\figurewidth}{5.1cm}
	\setlength{\figureheight}{5.1cm}
%
%
\begin{tikzpicture}[scale=0.55]

\begin{axis}[%
width=\figurewidth,
height=\figureheight,
at={(0.92in,0.642in)},
scale only axis,
point meta min=1e-08,
point meta max=0.0401068360401745,
axis on top,
xmin=-1.00200400801603,
xmax=1.00200400801603,
xlabel style={font=\color{white!15!black}},
xlabel={$\text{South }\leftarrow\text{ Direction }\rightarrow\text{ North}$},
y dir=reverse,
ymin=-1.00200400801603,
ymax=1.00200400801603,
ylabel style={font=\color{white!15!black}},
ylabel={$\text{East }\leftarrow\text{ Direction }\rightarrow\text{ West}$},
axis background/.style={fill=white},
legend style={legend cell align=left, align=left, draw=white!15!black},
colormap={mymap}{[1pt] rgb(0pt)=(0.2422,0.1504,0.6603); rgb(1pt)=(0.25039,0.164995,0.707614); rgb(2pt)=(0.257771,0.181781,0.751138); rgb(3pt)=(0.264729,0.197757,0.795214); rgb(4pt)=(0.270648,0.214676,0.836371); rgb(5pt)=(0.275114,0.234238,0.870986); rgb(6pt)=(0.2783,0.255871,0.899071); rgb(7pt)=(0.280333,0.278233,0.9221); rgb(8pt)=(0.281338,0.300595,0.941376); rgb(9pt)=(0.281014,0.322757,0.957886); rgb(10pt)=(0.279467,0.344671,0.971676); rgb(11pt)=(0.275971,0.366681,0.982905); rgb(12pt)=(0.269914,0.3892,0.9906); rgb(13pt)=(0.260243,0.412329,0.995157); rgb(14pt)=(0.244033,0.435833,0.998833); rgb(15pt)=(0.220643,0.460257,0.997286); rgb(16pt)=(0.196333,0.484719,0.989152); rgb(17pt)=(0.183405,0.507371,0.979795); rgb(18pt)=(0.178643,0.528857,0.968157); rgb(19pt)=(0.176438,0.549905,0.952019); rgb(20pt)=(0.168743,0.570262,0.935871); rgb(21pt)=(0.154,0.5902,0.9218); rgb(22pt)=(0.146029,0.609119,0.907857); rgb(23pt)=(0.138024,0.627629,0.89729); rgb(24pt)=(0.124814,0.645929,0.888343); rgb(25pt)=(0.111252,0.6635,0.876314); rgb(26pt)=(0.0952095,0.679829,0.859781); rgb(27pt)=(0.0688714,0.694771,0.839357); rgb(28pt)=(0.0296667,0.708167,0.816333); rgb(29pt)=(0.00357143,0.720267,0.7917); rgb(30pt)=(0.00665714,0.731214,0.766014); rgb(31pt)=(0.0433286,0.741095,0.73941); rgb(32pt)=(0.0963952,0.75,0.712038); rgb(33pt)=(0.140771,0.7584,0.684157); rgb(34pt)=(0.1717,0.766962,0.655443); rgb(35pt)=(0.193767,0.775767,0.6251); rgb(36pt)=(0.216086,0.7843,0.5923); rgb(37pt)=(0.246957,0.791795,0.556743); rgb(38pt)=(0.290614,0.79729,0.518829); rgb(39pt)=(0.340643,0.8008,0.478857); rgb(40pt)=(0.3909,0.802871,0.435448); rgb(41pt)=(0.445629,0.802419,0.390919); rgb(42pt)=(0.5044,0.7993,0.348); rgb(43pt)=(0.561562,0.794233,0.304481); rgb(44pt)=(0.617395,0.787619,0.261238); rgb(45pt)=(0.671986,0.779271,0.2227); rgb(46pt)=(0.7242,0.769843,0.191029); rgb(47pt)=(0.773833,0.759805,0.16461); rgb(48pt)=(0.820314,0.749814,0.153529); rgb(49pt)=(0.863433,0.7406,0.159633); rgb(50pt)=(0.903543,0.733029,0.177414); rgb(51pt)=(0.939257,0.728786,0.209957); rgb(52pt)=(0.972757,0.729771,0.239443); rgb(53pt)=(0.995648,0.743371,0.237148); rgb(54pt)=(0.996986,0.765857,0.219943); rgb(55pt)=(0.995205,0.789252,0.202762); rgb(56pt)=(0.9892,0.813567,0.188533); rgb(57pt)=(0.978629,0.838629,0.176557); rgb(58pt)=(0.967648,0.8639,0.16429); rgb(59pt)=(0.96101,0.889019,0.153676); rgb(60pt)=(0.959671,0.913457,0.142257); rgb(61pt)=(0.962795,0.937338,0.12651); rgb(62pt)=(0.969114,0.960629,0.106362); rgb(63pt)=(0.9769,0.9839,0.0805)},
colorbar
]
\addplot [forget plot] graphics [xmin=-1.00200400801603, xmax=1.00200400801603, ymin=-1.00200400801603, ymax=1.00200400801603] {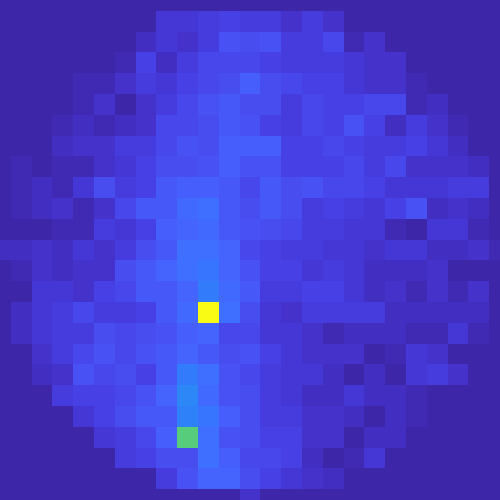};
\end{axis}
\end{tikzpicture}%
	\caption{$64$-bit Data}
	\label{fig:LOFAR:mle:sources:64bit}
\end{subfigure}%
    ~ 
\begin{subfigure}[t]{0.25\textwidth}
\centering
	\setlength{\figurewidth}{5.1cm}
	\setlength{\figureheight}{5.1cm}
%
%
\begin{tikzpicture}[scale=0.55]

\begin{axis}[%
width=\figurewidth,
height=\figureheight,
at={(0.92in,0.642in)},
scale only axis,
point meta min=1e-08,
point meta max=0.0441050164637363,
axis on top,
xmin=-1.00200400801603,
xmax=1.00200400801603,
xlabel style={font=\color{white!15!black}},
xlabel={$\text{South }\leftarrow\text{ Direction }\rightarrow\text{ North}$},
y dir=reverse,
ymin=-1.00200400801603,
ymax=1.00200400801603,
axis background/.style={fill=white},
legend style={legend cell align=left, align=left, draw=white!15!black},
colormap={mymap}{[1pt] rgb(0pt)=(0.2422,0.1504,0.6603); rgb(1pt)=(0.25039,0.164995,0.707614); rgb(2pt)=(0.257771,0.181781,0.751138); rgb(3pt)=(0.264729,0.197757,0.795214); rgb(4pt)=(0.270648,0.214676,0.836371); rgb(5pt)=(0.275114,0.234238,0.870986); rgb(6pt)=(0.2783,0.255871,0.899071); rgb(7pt)=(0.280333,0.278233,0.9221); rgb(8pt)=(0.281338,0.300595,0.941376); rgb(9pt)=(0.281014,0.322757,0.957886); rgb(10pt)=(0.279467,0.344671,0.971676); rgb(11pt)=(0.275971,0.366681,0.982905); rgb(12pt)=(0.269914,0.3892,0.9906); rgb(13pt)=(0.260243,0.412329,0.995157); rgb(14pt)=(0.244033,0.435833,0.998833); rgb(15pt)=(0.220643,0.460257,0.997286); rgb(16pt)=(0.196333,0.484719,0.989152); rgb(17pt)=(0.183405,0.507371,0.979795); rgb(18pt)=(0.178643,0.528857,0.968157); rgb(19pt)=(0.176438,0.549905,0.952019); rgb(20pt)=(0.168743,0.570262,0.935871); rgb(21pt)=(0.154,0.5902,0.9218); rgb(22pt)=(0.146029,0.609119,0.907857); rgb(23pt)=(0.138024,0.627629,0.89729); rgb(24pt)=(0.124814,0.645929,0.888343); rgb(25pt)=(0.111252,0.6635,0.876314); rgb(26pt)=(0.0952095,0.679829,0.859781); rgb(27pt)=(0.0688714,0.694771,0.839357); rgb(28pt)=(0.0296667,0.708167,0.816333); rgb(29pt)=(0.00357143,0.720267,0.7917); rgb(30pt)=(0.00665714,0.731214,0.766014); rgb(31pt)=(0.0433286,0.741095,0.73941); rgb(32pt)=(0.0963952,0.75,0.712038); rgb(33pt)=(0.140771,0.7584,0.684157); rgb(34pt)=(0.1717,0.766962,0.655443); rgb(35pt)=(0.193767,0.775767,0.6251); rgb(36pt)=(0.216086,0.7843,0.5923); rgb(37pt)=(0.246957,0.791795,0.556743); rgb(38pt)=(0.290614,0.79729,0.518829); rgb(39pt)=(0.340643,0.8008,0.478857); rgb(40pt)=(0.3909,0.802871,0.435448); rgb(41pt)=(0.445629,0.802419,0.390919); rgb(42pt)=(0.5044,0.7993,0.348); rgb(43pt)=(0.561562,0.794233,0.304481); rgb(44pt)=(0.617395,0.787619,0.261238); rgb(45pt)=(0.671986,0.779271,0.2227); rgb(46pt)=(0.7242,0.769843,0.191029); rgb(47pt)=(0.773833,0.759805,0.16461); rgb(48pt)=(0.820314,0.749814,0.153529); rgb(49pt)=(0.863433,0.7406,0.159633); rgb(50pt)=(0.903543,0.733029,0.177414); rgb(51pt)=(0.939257,0.728786,0.209957); rgb(52pt)=(0.972757,0.729771,0.239443); rgb(53pt)=(0.995648,0.743371,0.237148); rgb(54pt)=(0.996986,0.765857,0.219943); rgb(55pt)=(0.995205,0.789252,0.202762); rgb(56pt)=(0.9892,0.813567,0.188533); rgb(57pt)=(0.978629,0.838629,0.176557); rgb(58pt)=(0.967648,0.8639,0.16429); rgb(59pt)=(0.96101,0.889019,0.153676); rgb(60pt)=(0.959671,0.913457,0.142257); rgb(61pt)=(0.962795,0.937338,0.12651); rgb(62pt)=(0.969114,0.960629,0.106362); rgb(63pt)=(0.9769,0.9839,0.0805)},
colorbar
]
\addplot [forget plot] graphics [xmin=-1.00200400801603, xmax=1.00200400801603, ymin=-1.00200400801603, ymax=1.00200400801603] {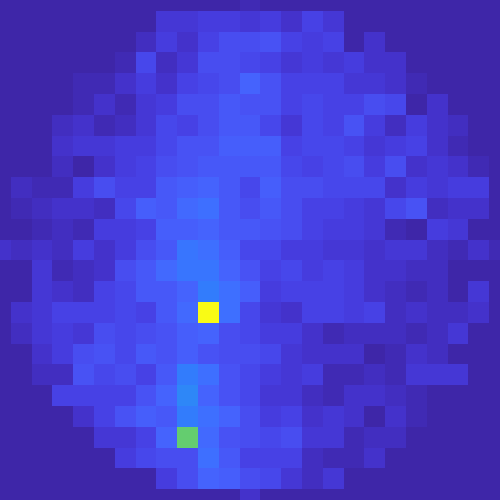};
\end{axis}
\end{tikzpicture}%
	\caption{$1$-bit Data}
	\label{fig:LOFAR:cmle:sources:1bit}
\end{subfigure}%
\caption{CS302 - Norm. Power Estimates ($M=47, D_\text{S}=441$)}
\hspace{0.5cm}
\begin{subfigure}[t]{0.24\textwidth}
    \centering
	\setlength{\figurewidth}{5.1cm}
	\setlength{\figureheight}{5.1cm}
%
%
\begin{tikzpicture}[scale=0.55]

\begin{axis}[%
width=\figurewidth,
height=\figureheight,
at={(0.92in,0.642in)},
scale only axis,
point meta min=1e-08,
point meta max=0.0372746234790098,
axis on top,
xmin=-1.00200400801603,
xmax=1.00200400801603,
xlabel style={font=\color{white!15!black}},
xlabel={$\text{South }\leftarrow\text{ Direction }\rightarrow\text{ North}$},
y dir=reverse,
ymin=-1.00200400801603,
ymax=1.00200400801603,
ylabel style={font=\color{white!15!black}},
ylabel={$\text{East }\leftarrow\text{ Direction }\rightarrow\text{ West}$},
axis background/.style={fill=white},
legend style={legend cell align=left, align=left, draw=white!15!black},
colormap={mymap}{[1pt] rgb(0pt)=(0.2422,0.1504,0.6603); rgb(1pt)=(0.25039,0.164995,0.707614); rgb(2pt)=(0.257771,0.181781,0.751138); rgb(3pt)=(0.264729,0.197757,0.795214); rgb(4pt)=(0.270648,0.214676,0.836371); rgb(5pt)=(0.275114,0.234238,0.870986); rgb(6pt)=(0.2783,0.255871,0.899071); rgb(7pt)=(0.280333,0.278233,0.9221); rgb(8pt)=(0.281338,0.300595,0.941376); rgb(9pt)=(0.281014,0.322757,0.957886); rgb(10pt)=(0.279467,0.344671,0.971676); rgb(11pt)=(0.275971,0.366681,0.982905); rgb(12pt)=(0.269914,0.3892,0.9906); rgb(13pt)=(0.260243,0.412329,0.995157); rgb(14pt)=(0.244033,0.435833,0.998833); rgb(15pt)=(0.220643,0.460257,0.997286); rgb(16pt)=(0.196333,0.484719,0.989152); rgb(17pt)=(0.183405,0.507371,0.979795); rgb(18pt)=(0.178643,0.528857,0.968157); rgb(19pt)=(0.176438,0.549905,0.952019); rgb(20pt)=(0.168743,0.570262,0.935871); rgb(21pt)=(0.154,0.5902,0.9218); rgb(22pt)=(0.146029,0.609119,0.907857); rgb(23pt)=(0.138024,0.627629,0.89729); rgb(24pt)=(0.124814,0.645929,0.888343); rgb(25pt)=(0.111252,0.6635,0.876314); rgb(26pt)=(0.0952095,0.679829,0.859781); rgb(27pt)=(0.0688714,0.694771,0.839357); rgb(28pt)=(0.0296667,0.708167,0.816333); rgb(29pt)=(0.00357143,0.720267,0.7917); rgb(30pt)=(0.00665714,0.731214,0.766014); rgb(31pt)=(0.0433286,0.741095,0.73941); rgb(32pt)=(0.0963952,0.75,0.712038); rgb(33pt)=(0.140771,0.7584,0.684157); rgb(34pt)=(0.1717,0.766962,0.655443); rgb(35pt)=(0.193767,0.775767,0.6251); rgb(36pt)=(0.216086,0.7843,0.5923); rgb(37pt)=(0.246957,0.791795,0.556743); rgb(38pt)=(0.290614,0.79729,0.518829); rgb(39pt)=(0.340643,0.8008,0.478857); rgb(40pt)=(0.3909,0.802871,0.435448); rgb(41pt)=(0.445629,0.802419,0.390919); rgb(42pt)=(0.5044,0.7993,0.348); rgb(43pt)=(0.561562,0.794233,0.304481); rgb(44pt)=(0.617395,0.787619,0.261238); rgb(45pt)=(0.671986,0.779271,0.2227); rgb(46pt)=(0.7242,0.769843,0.191029); rgb(47pt)=(0.773833,0.759805,0.16461); rgb(48pt)=(0.820314,0.749814,0.153529); rgb(49pt)=(0.863433,0.7406,0.159633); rgb(50pt)=(0.903543,0.733029,0.177414); rgb(51pt)=(0.939257,0.728786,0.209957); rgb(52pt)=(0.972757,0.729771,0.239443); rgb(53pt)=(0.995648,0.743371,0.237148); rgb(54pt)=(0.996986,0.765857,0.219943); rgb(55pt)=(0.995205,0.789252,0.202762); rgb(56pt)=(0.9892,0.813567,0.188533); rgb(57pt)=(0.978629,0.838629,0.176557); rgb(58pt)=(0.967648,0.8639,0.16429); rgb(59pt)=(0.96101,0.889019,0.153676); rgb(60pt)=(0.959671,0.913457,0.142257); rgb(61pt)=(0.962795,0.937338,0.12651); rgb(62pt)=(0.969114,0.960629,0.106362); rgb(63pt)=(0.9769,0.9839,0.0805)},
colorbar
]
\addplot [forget plot] graphics [xmin=-1.00200400801603, xmax=1.00200400801603, ymin=-1.00200400801603, ymax=1.00200400801603] {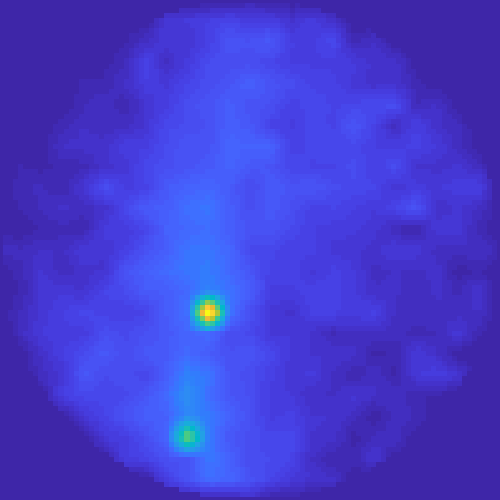};
\end{axis}
\end{tikzpicture}%
	\caption{$64$-bit Data}
	\label{fig:LOFAR:mle:image:64bit}
\end{subfigure}%
    ~ 
\begin{subfigure}[t]{0.25\textwidth}
\centering
	\setlength{\figurewidth}{5.1cm}
	\setlength{\figureheight}{5.1cm}
%
%
\begin{tikzpicture}[scale=0.55]

\begin{axis}[%
width=\figurewidth,
height=\figureheight,
at={(0.92in,0.642in)},
scale only axis,
point meta min=1e-08,
point meta max=0.040996999245704,
axis on top,
xmin=-1.00200400801603,
xmax=1.00200400801603,
xlabel style={font=\color{white!15!black}},
xlabel={$\text{South }\leftarrow\text{ Direction }\rightarrow\text{ North}$},
y dir=reverse,
ymin=-1.00200400801603,
ymax=1.00200400801603,
axis background/.style={fill=white},
legend style={legend cell align=left, align=left, draw=white!15!black},
colormap={mymap}{[1pt] rgb(0pt)=(0.2422,0.1504,0.6603); rgb(1pt)=(0.25039,0.164995,0.707614); rgb(2pt)=(0.257771,0.181781,0.751138); rgb(3pt)=(0.264729,0.197757,0.795214); rgb(4pt)=(0.270648,0.214676,0.836371); rgb(5pt)=(0.275114,0.234238,0.870986); rgb(6pt)=(0.2783,0.255871,0.899071); rgb(7pt)=(0.280333,0.278233,0.9221); rgb(8pt)=(0.281338,0.300595,0.941376); rgb(9pt)=(0.281014,0.322757,0.957886); rgb(10pt)=(0.279467,0.344671,0.971676); rgb(11pt)=(0.275971,0.366681,0.982905); rgb(12pt)=(0.269914,0.3892,0.9906); rgb(13pt)=(0.260243,0.412329,0.995157); rgb(14pt)=(0.244033,0.435833,0.998833); rgb(15pt)=(0.220643,0.460257,0.997286); rgb(16pt)=(0.196333,0.484719,0.989152); rgb(17pt)=(0.183405,0.507371,0.979795); rgb(18pt)=(0.178643,0.528857,0.968157); rgb(19pt)=(0.176438,0.549905,0.952019); rgb(20pt)=(0.168743,0.570262,0.935871); rgb(21pt)=(0.154,0.5902,0.9218); rgb(22pt)=(0.146029,0.609119,0.907857); rgb(23pt)=(0.138024,0.627629,0.89729); rgb(24pt)=(0.124814,0.645929,0.888343); rgb(25pt)=(0.111252,0.6635,0.876314); rgb(26pt)=(0.0952095,0.679829,0.859781); rgb(27pt)=(0.0688714,0.694771,0.839357); rgb(28pt)=(0.0296667,0.708167,0.816333); rgb(29pt)=(0.00357143,0.720267,0.7917); rgb(30pt)=(0.00665714,0.731214,0.766014); rgb(31pt)=(0.0433286,0.741095,0.73941); rgb(32pt)=(0.0963952,0.75,0.712038); rgb(33pt)=(0.140771,0.7584,0.684157); rgb(34pt)=(0.1717,0.766962,0.655443); rgb(35pt)=(0.193767,0.775767,0.6251); rgb(36pt)=(0.216086,0.7843,0.5923); rgb(37pt)=(0.246957,0.791795,0.556743); rgb(38pt)=(0.290614,0.79729,0.518829); rgb(39pt)=(0.340643,0.8008,0.478857); rgb(40pt)=(0.3909,0.802871,0.435448); rgb(41pt)=(0.445629,0.802419,0.390919); rgb(42pt)=(0.5044,0.7993,0.348); rgb(43pt)=(0.561562,0.794233,0.304481); rgb(44pt)=(0.617395,0.787619,0.261238); rgb(45pt)=(0.671986,0.779271,0.2227); rgb(46pt)=(0.7242,0.769843,0.191029); rgb(47pt)=(0.773833,0.759805,0.16461); rgb(48pt)=(0.820314,0.749814,0.153529); rgb(49pt)=(0.863433,0.7406,0.159633); rgb(50pt)=(0.903543,0.733029,0.177414); rgb(51pt)=(0.939257,0.728786,0.209957); rgb(52pt)=(0.972757,0.729771,0.239443); rgb(53pt)=(0.995648,0.743371,0.237148); rgb(54pt)=(0.996986,0.765857,0.219943); rgb(55pt)=(0.995205,0.789252,0.202762); rgb(56pt)=(0.9892,0.813567,0.188533); rgb(57pt)=(0.978629,0.838629,0.176557); rgb(58pt)=(0.967648,0.8639,0.16429); rgb(59pt)=(0.96101,0.889019,0.153676); rgb(60pt)=(0.959671,0.913457,0.142257); rgb(61pt)=(0.962795,0.937338,0.12651); rgb(62pt)=(0.969114,0.960629,0.106362); rgb(63pt)=(0.9769,0.9839,0.0805)},
colorbar
]
\addplot [forget plot] graphics [xmin=-1.00200400801603, xmax=1.00200400801603, ymin=-1.00200400801603, ymax=1.00200400801603] {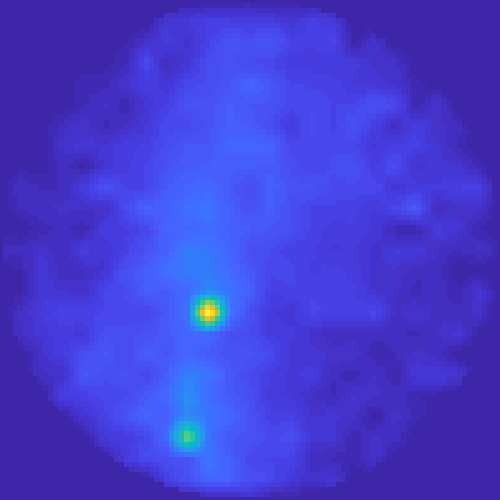};
\end{axis}
\end{tikzpicture}%
	\caption{$1$-bit Data}
	\label{fig:LOFAR:cmle:image:1bit}
\end{subfigure}%
\caption{CS302 - All-Sky Image ($M=47, D_\text{S}=441$)}
\hspace{0.5cm}
\begin{subfigure}[t]{0.25\textwidth}
    \centering
	\setlength{\figurewidth}{5.1cm}
	\setlength{\figureheight}{5.1cm}
%
%
\begin{tikzpicture}[scale=0.55]

\begin{axis}[%
width=\figurewidth,
height=\figureheight,
at={(0.92in,0.642in)},
scale only axis,
point meta min=0.141453682573754,
point meta max=0.149509255427503,
axis on top,
xmin=-1.00200400801603,
xmax=1.00200400801603,
xlabel style={font=\color{white!15!black}},
xlabel={$\text{South }\leftarrow\text{ Direction }\rightarrow\text{ North}$},
y dir=reverse,
ymin=-1.00200400801603,
ymax=1.00200400801603,
ylabel style={font=\color{white!15!black}},
ylabel={$\text{East }\leftarrow\text{ Direction }\rightarrow\text{ West}$},
axis background/.style={fill=white},
legend style={legend cell align=left, align=left, draw=white!15!black},
colormap={mymap}{[1pt] rgb(0pt)=(0.2422,0.1504,0.6603); rgb(1pt)=(0.25039,0.164995,0.707614); rgb(2pt)=(0.257771,0.181781,0.751138); rgb(3pt)=(0.264729,0.197757,0.795214); rgb(4pt)=(0.270648,0.214676,0.836371); rgb(5pt)=(0.275114,0.234238,0.870986); rgb(6pt)=(0.2783,0.255871,0.899071); rgb(7pt)=(0.280333,0.278233,0.9221); rgb(8pt)=(0.281338,0.300595,0.941376); rgb(9pt)=(0.281014,0.322757,0.957886); rgb(10pt)=(0.279467,0.344671,0.971676); rgb(11pt)=(0.275971,0.366681,0.982905); rgb(12pt)=(0.269914,0.3892,0.9906); rgb(13pt)=(0.260243,0.412329,0.995157); rgb(14pt)=(0.244033,0.435833,0.998833); rgb(15pt)=(0.220643,0.460257,0.997286); rgb(16pt)=(0.196333,0.484719,0.989152); rgb(17pt)=(0.183405,0.507371,0.979795); rgb(18pt)=(0.178643,0.528857,0.968157); rgb(19pt)=(0.176438,0.549905,0.952019); rgb(20pt)=(0.168743,0.570262,0.935871); rgb(21pt)=(0.154,0.5902,0.9218); rgb(22pt)=(0.146029,0.609119,0.907857); rgb(23pt)=(0.138024,0.627629,0.89729); rgb(24pt)=(0.124814,0.645929,0.888343); rgb(25pt)=(0.111252,0.6635,0.876314); rgb(26pt)=(0.0952095,0.679829,0.859781); rgb(27pt)=(0.0688714,0.694771,0.839357); rgb(28pt)=(0.0296667,0.708167,0.816333); rgb(29pt)=(0.00357143,0.720267,0.7917); rgb(30pt)=(0.00665714,0.731214,0.766014); rgb(31pt)=(0.0433286,0.741095,0.73941); rgb(32pt)=(0.0963952,0.75,0.712038); rgb(33pt)=(0.140771,0.7584,0.684157); rgb(34pt)=(0.1717,0.766962,0.655443); rgb(35pt)=(0.193767,0.775767,0.6251); rgb(36pt)=(0.216086,0.7843,0.5923); rgb(37pt)=(0.246957,0.791795,0.556743); rgb(38pt)=(0.290614,0.79729,0.518829); rgb(39pt)=(0.340643,0.8008,0.478857); rgb(40pt)=(0.3909,0.802871,0.435448); rgb(41pt)=(0.445629,0.802419,0.390919); rgb(42pt)=(0.5044,0.7993,0.348); rgb(43pt)=(0.561562,0.794233,0.304481); rgb(44pt)=(0.617395,0.787619,0.261238); rgb(45pt)=(0.671986,0.779271,0.2227); rgb(46pt)=(0.7242,0.769843,0.191029); rgb(47pt)=(0.773833,0.759805,0.16461); rgb(48pt)=(0.820314,0.749814,0.153529); rgb(49pt)=(0.863433,0.7406,0.159633); rgb(50pt)=(0.903543,0.733029,0.177414); rgb(51pt)=(0.939257,0.728786,0.209957); rgb(52pt)=(0.972757,0.729771,0.239443); rgb(53pt)=(0.995648,0.743371,0.237148); rgb(54pt)=(0.996986,0.765857,0.219943); rgb(55pt)=(0.995205,0.789252,0.202762); rgb(56pt)=(0.9892,0.813567,0.188533); rgb(57pt)=(0.978629,0.838629,0.176557); rgb(58pt)=(0.967648,0.8639,0.16429); rgb(59pt)=(0.96101,0.889019,0.153676); rgb(60pt)=(0.959671,0.913457,0.142257); rgb(61pt)=(0.962795,0.937338,0.12651); rgb(62pt)=(0.969114,0.960629,0.106362); rgb(63pt)=(0.9769,0.9839,0.0805)},
colorbar
]
\addplot [forget plot] graphics [xmin=-1.00200400801603, xmax=1.00200400801603, ymin=-1.00200400801603, ymax=1.00200400801603] {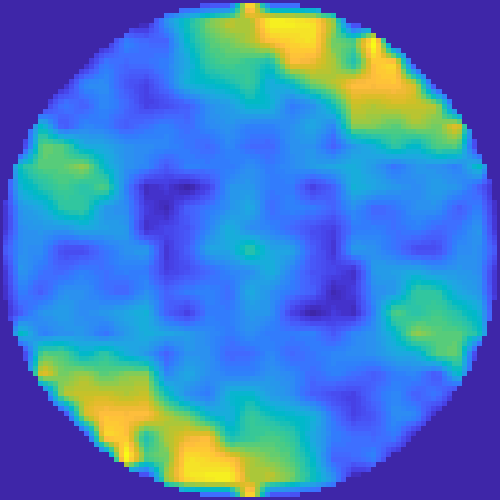};
\end{axis}
\end{tikzpicture}%
	\caption{$64$-bit Data}
	\label{fig:LOFAR:mle:uncert:64bit}
\end{subfigure}%
    ~ 
\begin{subfigure}[t]{0.25\textwidth}
\centering
	\setlength{\figurewidth}{5.1cm}
	\setlength{\figureheight}{5.1cm}
%
%
\begin{tikzpicture}[scale=0.55]

\begin{axis}[%
width=\figurewidth,
height=\figureheight,
at={(0.92in,0.642in)},
scale only axis,
point meta min=0.225213315546567,
point meta max=0.235165286162668,
axis on top,
xmin=-1.00200400801603,
xmax=1.00200400801603,
xlabel style={font=\color{white!15!black}},
xlabel={$\text{South }\leftarrow\text{ Direction }\rightarrow\text{ North}$},
y dir=reverse,
ymin=-1.00200400801603,
ymax=1.00200400801603,
axis background/.style={fill=white},
legend style={legend cell align=left, align=left, draw=white!15!black},
colormap={mymap}{[1pt] rgb(0pt)=(0.2422,0.1504,0.6603); rgb(1pt)=(0.25039,0.164995,0.707614); rgb(2pt)=(0.257771,0.181781,0.751138); rgb(3pt)=(0.264729,0.197757,0.795214); rgb(4pt)=(0.270648,0.214676,0.836371); rgb(5pt)=(0.275114,0.234238,0.870986); rgb(6pt)=(0.2783,0.255871,0.899071); rgb(7pt)=(0.280333,0.278233,0.9221); rgb(8pt)=(0.281338,0.300595,0.941376); rgb(9pt)=(0.281014,0.322757,0.957886); rgb(10pt)=(0.279467,0.344671,0.971676); rgb(11pt)=(0.275971,0.366681,0.982905); rgb(12pt)=(0.269914,0.3892,0.9906); rgb(13pt)=(0.260243,0.412329,0.995157); rgb(14pt)=(0.244033,0.435833,0.998833); rgb(15pt)=(0.220643,0.460257,0.997286); rgb(16pt)=(0.196333,0.484719,0.989152); rgb(17pt)=(0.183405,0.507371,0.979795); rgb(18pt)=(0.178643,0.528857,0.968157); rgb(19pt)=(0.176438,0.549905,0.952019); rgb(20pt)=(0.168743,0.570262,0.935871); rgb(21pt)=(0.154,0.5902,0.9218); rgb(22pt)=(0.146029,0.609119,0.907857); rgb(23pt)=(0.138024,0.627629,0.89729); rgb(24pt)=(0.124814,0.645929,0.888343); rgb(25pt)=(0.111252,0.6635,0.876314); rgb(26pt)=(0.0952095,0.679829,0.859781); rgb(27pt)=(0.0688714,0.694771,0.839357); rgb(28pt)=(0.0296667,0.708167,0.816333); rgb(29pt)=(0.00357143,0.720267,0.7917); rgb(30pt)=(0.00665714,0.731214,0.766014); rgb(31pt)=(0.0433286,0.741095,0.73941); rgb(32pt)=(0.0963952,0.75,0.712038); rgb(33pt)=(0.140771,0.7584,0.684157); rgb(34pt)=(0.1717,0.766962,0.655443); rgb(35pt)=(0.193767,0.775767,0.6251); rgb(36pt)=(0.216086,0.7843,0.5923); rgb(37pt)=(0.246957,0.791795,0.556743); rgb(38pt)=(0.290614,0.79729,0.518829); rgb(39pt)=(0.340643,0.8008,0.478857); rgb(40pt)=(0.3909,0.802871,0.435448); rgb(41pt)=(0.445629,0.802419,0.390919); rgb(42pt)=(0.5044,0.7993,0.348); rgb(43pt)=(0.561562,0.794233,0.304481); rgb(44pt)=(0.617395,0.787619,0.261238); rgb(45pt)=(0.671986,0.779271,0.2227); rgb(46pt)=(0.7242,0.769843,0.191029); rgb(47pt)=(0.773833,0.759805,0.16461); rgb(48pt)=(0.820314,0.749814,0.153529); rgb(49pt)=(0.863433,0.7406,0.159633); rgb(50pt)=(0.903543,0.733029,0.177414); rgb(51pt)=(0.939257,0.728786,0.209957); rgb(52pt)=(0.972757,0.729771,0.239443); rgb(53pt)=(0.995648,0.743371,0.237148); rgb(54pt)=(0.996986,0.765857,0.219943); rgb(55pt)=(0.995205,0.789252,0.202762); rgb(56pt)=(0.9892,0.813567,0.188533); rgb(57pt)=(0.978629,0.838629,0.176557); rgb(58pt)=(0.967648,0.8639,0.16429); rgb(59pt)=(0.96101,0.889019,0.153676); rgb(60pt)=(0.959671,0.913457,0.142257); rgb(61pt)=(0.962795,0.937338,0.12651); rgb(62pt)=(0.969114,0.960629,0.106362); rgb(63pt)=(0.9769,0.9839,0.0805)},
colorbar
]
\addplot [forget plot] graphics [xmin=-1.00200400801603, xmax=1.00200400801603, ymin=-1.00200400801603, ymax=1.00200400801603] {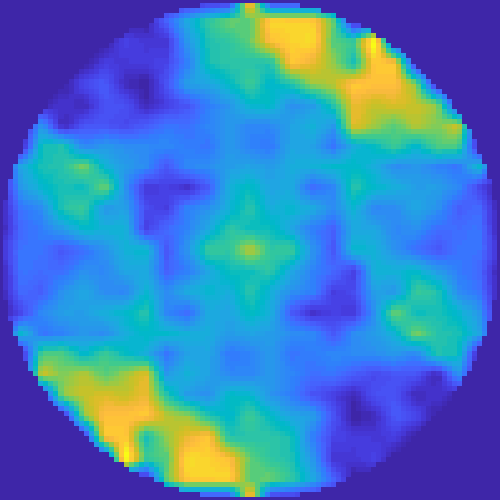};
\end{axis}
\end{tikzpicture}%
	\caption{$1$-bit Data}
	\label{fig:LOFAR:cmle:uncert:1bit}
\end{subfigure}%
\caption{CS302 - Uncertainty ($M=47, D_\text{S}=441$)}
\label{fig:LOFAR:uncert}
\end{figure}

For interferometric imaging with scoring \eqref{update:scoring:rule} and $64$-bit LOFAR radio data, we initialize the noise power estimate by
\begin{align}\label{noise:estimate:beam}
\hat{\theta}_{0}^{(0)}&= \frac{1}{2M} \vecT{\ve{I}}\ve{\hat{r}}_{\ve{y}}(\ve{Y})
\end{align}
and the intensity estimates by $\hat{\theta}_{d}^{(0)}=\theta_\Delta$. The imaging result with the $D_\text{S}=441$ source power levels (corresponds to $25$ pixels) and $I=10$ iterations is depicted in Fig. \ref{fig:LOFAR:mle:sources:64bit} after normalizing with the final noise power estimate $\hat{\theta}_0$. Note that each considered source direction lies in the center of the pixels visualized in Fig. \ref{fig:LOFAR:mle:sources:64bit} while the dominant intensity value corresponds to a power level of approximately $\SI{-14}{\decibel}$.
For mimicking the sensor data stream of a radio telescope with $1$-bit A/D conversion at the sensors, we hard-limit the LOFAR dataset according to \eqref{binarization}. The intensity estimates after $I=10$ scoring iterations are depicted in Fig. \ref{fig:LOFAR:cmle:sources:1bit}. As for the binary telescope $\theta_0=1$, normalization of the final estimates $\ve{\hat{\theta}}$ is not required. While the two likelihood-oriented algorithms \eqref{update:scoring:rule} and \eqref{scoring:rule:quant} produce slightly different numerical results for the spatial power distribution, the difference can hardly be noticed by visual inspection. Fig. \ref{fig:LOFAR:mle:image:64bit} and Fig. \ref{fig:LOFAR:cmle:image:1bit} show the all-sky images after interpolation. Note, that Fig. \ref{fig:LOFAR:mle:image:64bit} is produced with data which is $64$ times larger than the measurements used for reconstructing the interferometric image in Fig. \ref{fig:LOFAR:cmle:image:1bit}.

To assess the achievable imaging sensitivity under both signal acquisition approaches in a quantitative manner, we consider a situation where emissions from $D_\text{S}=441$ directions are received at CS302 with equal intensity $\theta_{d, \text{qual}}=\SI{-30}{\decibel}, \forall d$. For this setting the element-wise normalized square-root of the diagonal of the inverse Fisher matrices
\begin{align}
\ve{\hat{\sigma}}_{\ve{Y}}(\ve{\theta}_{\text{qual}})&=\sqrt{\diagvec{\ve{F}_{\ve{Y}}^{-1}(\ve{\theta}_{\text{qual}})} ./ \ve{\theta}_{\text{qual}}},\\
\ve{\hat{\sigma}}_{\ve{Z}}(\ve{\theta}_{\text{qual}})&=\sqrt{\diagvec{\ve{\tilde{F}}_{\ve{Z}}^{-1}(\ve{\theta}_{\text{qual}})} ./ \ve{\theta}_{\text{qual}}},
\end{align}
are calculate, where the operator $\diagvec{\ve{U}}$ returns the vectorized diagonal of a square matrix $\ve{U}$ and $./$ denotes an element-wise division. In both cases, this provides a quantitative measure for the relative standard deviation of the intensity estimates when performing interferometric processing with \eqref{update:scoring:rule} and \eqref{scoring:rule:quant} on a large number of independent radio datasets.

In Fig. \ref{fig:LOFAR:mle:uncert:64bit} and Fig. \ref{fig:LOFAR:cmle:uncert:1bit}, it can be observed that the average relative imaging uncertainty of the $64$-bit LOFAR station is $\SI{14.47}{\percent}$ while the $1$-bit telescope operates at an average uncertainty level of $\SI{22.92}{\percent}$ with a similar spatial performance structure. From Fig. \ref{fig:LOFAR:uncert} it becomes obvious that hard-limiting the LOFAR radio measurements significantly increases the variance of the intensity estimates. To recover this loss in interferometric imaging performance, one can increase the observation time $T$ of the binary system by $\big(\frac{22,92}{14,47}\big)^{2} \approx 2.5$.

\begin{figure}[!h]
\centering
    \begin{subfigure}[t]{0.24\textwidth}
    \centering
	\setlength{\figurewidth}{5.1cm}
	\setlength{\figureheight}{5.1cm}
%
%
\begin{tikzpicture}[scale=0.55]

\begin{axis}[%
width=\figurewidth,
height=\figureheight,
at={(0.92in,0.642in)},
scale only axis,
point meta min=0.931281527456769,
point meta max=1.71906576454531,
axis on top,
xmin=-1.00200400801603,
xmax=1.00200400801603,
xlabel style={font=\color{white!15!black}},
xlabel={$\text{South }\leftarrow\text{ Direction }\rightarrow\text{ North}$},
y dir=reverse,
ymin=-1.00200400801603,
ymax=1.00200400801603,
ylabel style={font=\color{white!15!black}},
ylabel={$\text{East }\leftarrow\text{ Direction }\rightarrow\text{ West}$},
axis background/.style={fill=white},
legend style={legend cell align=left, align=left, draw=white!15!black},
colormap={mymap}{[1pt] rgb(0pt)=(0.2422,0.1504,0.6603); rgb(1pt)=(0.25039,0.164995,0.707614); rgb(2pt)=(0.257771,0.181781,0.751138); rgb(3pt)=(0.264729,0.197757,0.795214); rgb(4pt)=(0.270648,0.214676,0.836371); rgb(5pt)=(0.275114,0.234238,0.870986); rgb(6pt)=(0.2783,0.255871,0.899071); rgb(7pt)=(0.280333,0.278233,0.9221); rgb(8pt)=(0.281338,0.300595,0.941376); rgb(9pt)=(0.281014,0.322757,0.957886); rgb(10pt)=(0.279467,0.344671,0.971676); rgb(11pt)=(0.275971,0.366681,0.982905); rgb(12pt)=(0.269914,0.3892,0.9906); rgb(13pt)=(0.260243,0.412329,0.995157); rgb(14pt)=(0.244033,0.435833,0.998833); rgb(15pt)=(0.220643,0.460257,0.997286); rgb(16pt)=(0.196333,0.484719,0.989152); rgb(17pt)=(0.183405,0.507371,0.979795); rgb(18pt)=(0.178643,0.528857,0.968157); rgb(19pt)=(0.176438,0.549905,0.952019); rgb(20pt)=(0.168743,0.570262,0.935871); rgb(21pt)=(0.154,0.5902,0.9218); rgb(22pt)=(0.146029,0.609119,0.907857); rgb(23pt)=(0.138024,0.627629,0.89729); rgb(24pt)=(0.124814,0.645929,0.888343); rgb(25pt)=(0.111252,0.6635,0.876314); rgb(26pt)=(0.0952095,0.679829,0.859781); rgb(27pt)=(0.0688714,0.694771,0.839357); rgb(28pt)=(0.0296667,0.708167,0.816333); rgb(29pt)=(0.00357143,0.720267,0.7917); rgb(30pt)=(0.00665714,0.731214,0.766014); rgb(31pt)=(0.0433286,0.741095,0.73941); rgb(32pt)=(0.0963952,0.75,0.712038); rgb(33pt)=(0.140771,0.7584,0.684157); rgb(34pt)=(0.1717,0.766962,0.655443); rgb(35pt)=(0.193767,0.775767,0.6251); rgb(36pt)=(0.216086,0.7843,0.5923); rgb(37pt)=(0.246957,0.791795,0.556743); rgb(38pt)=(0.290614,0.79729,0.518829); rgb(39pt)=(0.340643,0.8008,0.478857); rgb(40pt)=(0.3909,0.802871,0.435448); rgb(41pt)=(0.445629,0.802419,0.390919); rgb(42pt)=(0.5044,0.7993,0.348); rgb(43pt)=(0.561562,0.794233,0.304481); rgb(44pt)=(0.617395,0.787619,0.261238); rgb(45pt)=(0.671986,0.779271,0.2227); rgb(46pt)=(0.7242,0.769843,0.191029); rgb(47pt)=(0.773833,0.759805,0.16461); rgb(48pt)=(0.820314,0.749814,0.153529); rgb(49pt)=(0.863433,0.7406,0.159633); rgb(50pt)=(0.903543,0.733029,0.177414); rgb(51pt)=(0.939257,0.728786,0.209957); rgb(52pt)=(0.972757,0.729771,0.239443); rgb(53pt)=(0.995648,0.743371,0.237148); rgb(54pt)=(0.996986,0.765857,0.219943); rgb(55pt)=(0.995205,0.789252,0.202762); rgb(56pt)=(0.9892,0.813567,0.188533); rgb(57pt)=(0.978629,0.838629,0.176557); rgb(58pt)=(0.967648,0.8639,0.16429); rgb(59pt)=(0.96101,0.889019,0.153676); rgb(60pt)=(0.959671,0.913457,0.142257); rgb(61pt)=(0.962795,0.937338,0.12651); rgb(62pt)=(0.969114,0.960629,0.106362); rgb(63pt)=(0.9769,0.9839,0.0805)},
colorbar
]
\addplot [forget plot] graphics [xmin=-1.00200400801603, xmax=1.00200400801603, ymin=-1.00200400801603, ymax=1.00200400801603] {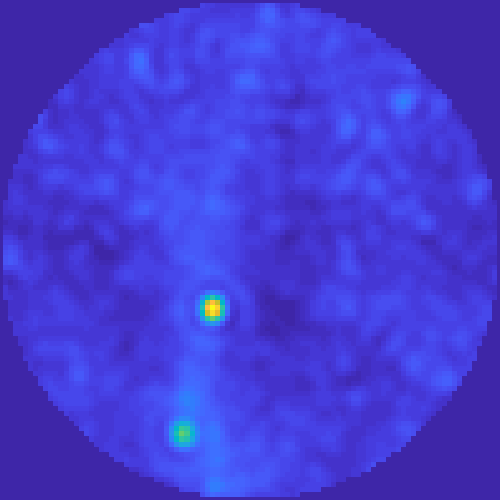};
\end{axis}
\end{tikzpicture}%
	\caption{$64$-bit Data}
	\label{fig:LOFAR:beam:64bit}
\end{subfigure}%
    ~ 
\begin{subfigure}[t]{0.25\textwidth}
\centering
	\setlength{\figurewidth}{5.1cm}
	\setlength{\figureheight}{5.1cm}
%
%
\begin{tikzpicture}[scale=0.55]

\begin{axis}[%
width=\figurewidth,
height=\figureheight,
at={(0.92in,0.642in)},
scale only axis,
point meta min=0.955923545245238,
point meta max=1.45736669432812,
axis on top,
xmin=-1.00200400801603,
xmax=1.00200400801603,
xlabel style={font=\color{white!15!black}},
xlabel={$\text{South }\leftarrow\text{ Direction }\rightarrow\text{ North}$},
y dir=reverse,
ymin=-1.00200400801603,
ymax=1.00200400801603,
axis background/.style={fill=white},
legend style={legend cell align=left, align=left, draw=white!15!black},
colormap={mymap}{[1pt] rgb(0pt)=(0.2422,0.1504,0.6603); rgb(1pt)=(0.25039,0.164995,0.707614); rgb(2pt)=(0.257771,0.181781,0.751138); rgb(3pt)=(0.264729,0.197757,0.795214); rgb(4pt)=(0.270648,0.214676,0.836371); rgb(5pt)=(0.275114,0.234238,0.870986); rgb(6pt)=(0.2783,0.255871,0.899071); rgb(7pt)=(0.280333,0.278233,0.9221); rgb(8pt)=(0.281338,0.300595,0.941376); rgb(9pt)=(0.281014,0.322757,0.957886); rgb(10pt)=(0.279467,0.344671,0.971676); rgb(11pt)=(0.275971,0.366681,0.982905); rgb(12pt)=(0.269914,0.3892,0.9906); rgb(13pt)=(0.260243,0.412329,0.995157); rgb(14pt)=(0.244033,0.435833,0.998833); rgb(15pt)=(0.220643,0.460257,0.997286); rgb(16pt)=(0.196333,0.484719,0.989152); rgb(17pt)=(0.183405,0.507371,0.979795); rgb(18pt)=(0.178643,0.528857,0.968157); rgb(19pt)=(0.176438,0.549905,0.952019); rgb(20pt)=(0.168743,0.570262,0.935871); rgb(21pt)=(0.154,0.5902,0.9218); rgb(22pt)=(0.146029,0.609119,0.907857); rgb(23pt)=(0.138024,0.627629,0.89729); rgb(24pt)=(0.124814,0.645929,0.888343); rgb(25pt)=(0.111252,0.6635,0.876314); rgb(26pt)=(0.0952095,0.679829,0.859781); rgb(27pt)=(0.0688714,0.694771,0.839357); rgb(28pt)=(0.0296667,0.708167,0.816333); rgb(29pt)=(0.00357143,0.720267,0.7917); rgb(30pt)=(0.00665714,0.731214,0.766014); rgb(31pt)=(0.0433286,0.741095,0.73941); rgb(32pt)=(0.0963952,0.75,0.712038); rgb(33pt)=(0.140771,0.7584,0.684157); rgb(34pt)=(0.1717,0.766962,0.655443); rgb(35pt)=(0.193767,0.775767,0.6251); rgb(36pt)=(0.216086,0.7843,0.5923); rgb(37pt)=(0.246957,0.791795,0.556743); rgb(38pt)=(0.290614,0.79729,0.518829); rgb(39pt)=(0.340643,0.8008,0.478857); rgb(40pt)=(0.3909,0.802871,0.435448); rgb(41pt)=(0.445629,0.802419,0.390919); rgb(42pt)=(0.5044,0.7993,0.348); rgb(43pt)=(0.561562,0.794233,0.304481); rgb(44pt)=(0.617395,0.787619,0.261238); rgb(45pt)=(0.671986,0.779271,0.2227); rgb(46pt)=(0.7242,0.769843,0.191029); rgb(47pt)=(0.773833,0.759805,0.16461); rgb(48pt)=(0.820314,0.749814,0.153529); rgb(49pt)=(0.863433,0.7406,0.159633); rgb(50pt)=(0.903543,0.733029,0.177414); rgb(51pt)=(0.939257,0.728786,0.209957); rgb(52pt)=(0.972757,0.729771,0.239443); rgb(53pt)=(0.995648,0.743371,0.237148); rgb(54pt)=(0.996986,0.765857,0.219943); rgb(55pt)=(0.995205,0.789252,0.202762); rgb(56pt)=(0.9892,0.813567,0.188533); rgb(57pt)=(0.978629,0.838629,0.176557); rgb(58pt)=(0.967648,0.8639,0.16429); rgb(59pt)=(0.96101,0.889019,0.153676); rgb(60pt)=(0.959671,0.913457,0.142257); rgb(61pt)=(0.962795,0.937338,0.12651); rgb(62pt)=(0.969114,0.960629,0.106362); rgb(63pt)=(0.9769,0.9839,0.0805)},
colorbar
]
\addplot [forget plot] graphics [xmin=-1.00200400801603, xmax=1.00200400801603, ymin=-1.00200400801603, ymax=1.00200400801603] {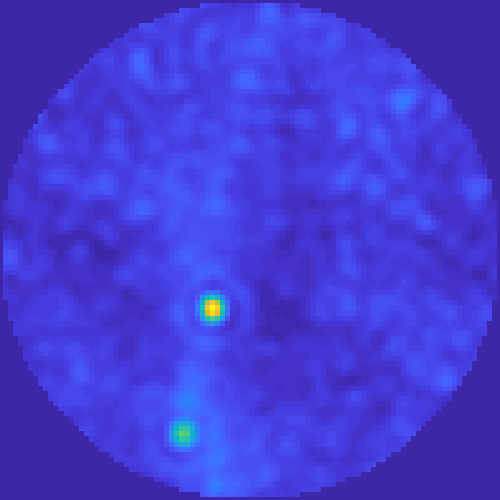};
\end{axis}
\end{tikzpicture}%
	\caption{$1$-bit Data}
	\label{fig:LOFAR:beam:1bit}
\end{subfigure}%
\caption{CS302 - Beam-forming Output ($M=47, D_\text{S}=441$)}
\end{figure}

As a visual comparison to iterative methods, in Fig. \ref{fig:LOFAR:beam:64bit} and Fig. \ref{fig:LOFAR:beam:1bit}, we show the non-iterative beam-forming outputs
\begin{align}\label{beam:forming:high}
\ve{\hat{\theta}}_{\text{BEAM}}(\ve{Y}) &=\frac{1}{2M} \ve{M}^{\T} \ve{\hat{r}}_{\ve{y}}(\ve{Y}),\\
\label{beam:forming:low}
\ve{\hat{\theta}}_{\text{BEAM}}(\ve{Z}) &=\frac{1}{2M} \ve{M}^{\T} \ve{\hat{r}}_{\ve{z}}(\ve{Z}),
\end{align}
with $D_\text{S}=7668$ after normalization by the according noise level $\hat{\theta}_{\text{BEAM},0}$.
While the numerical intensity values are not the same, also here the visual difference between the image produced with the high-resolution and the binary data is hardly notable. In comparison to the likelihood-based method \eqref{update:scoring:rule}, the numerical intensity values in Fig. \ref{fig:LOFAR:beam:64bit} show that with \eqref{beam:forming:high} it is not possible to separate the spatially white sensor noise from the source signals and produce consistent intensity estimates. Although beam-forming here uses significantly more steering directions, the spatial resolution, in comparison to the interpolated MLE result in Fig. \ref{fig:LOFAR:mle:image:64bit}, does not increase. Also, the array pattern is visible as a processing artifact. Note that such effects of the instrument point spread function (dirty beam) on the $64$-bit beam-forming output \eqref{beam:forming:high} can to some extend be removed by deconvolution methods like CLEAN \cite{Hoegbom74}.
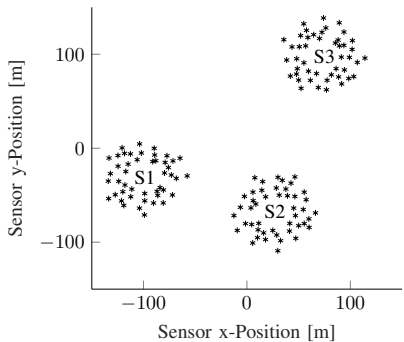
\begin{figure}[!ht]
\centering
	\setlength{\figurewidth}{5.5cm}
	\setlength{\figureheight}{5cm}
%
%
\begin{tikzpicture}[scale=0.75]

\begin{axis}[%
width=\figurewidth,
height=\figureheight,
at={(1.011in,0.642in)},
scale only axis,
xmin=-150,
xmax=150,
xlabel style={font=\color{white!15!black}},
xlabel={Sensor x-Position [m]},
ymin=-150,
ymax=150,
ylabel style={font=\color{white!15!black}},
ylabel={Sensor y-Position [m]},
axis background/.style={fill=white},
axis x line*=bottom,
axis y line*=left,
legend style={legend cell align=left, align=left, draw=white!15!black}
]
\addplot[only marks, mark=asterisk, mark options={}, mark size=1.5000pt, draw=black] table[row sep=crcr]{%
x	y\\
-91.05108206589	-14.150780645552\\
-88.0970687269295	-13.303781879953\\
-79.680576725133	-26.3506805020188\\
-84.0158751473252	-41.8738436441672\\
-86.9017858642223	-45.86920741422\\
-98.8212335033317	-48.1524379522343\\
-111.285053876289	-43.5554009893713\\
-117.949093358304	-39.197679681125\\
-117.115702229128	-24.7941352067319\\
-114.867891011092	-16.9369861086634\\
-106.105978306936	-12.2204244526838\\
-102.039142507964	-5.10836880596253\\
-89.1952097598658	-7.20751864544945\\
-79.0683526609521	-15.1597133002979\\
-75.9033365390773	-20.5750444046025\\
-73.0746421431002	-28.5256249200356\\
-80.7411407572827	-44.4076511048515\\
-86.2159722652288	-50.208952015153\\
-98.8080567684799	-55.9906681369794\\
-112.333788770423	-51.6272824112129\\
-119.965156031458	-46.2991383046647\\
-123.951879572276	-35.2643855622773\\
-124.590633764068	-19.158349437992\\
-112.61539687392	-6.06585446035388\\
-103.788877474694	4.5626540254998\\
-89.646340657898	-0.142526055573398\\
-77.3381853289452	-8.02574666997316\\
-71.4934165459734	-13.497578956074\\
-68.2650457850949	-32.1193769223792\\
-72.0146448622061	-49.7745727350739\\
-80.9238624474571	-58.2454662256678\\
-90.2912064415179	-58.3095402162326\\
-103.874628398439	-63.8004547160854\\
-121.260665582459	-56.100340111534\\
-127.291770869448	-49.4643333208551\\
-134.065787885292	-35.0163841074136\\
-131.983191338251	-27.0553481652205\\
-124.694519824974	-7.7783858374496\\
-118.386289465948	-5.47348311439055\\
-120.678997297938	0.371743951571261\\
-64.6755450829734	-10.6270514376265\\
-57.7914906290805	-29.484365089218\\
-99.1072761116752	-71.0097559605529\\
-119.068792412424	-60.8989147417051\\
-133.728545822434	-53.6103840557662\\
-133.348341968038	-10.4476964561072\\
24.9897180053238	-87.5337978060161\\
11.4336336295183	-79.9643949590427\\
8.64567295764181	-59.3393503617299\\
18.0998522456301	-51.7273491177511\\
31.262153363856	-49.9371172575084\\
39.1924658459052	-50.6559171633598\\
45.2170142096273	-63.7635946971168\\
46.9215966446484	-71.7561945057278\\
41.3999902374698	-80.0351891251084\\
41.1488476944625	-88.2235315063772\\
28.8239431912789	-92.3994470363286\\
16.1472352282937	-90.1331332846348\\
10.8104266733572	-86.8368940850523\\
4.58074218248746	-81.1458203139518\\
3.89443178965038	-63.522470387983\\
6.00341520865348	-55.8318002413762\\
14.4079509898739	-44.8145251827118\\
28.3977884337999	-42.3160353981693\\
37.6381085038342	-43.4374458472048\\
46.3382709418916	-51.3109502120678\\
54.4643909006451	-65.2304655712833\\
50.0360027935146	-82.4137384503883\\
47.2331570633455	-95.9413945915912\\
32.5385009302997	-98.4267756290537\\
17.9698696133071	-97.2433812230629\\
10.2394880182796	-95.1562343024452\\
-1.35353439352116	-80.2305317473401\\
-6.33176001933299	-62.8813197239106\\
-2.4421859703159	-51.2196221024533\\
5.79908154888273	-46.7644362485771\\
15.2146448218143	-35.5405650193607\\
34.1819775660324	-34.1764085026261\\
42.6217092450381	-37.2033498975252\\
55.3843939798961	-46.7814449210455\\
57.2841313708489	-54.7871806569201\\
59.8982900754871	-75.2295548644652\\
55.4102879664864	-80.2266010314961\\
60.1787110644752	-84.3114665116918\\
29.8503862401823	-109.156571533884\\
5.56668905312256	-100.891584006491\\
-9.36313431068194	-87.4738823421103\\
-12.5617748944358	-71.7082364399721\\
7.62003207897692	-31.4126460416046\\
29.9929408980807	-30.9692156094861\\
46.3571573730465	-30.5209532998306\\
66.2866841756729	-68.8104437085603\\
56.8051707663848	108.896598841176\\
70.1379760596096	116.852608806566\\
85.5001374785581	111.97833919909\\
89.3935662663401	108.954275039603\\
91.2585816604645	96.960963192043\\
86.2283881651862	84.6665152740861\\
81.6414354962481	78.1582033886771\\
67.2767992472321	79.4945786074932\\
59.5024593871773	82.0143168742034\\
55.0931960022362	90.9370048303354\\
48.1285351992506	95.2480804977204\\
50.6746106404277	108.010202959293\\
58.9745267465381	117.854506727421\\
64.4969712785901	120.829481620454\\
72.5419560065801	123.37830264412\\
88.1469320675952	115.161457577367\\
93.7537703736014	109.487868345455\\
99.0911322754223	96.702229733226\\
94.2602181152085	83.3376007085863\\
88.6689118831561	75.8953159215125\\
77.5014458790428	72.2974908120195\\
61.3823731650968	72.2202719646702\\
48.7151360562986	84.6454897980402\\
38.4012966822134	93.838673090675\\
43.597342864412	107.806327746828\\
51.9055187495601	119.832662348283\\
57.5772393695609	125.482225632408\\
76.3016219582229	128.059391751067\\
93.8149870652383	123.696607341171\\
101.969753133537	114.49721678218\\
101.707281141379	105.132158524004\\
106.720669416426	91.3662250555317\\
98.4187389902494	74.2585826317045\\
91.576639368054	68.463352254239\\
76.900479267035	62.198055161276\\
69.0169407719576	64.5565033146377\\
50.006732611132	72.5134155788655\\
47.9219800421171	78.8984193355973\\
42.0013328831297	76.8112252402316\\
35.6483622434449	115.499318350219\\
54.9481951262048	132.396538462137\\
74.0348569204055	138.618367467188\\
89.2853673784418	133.504884838239\\
114.091126935388	95.8782047094974\\
103.289771547203	76.2823299426782\\
52.3716116139312	63.7717904491547\\
};

\node[right, align=left]
at (axis cs:-100.394-15,-30.608) {S1};
\node[right, align=left]
at (axis cs:27.335-17,-67.161) {S2};
\node[right, align=left]
at (axis cs:73.058-15,97.768) {S3};
\end{axis}
\end{tikzpicture}%
\caption{LOFAR - AARTFAAC Stations ($M_{\text{S}}=3, M_{\text{A}}=46$)}
\label{fig:LOFAR:aartfaac:pos}
\end{figure}
\subsection{LOFAR Dataset 2 - AARTFAAC Stations}
As a second dataset, radio measurements from the Amsterdam-ASTRON Radio Transients Facility and Analysis Center (AARTFAAC) all-sky monitor are considered. The measurements include radio sensor data from $M_{\text{S}}=3$ LOFAR superterp stations, each with $M_{\text{A}}=46$ active single-polarized LBAs. Fig. \ref{fig:LOFAR:aartfaac:pos} shows the spatial arrangement of the three considered LOFAR stations and the $M=138$ sensor elements.
\begin{figure}
\centering
    \begin{subfigure}[t]{0.25\textwidth}
    \centering
	\setlength{\figurewidth}{5.1cm}
	\setlength{\figureheight}{5.1cm}
%
%
\begin{tikzpicture}[scale=0.55]

\begin{axis}[%
width=\figurewidth,
height=\figureheight,
at={(0.92in,0.642in)},
scale only axis,
point meta min=1e-08,
point meta max=0.00410480590515569,
axis on top,
xmin=-1.00200400801603,
xmax=1.00200400801603,
xlabel style={font=\color{white!15!black}},
xlabel={$\text{South }\leftarrow\text{ Direction }\rightarrow\text{ North}$},
y dir=reverse,
ymin=-1.00200400801603,
ymax=1.00200400801603,
ylabel style={font=\color{white!15!black}},
ylabel={$\text{East }\leftarrow\text{ Direction }\rightarrow\text{ West}$},
axis background/.style={fill=white},
legend style={legend cell align=left, align=left, draw=white!15!black},
colormap={mymap}{[1pt] rgb(0pt)=(0.2422,0.1504,0.6603); rgb(1pt)=(0.25039,0.164995,0.707614); rgb(2pt)=(0.257771,0.181781,0.751138); rgb(3pt)=(0.264729,0.197757,0.795214); rgb(4pt)=(0.270648,0.214676,0.836371); rgb(5pt)=(0.275114,0.234238,0.870986); rgb(6pt)=(0.2783,0.255871,0.899071); rgb(7pt)=(0.280333,0.278233,0.9221); rgb(8pt)=(0.281338,0.300595,0.941376); rgb(9pt)=(0.281014,0.322757,0.957886); rgb(10pt)=(0.279467,0.344671,0.971676); rgb(11pt)=(0.275971,0.366681,0.982905); rgb(12pt)=(0.269914,0.3892,0.9906); rgb(13pt)=(0.260243,0.412329,0.995157); rgb(14pt)=(0.244033,0.435833,0.998833); rgb(15pt)=(0.220643,0.460257,0.997286); rgb(16pt)=(0.196333,0.484719,0.989152); rgb(17pt)=(0.183405,0.507371,0.979795); rgb(18pt)=(0.178643,0.528857,0.968157); rgb(19pt)=(0.176438,0.549905,0.952019); rgb(20pt)=(0.168743,0.570262,0.935871); rgb(21pt)=(0.154,0.5902,0.9218); rgb(22pt)=(0.146029,0.609119,0.907857); rgb(23pt)=(0.138024,0.627629,0.89729); rgb(24pt)=(0.124814,0.645929,0.888343); rgb(25pt)=(0.111252,0.6635,0.876314); rgb(26pt)=(0.0952095,0.679829,0.859781); rgb(27pt)=(0.0688714,0.694771,0.839357); rgb(28pt)=(0.0296667,0.708167,0.816333); rgb(29pt)=(0.00357143,0.720267,0.7917); rgb(30pt)=(0.00665714,0.731214,0.766014); rgb(31pt)=(0.0433286,0.741095,0.73941); rgb(32pt)=(0.0963952,0.75,0.712038); rgb(33pt)=(0.140771,0.7584,0.684157); rgb(34pt)=(0.1717,0.766962,0.655443); rgb(35pt)=(0.193767,0.775767,0.6251); rgb(36pt)=(0.216086,0.7843,0.5923); rgb(37pt)=(0.246957,0.791795,0.556743); rgb(38pt)=(0.290614,0.79729,0.518829); rgb(39pt)=(0.340643,0.8008,0.478857); rgb(40pt)=(0.3909,0.802871,0.435448); rgb(41pt)=(0.445629,0.802419,0.390919); rgb(42pt)=(0.5044,0.7993,0.348); rgb(43pt)=(0.561562,0.794233,0.304481); rgb(44pt)=(0.617395,0.787619,0.261238); rgb(45pt)=(0.671986,0.779271,0.2227); rgb(46pt)=(0.7242,0.769843,0.191029); rgb(47pt)=(0.773833,0.759805,0.16461); rgb(48pt)=(0.820314,0.749814,0.153529); rgb(49pt)=(0.863433,0.7406,0.159633); rgb(50pt)=(0.903543,0.733029,0.177414); rgb(51pt)=(0.939257,0.728786,0.209957); rgb(52pt)=(0.972757,0.729771,0.239443); rgb(53pt)=(0.995648,0.743371,0.237148); rgb(54pt)=(0.996986,0.765857,0.219943); rgb(55pt)=(0.995205,0.789252,0.202762); rgb(56pt)=(0.9892,0.813567,0.188533); rgb(57pt)=(0.978629,0.838629,0.176557); rgb(58pt)=(0.967648,0.8639,0.16429); rgb(59pt)=(0.96101,0.889019,0.153676); rgb(60pt)=(0.959671,0.913457,0.142257); rgb(61pt)=(0.962795,0.937338,0.12651); rgb(62pt)=(0.969114,0.960629,0.106362); rgb(63pt)=(0.9769,0.9839,0.0805)},
colorbar
]
\addplot [forget plot] graphics [xmin=-1.00200400801603, xmax=1.00200400801603, ymin=-1.00200400801603, ymax=1.00200400801603] {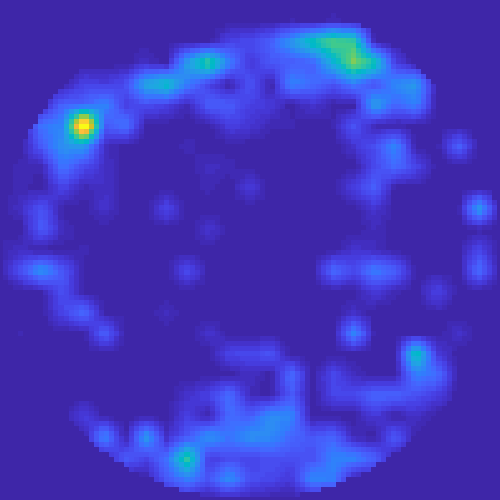};
\end{axis}
\end{tikzpicture}%
	\caption{$64$-bit Data (S1)}
	\label{fig:AARTFAAC:s3:mle:image:64bit}
\end{subfigure}%
    ~ 
\begin{subfigure}[t]{0.25\textwidth}
\centering
	\setlength{\figurewidth}{5.1cm}
	\setlength{\figureheight}{5.1cm}
%
%
\begin{tikzpicture}[scale=0.55]

\begin{axis}[%
width=\figurewidth,
height=\figureheight,
at={(0.92in,0.642in)},
scale only axis,
point meta min=1e-08,
point meta max=0.00423018821887344,
axis on top,
xmin=-1.00200400801603,
xmax=1.00200400801603,
xlabel style={font=\color{white!15!black}},
xlabel={$\text{South }\leftarrow\text{ Direction }\rightarrow\text{ North}$},
y dir=reverse,
ymin=-1.00200400801603,
ymax=1.00200400801603,
ylabel style={font=\color{white!15!black}},
axis background/.style={fill=white},
legend style={legend cell align=left, align=left, draw=white!15!black},
colormap={mymap}{[1pt] rgb(0pt)=(0.2422,0.1504,0.6603); rgb(1pt)=(0.25039,0.164995,0.707614); rgb(2pt)=(0.257771,0.181781,0.751138); rgb(3pt)=(0.264729,0.197757,0.795214); rgb(4pt)=(0.270648,0.214676,0.836371); rgb(5pt)=(0.275114,0.234238,0.870986); rgb(6pt)=(0.2783,0.255871,0.899071); rgb(7pt)=(0.280333,0.278233,0.9221); rgb(8pt)=(0.281338,0.300595,0.941376); rgb(9pt)=(0.281014,0.322757,0.957886); rgb(10pt)=(0.279467,0.344671,0.971676); rgb(11pt)=(0.275971,0.366681,0.982905); rgb(12pt)=(0.269914,0.3892,0.9906); rgb(13pt)=(0.260243,0.412329,0.995157); rgb(14pt)=(0.244033,0.435833,0.998833); rgb(15pt)=(0.220643,0.460257,0.997286); rgb(16pt)=(0.196333,0.484719,0.989152); rgb(17pt)=(0.183405,0.507371,0.979795); rgb(18pt)=(0.178643,0.528857,0.968157); rgb(19pt)=(0.176438,0.549905,0.952019); rgb(20pt)=(0.168743,0.570262,0.935871); rgb(21pt)=(0.154,0.5902,0.9218); rgb(22pt)=(0.146029,0.609119,0.907857); rgb(23pt)=(0.138024,0.627629,0.89729); rgb(24pt)=(0.124814,0.645929,0.888343); rgb(25pt)=(0.111252,0.6635,0.876314); rgb(26pt)=(0.0952095,0.679829,0.859781); rgb(27pt)=(0.0688714,0.694771,0.839357); rgb(28pt)=(0.0296667,0.708167,0.816333); rgb(29pt)=(0.00357143,0.720267,0.7917); rgb(30pt)=(0.00665714,0.731214,0.766014); rgb(31pt)=(0.0433286,0.741095,0.73941); rgb(32pt)=(0.0963952,0.75,0.712038); rgb(33pt)=(0.140771,0.7584,0.684157); rgb(34pt)=(0.1717,0.766962,0.655443); rgb(35pt)=(0.193767,0.775767,0.6251); rgb(36pt)=(0.216086,0.7843,0.5923); rgb(37pt)=(0.246957,0.791795,0.556743); rgb(38pt)=(0.290614,0.79729,0.518829); rgb(39pt)=(0.340643,0.8008,0.478857); rgb(40pt)=(0.3909,0.802871,0.435448); rgb(41pt)=(0.445629,0.802419,0.390919); rgb(42pt)=(0.5044,0.7993,0.348); rgb(43pt)=(0.561562,0.794233,0.304481); rgb(44pt)=(0.617395,0.787619,0.261238); rgb(45pt)=(0.671986,0.779271,0.2227); rgb(46pt)=(0.7242,0.769843,0.191029); rgb(47pt)=(0.773833,0.759805,0.16461); rgb(48pt)=(0.820314,0.749814,0.153529); rgb(49pt)=(0.863433,0.7406,0.159633); rgb(50pt)=(0.903543,0.733029,0.177414); rgb(51pt)=(0.939257,0.728786,0.209957); rgb(52pt)=(0.972757,0.729771,0.239443); rgb(53pt)=(0.995648,0.743371,0.237148); rgb(54pt)=(0.996986,0.765857,0.219943); rgb(55pt)=(0.995205,0.789252,0.202762); rgb(56pt)=(0.9892,0.813567,0.188533); rgb(57pt)=(0.978629,0.838629,0.176557); rgb(58pt)=(0.967648,0.8639,0.16429); rgb(59pt)=(0.96101,0.889019,0.153676); rgb(60pt)=(0.959671,0.913457,0.142257); rgb(61pt)=(0.962795,0.937338,0.12651); rgb(62pt)=(0.969114,0.960629,0.106362); rgb(63pt)=(0.9769,0.9839,0.0805)},
colorbar
]
\addplot [forget plot] graphics [xmin=-1.00200400801603, xmax=1.00200400801603, ymin=-1.00200400801603, ymax=1.00200400801603] {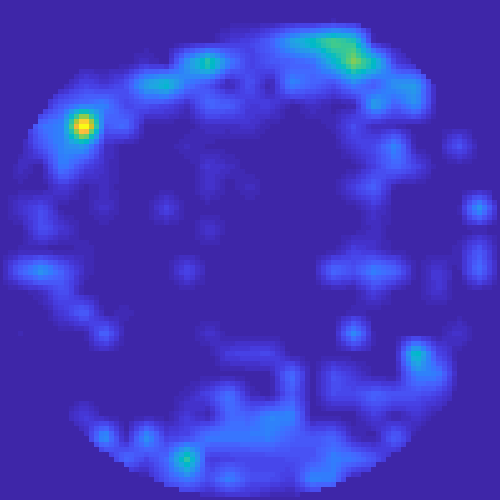};
\end{axis}
\end{tikzpicture}%
	\caption{$1$-bit Data (S1)}
	\label{fig:AARTFAAC:s3:cmle:image:1bit}
\end{subfigure}

\begin{subfigure}[t]{0.25\textwidth}
    \centering
	\setlength{\figurewidth}{5.1cm}
	\setlength{\figureheight}{5.1cm}
%
%
\begin{tikzpicture}[scale=0.55]

\begin{axis}[%
width=\figurewidth,
height=\figureheight,
at={(0.92in,0.642in)},
scale only axis,
point meta min=1e-08,
point meta max=0.00392743655944806,
axis on top,
xmin=-1.00200400801603,
xmax=1.00200400801603,
xlabel style={font=\color{white!15!black}},
xlabel={$\text{South }\leftarrow\text{ Direction }\rightarrow\text{ North}$},
y dir=reverse,
ymin=-1.00200400801603,
ymax=1.00200400801603,
ylabel style={font=\color{white!15!black}},
ylabel={$\text{East }\leftarrow\text{ Direction }\rightarrow\text{ West}$},
axis background/.style={fill=white},
legend style={legend cell align=left, align=left, draw=white!15!black},
colormap={mymap}{[1pt] rgb(0pt)=(0.2422,0.1504,0.6603); rgb(1pt)=(0.25039,0.164995,0.707614); rgb(2pt)=(0.257771,0.181781,0.751138); rgb(3pt)=(0.264729,0.197757,0.795214); rgb(4pt)=(0.270648,0.214676,0.836371); rgb(5pt)=(0.275114,0.234238,0.870986); rgb(6pt)=(0.2783,0.255871,0.899071); rgb(7pt)=(0.280333,0.278233,0.9221); rgb(8pt)=(0.281338,0.300595,0.941376); rgb(9pt)=(0.281014,0.322757,0.957886); rgb(10pt)=(0.279467,0.344671,0.971676); rgb(11pt)=(0.275971,0.366681,0.982905); rgb(12pt)=(0.269914,0.3892,0.9906); rgb(13pt)=(0.260243,0.412329,0.995157); rgb(14pt)=(0.244033,0.435833,0.998833); rgb(15pt)=(0.220643,0.460257,0.997286); rgb(16pt)=(0.196333,0.484719,0.989152); rgb(17pt)=(0.183405,0.507371,0.979795); rgb(18pt)=(0.178643,0.528857,0.968157); rgb(19pt)=(0.176438,0.549905,0.952019); rgb(20pt)=(0.168743,0.570262,0.935871); rgb(21pt)=(0.154,0.5902,0.9218); rgb(22pt)=(0.146029,0.609119,0.907857); rgb(23pt)=(0.138024,0.627629,0.89729); rgb(24pt)=(0.124814,0.645929,0.888343); rgb(25pt)=(0.111252,0.6635,0.876314); rgb(26pt)=(0.0952095,0.679829,0.859781); rgb(27pt)=(0.0688714,0.694771,0.839357); rgb(28pt)=(0.0296667,0.708167,0.816333); rgb(29pt)=(0.00357143,0.720267,0.7917); rgb(30pt)=(0.00665714,0.731214,0.766014); rgb(31pt)=(0.0433286,0.741095,0.73941); rgb(32pt)=(0.0963952,0.75,0.712038); rgb(33pt)=(0.140771,0.7584,0.684157); rgb(34pt)=(0.1717,0.766962,0.655443); rgb(35pt)=(0.193767,0.775767,0.6251); rgb(36pt)=(0.216086,0.7843,0.5923); rgb(37pt)=(0.246957,0.791795,0.556743); rgb(38pt)=(0.290614,0.79729,0.518829); rgb(39pt)=(0.340643,0.8008,0.478857); rgb(40pt)=(0.3909,0.802871,0.435448); rgb(41pt)=(0.445629,0.802419,0.390919); rgb(42pt)=(0.5044,0.7993,0.348); rgb(43pt)=(0.561562,0.794233,0.304481); rgb(44pt)=(0.617395,0.787619,0.261238); rgb(45pt)=(0.671986,0.779271,0.2227); rgb(46pt)=(0.7242,0.769843,0.191029); rgb(47pt)=(0.773833,0.759805,0.16461); rgb(48pt)=(0.820314,0.749814,0.153529); rgb(49pt)=(0.863433,0.7406,0.159633); rgb(50pt)=(0.903543,0.733029,0.177414); rgb(51pt)=(0.939257,0.728786,0.209957); rgb(52pt)=(0.972757,0.729771,0.239443); rgb(53pt)=(0.995648,0.743371,0.237148); rgb(54pt)=(0.996986,0.765857,0.219943); rgb(55pt)=(0.995205,0.789252,0.202762); rgb(56pt)=(0.9892,0.813567,0.188533); rgb(57pt)=(0.978629,0.838629,0.176557); rgb(58pt)=(0.967648,0.8639,0.16429); rgb(59pt)=(0.96101,0.889019,0.153676); rgb(60pt)=(0.959671,0.913457,0.142257); rgb(61pt)=(0.962795,0.937338,0.12651); rgb(62pt)=(0.969114,0.960629,0.106362); rgb(63pt)=(0.9769,0.9839,0.0805)},
colorbar
]
\addplot [forget plot] graphics [xmin=-1.00200400801603, xmax=1.00200400801603, ymin=-1.00200400801603, ymax=1.00200400801603] {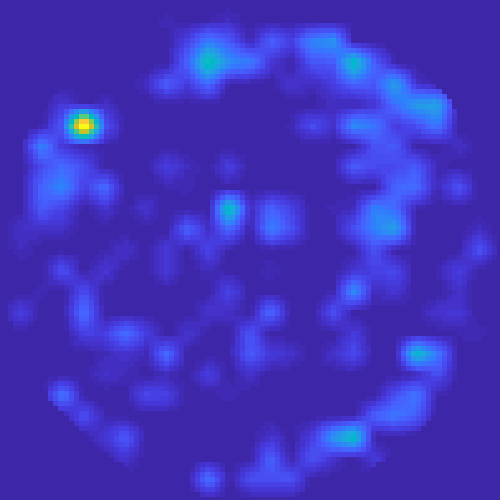};
\end{axis}
\end{tikzpicture}%
	\caption{$64$-bit Data (S2)}
	\label{fig:AARTFAAC:s4:mle:image:64bit}
\end{subfigure}%
    ~ 
\begin{subfigure}[t]{0.25\textwidth}
\centering
	\setlength{\figurewidth}{5.1cm}
	\setlength{\figureheight}{5.1cm}
%
%
\begin{tikzpicture}[scale=0.55]

\begin{axis}[%
width=\figurewidth,
height=\figureheight,
at={(0.92in,0.642in)},
scale only axis,
point meta min=1e-08,
point meta max=0.00382110616864291,
axis on top,
xmin=-1.00200400801603,
xmax=1.00200400801603,
xlabel style={font=\color{white!15!black}},
xlabel={$\text{South }\leftarrow\text{ Direction }\rightarrow\text{ North}$},
y dir=reverse,
ymin=-1.00200400801603,
ymax=1.00200400801603,
ylabel style={font=\color{white!15!black}},
axis background/.style={fill=white},
legend style={legend cell align=left, align=left, draw=white!15!black},
colormap={mymap}{[1pt] rgb(0pt)=(0.2422,0.1504,0.6603); rgb(1pt)=(0.25039,0.164995,0.707614); rgb(2pt)=(0.257771,0.181781,0.751138); rgb(3pt)=(0.264729,0.197757,0.795214); rgb(4pt)=(0.270648,0.214676,0.836371); rgb(5pt)=(0.275114,0.234238,0.870986); rgb(6pt)=(0.2783,0.255871,0.899071); rgb(7pt)=(0.280333,0.278233,0.9221); rgb(8pt)=(0.281338,0.300595,0.941376); rgb(9pt)=(0.281014,0.322757,0.957886); rgb(10pt)=(0.279467,0.344671,0.971676); rgb(11pt)=(0.275971,0.366681,0.982905); rgb(12pt)=(0.269914,0.3892,0.9906); rgb(13pt)=(0.260243,0.412329,0.995157); rgb(14pt)=(0.244033,0.435833,0.998833); rgb(15pt)=(0.220643,0.460257,0.997286); rgb(16pt)=(0.196333,0.484719,0.989152); rgb(17pt)=(0.183405,0.507371,0.979795); rgb(18pt)=(0.178643,0.528857,0.968157); rgb(19pt)=(0.176438,0.549905,0.952019); rgb(20pt)=(0.168743,0.570262,0.935871); rgb(21pt)=(0.154,0.5902,0.9218); rgb(22pt)=(0.146029,0.609119,0.907857); rgb(23pt)=(0.138024,0.627629,0.89729); rgb(24pt)=(0.124814,0.645929,0.888343); rgb(25pt)=(0.111252,0.6635,0.876314); rgb(26pt)=(0.0952095,0.679829,0.859781); rgb(27pt)=(0.0688714,0.694771,0.839357); rgb(28pt)=(0.0296667,0.708167,0.816333); rgb(29pt)=(0.00357143,0.720267,0.7917); rgb(30pt)=(0.00665714,0.731214,0.766014); rgb(31pt)=(0.0433286,0.741095,0.73941); rgb(32pt)=(0.0963952,0.75,0.712038); rgb(33pt)=(0.140771,0.7584,0.684157); rgb(34pt)=(0.1717,0.766962,0.655443); rgb(35pt)=(0.193767,0.775767,0.6251); rgb(36pt)=(0.216086,0.7843,0.5923); rgb(37pt)=(0.246957,0.791795,0.556743); rgb(38pt)=(0.290614,0.79729,0.518829); rgb(39pt)=(0.340643,0.8008,0.478857); rgb(40pt)=(0.3909,0.802871,0.435448); rgb(41pt)=(0.445629,0.802419,0.390919); rgb(42pt)=(0.5044,0.7993,0.348); rgb(43pt)=(0.561562,0.794233,0.304481); rgb(44pt)=(0.617395,0.787619,0.261238); rgb(45pt)=(0.671986,0.779271,0.2227); rgb(46pt)=(0.7242,0.769843,0.191029); rgb(47pt)=(0.773833,0.759805,0.16461); rgb(48pt)=(0.820314,0.749814,0.153529); rgb(49pt)=(0.863433,0.7406,0.159633); rgb(50pt)=(0.903543,0.733029,0.177414); rgb(51pt)=(0.939257,0.728786,0.209957); rgb(52pt)=(0.972757,0.729771,0.239443); rgb(53pt)=(0.995648,0.743371,0.237148); rgb(54pt)=(0.996986,0.765857,0.219943); rgb(55pt)=(0.995205,0.789252,0.202762); rgb(56pt)=(0.9892,0.813567,0.188533); rgb(57pt)=(0.978629,0.838629,0.176557); rgb(58pt)=(0.967648,0.8639,0.16429); rgb(59pt)=(0.96101,0.889019,0.153676); rgb(60pt)=(0.959671,0.913457,0.142257); rgb(61pt)=(0.962795,0.937338,0.12651); rgb(62pt)=(0.969114,0.960629,0.106362); rgb(63pt)=(0.9769,0.9839,0.0805)},
colorbar
]
\addplot [forget plot] graphics [xmin=-1.00200400801603, xmax=1.00200400801603, ymin=-1.00200400801603, ymax=1.00200400801603] {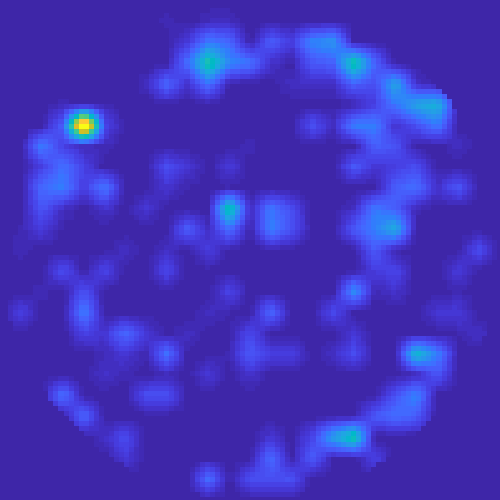};
\end{axis}
\end{tikzpicture}%
	\caption{$1$-bit Data (S2)}
	\label{fig:AARTFAAC:s4:cmle:image:1bit}
\end{subfigure}%

\begin{subfigure}[t]{0.25\textwidth}
    \centering
	\setlength{\figurewidth}{5.1cm}
	\setlength{\figureheight}{5.1cm}
%
%
\begin{tikzpicture}[scale=0.55]

\begin{axis}[%
width=\figurewidth,
height=\figureheight,
at={(0.92in,0.642in)},
scale only axis,
point meta min=1e-08,
point meta max=0.00334891648598148,
axis on top,
xmin=-1.00200400801603,
xmax=1.00200400801603,
xlabel style={font=\color{white!15!black}},
xlabel={$\text{South }\leftarrow\text{ Direction }\rightarrow\text{ North}$},
y dir=reverse,
ymin=-1.00200400801603,
ymax=1.00200400801603,
ylabel style={font=\color{white!15!black}},
ylabel={$\text{East }\leftarrow\text{ Direction }\rightarrow\text{ West}$},
axis background/.style={fill=white},
legend style={legend cell align=left, align=left, draw=white!15!black},
colormap={mymap}{[1pt] rgb(0pt)=(0.2422,0.1504,0.6603); rgb(1pt)=(0.25039,0.164995,0.707614); rgb(2pt)=(0.257771,0.181781,0.751138); rgb(3pt)=(0.264729,0.197757,0.795214); rgb(4pt)=(0.270648,0.214676,0.836371); rgb(5pt)=(0.275114,0.234238,0.870986); rgb(6pt)=(0.2783,0.255871,0.899071); rgb(7pt)=(0.280333,0.278233,0.9221); rgb(8pt)=(0.281338,0.300595,0.941376); rgb(9pt)=(0.281014,0.322757,0.957886); rgb(10pt)=(0.279467,0.344671,0.971676); rgb(11pt)=(0.275971,0.366681,0.982905); rgb(12pt)=(0.269914,0.3892,0.9906); rgb(13pt)=(0.260243,0.412329,0.995157); rgb(14pt)=(0.244033,0.435833,0.998833); rgb(15pt)=(0.220643,0.460257,0.997286); rgb(16pt)=(0.196333,0.484719,0.989152); rgb(17pt)=(0.183405,0.507371,0.979795); rgb(18pt)=(0.178643,0.528857,0.968157); rgb(19pt)=(0.176438,0.549905,0.952019); rgb(20pt)=(0.168743,0.570262,0.935871); rgb(21pt)=(0.154,0.5902,0.9218); rgb(22pt)=(0.146029,0.609119,0.907857); rgb(23pt)=(0.138024,0.627629,0.89729); rgb(24pt)=(0.124814,0.645929,0.888343); rgb(25pt)=(0.111252,0.6635,0.876314); rgb(26pt)=(0.0952095,0.679829,0.859781); rgb(27pt)=(0.0688714,0.694771,0.839357); rgb(28pt)=(0.0296667,0.708167,0.816333); rgb(29pt)=(0.00357143,0.720267,0.7917); rgb(30pt)=(0.00665714,0.731214,0.766014); rgb(31pt)=(0.0433286,0.741095,0.73941); rgb(32pt)=(0.0963952,0.75,0.712038); rgb(33pt)=(0.140771,0.7584,0.684157); rgb(34pt)=(0.1717,0.766962,0.655443); rgb(35pt)=(0.193767,0.775767,0.6251); rgb(36pt)=(0.216086,0.7843,0.5923); rgb(37pt)=(0.246957,0.791795,0.556743); rgb(38pt)=(0.290614,0.79729,0.518829); rgb(39pt)=(0.340643,0.8008,0.478857); rgb(40pt)=(0.3909,0.802871,0.435448); rgb(41pt)=(0.445629,0.802419,0.390919); rgb(42pt)=(0.5044,0.7993,0.348); rgb(43pt)=(0.561562,0.794233,0.304481); rgb(44pt)=(0.617395,0.787619,0.261238); rgb(45pt)=(0.671986,0.779271,0.2227); rgb(46pt)=(0.7242,0.769843,0.191029); rgb(47pt)=(0.773833,0.759805,0.16461); rgb(48pt)=(0.820314,0.749814,0.153529); rgb(49pt)=(0.863433,0.7406,0.159633); rgb(50pt)=(0.903543,0.733029,0.177414); rgb(51pt)=(0.939257,0.728786,0.209957); rgb(52pt)=(0.972757,0.729771,0.239443); rgb(53pt)=(0.995648,0.743371,0.237148); rgb(54pt)=(0.996986,0.765857,0.219943); rgb(55pt)=(0.995205,0.789252,0.202762); rgb(56pt)=(0.9892,0.813567,0.188533); rgb(57pt)=(0.978629,0.838629,0.176557); rgb(58pt)=(0.967648,0.8639,0.16429); rgb(59pt)=(0.96101,0.889019,0.153676); rgb(60pt)=(0.959671,0.913457,0.142257); rgb(61pt)=(0.962795,0.937338,0.12651); rgb(62pt)=(0.969114,0.960629,0.106362); rgb(63pt)=(0.9769,0.9839,0.0805)},
colorbar
]
\addplot [forget plot] graphics [xmin=-1.00200400801603, xmax=1.00200400801603, ymin=-1.00200400801603, ymax=1.00200400801603] {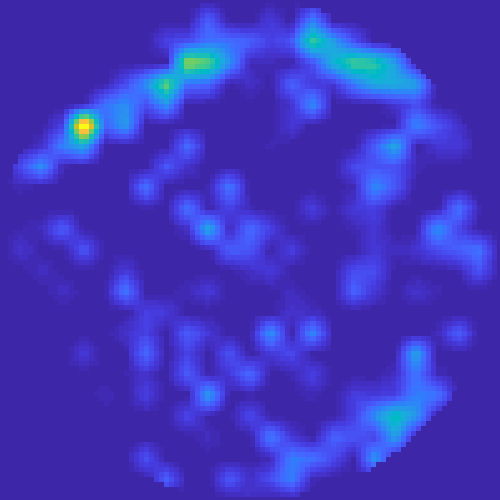};
\end{axis}
\end{tikzpicture}%
	\caption{$64$-bit Data (S3)}
	\label{fig:AARTFAAC:s6:mle:image:64bit}
\end{subfigure}%
    ~ 
\begin{subfigure}[t]{0.25\textwidth}
\centering
	\setlength{\figurewidth}{5.1cm}
	\setlength{\figureheight}{5.1cm}
%
%
\begin{tikzpicture}[scale=0.55]

\begin{axis}[%
width=\figurewidth,
height=\figureheight,
at={(0.92in,0.642in)},
scale only axis,
point meta min=1e-08,
point meta max=0.00327511217869635,
axis on top,
xmin=-1.00200400801603,
xmax=1.00200400801603,
xlabel style={font=\color{white!15!black}},
xlabel={$\text{South }\leftarrow\text{ Direction }\rightarrow\text{ North}$},
y dir=reverse,
ymin=-1.00200400801603,
ymax=1.00200400801603,
ylabel style={font=\color{white!15!black}},
axis background/.style={fill=white},
legend style={legend cell align=left, align=left, draw=white!15!black},
colormap={mymap}{[1pt] rgb(0pt)=(0.2422,0.1504,0.6603); rgb(1pt)=(0.25039,0.164995,0.707614); rgb(2pt)=(0.257771,0.181781,0.751138); rgb(3pt)=(0.264729,0.197757,0.795214); rgb(4pt)=(0.270648,0.214676,0.836371); rgb(5pt)=(0.275114,0.234238,0.870986); rgb(6pt)=(0.2783,0.255871,0.899071); rgb(7pt)=(0.280333,0.278233,0.9221); rgb(8pt)=(0.281338,0.300595,0.941376); rgb(9pt)=(0.281014,0.322757,0.957886); rgb(10pt)=(0.279467,0.344671,0.971676); rgb(11pt)=(0.275971,0.366681,0.982905); rgb(12pt)=(0.269914,0.3892,0.9906); rgb(13pt)=(0.260243,0.412329,0.995157); rgb(14pt)=(0.244033,0.435833,0.998833); rgb(15pt)=(0.220643,0.460257,0.997286); rgb(16pt)=(0.196333,0.484719,0.989152); rgb(17pt)=(0.183405,0.507371,0.979795); rgb(18pt)=(0.178643,0.528857,0.968157); rgb(19pt)=(0.176438,0.549905,0.952019); rgb(20pt)=(0.168743,0.570262,0.935871); rgb(21pt)=(0.154,0.5902,0.9218); rgb(22pt)=(0.146029,0.609119,0.907857); rgb(23pt)=(0.138024,0.627629,0.89729); rgb(24pt)=(0.124814,0.645929,0.888343); rgb(25pt)=(0.111252,0.6635,0.876314); rgb(26pt)=(0.0952095,0.679829,0.859781); rgb(27pt)=(0.0688714,0.694771,0.839357); rgb(28pt)=(0.0296667,0.708167,0.816333); rgb(29pt)=(0.00357143,0.720267,0.7917); rgb(30pt)=(0.00665714,0.731214,0.766014); rgb(31pt)=(0.0433286,0.741095,0.73941); rgb(32pt)=(0.0963952,0.75,0.712038); rgb(33pt)=(0.140771,0.7584,0.684157); rgb(34pt)=(0.1717,0.766962,0.655443); rgb(35pt)=(0.193767,0.775767,0.6251); rgb(36pt)=(0.216086,0.7843,0.5923); rgb(37pt)=(0.246957,0.791795,0.556743); rgb(38pt)=(0.290614,0.79729,0.518829); rgb(39pt)=(0.340643,0.8008,0.478857); rgb(40pt)=(0.3909,0.802871,0.435448); rgb(41pt)=(0.445629,0.802419,0.390919); rgb(42pt)=(0.5044,0.7993,0.348); rgb(43pt)=(0.561562,0.794233,0.304481); rgb(44pt)=(0.617395,0.787619,0.261238); rgb(45pt)=(0.671986,0.779271,0.2227); rgb(46pt)=(0.7242,0.769843,0.191029); rgb(47pt)=(0.773833,0.759805,0.16461); rgb(48pt)=(0.820314,0.749814,0.153529); rgb(49pt)=(0.863433,0.7406,0.159633); rgb(50pt)=(0.903543,0.733029,0.177414); rgb(51pt)=(0.939257,0.728786,0.209957); rgb(52pt)=(0.972757,0.729771,0.239443); rgb(53pt)=(0.995648,0.743371,0.237148); rgb(54pt)=(0.996986,0.765857,0.219943); rgb(55pt)=(0.995205,0.789252,0.202762); rgb(56pt)=(0.9892,0.813567,0.188533); rgb(57pt)=(0.978629,0.838629,0.176557); rgb(58pt)=(0.967648,0.8639,0.16429); rgb(59pt)=(0.96101,0.889019,0.153676); rgb(60pt)=(0.959671,0.913457,0.142257); rgb(61pt)=(0.962795,0.937338,0.12651); rgb(62pt)=(0.969114,0.960629,0.106362); rgb(63pt)=(0.9769,0.9839,0.0805)},
colorbar
]
\addplot [forget plot] graphics [xmin=-1.00200400801603, xmax=1.00200400801603, ymin=-1.00200400801603, ymax=1.00200400801603] {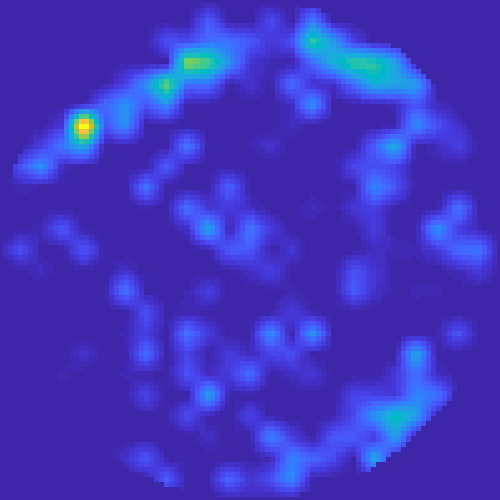};
\end{axis}
\end{tikzpicture}%
	\caption{$1$-bit Data (S3)}
	\label{fig:AARTFAAC:s6:cmle:image:1bit}
\end{subfigure}%
\caption{AARTFAAC - All-Sky Image ($M=46, D_\text{S}=441$)}
\label{fig:LOFAR:aartfaac:indi:image}
\end{figure}
\begin{figure}
\centering
    \begin{subfigure}[t]{0.25\textwidth}
    \centering
	\setlength{\figurewidth}{5.1cm}
	\setlength{\figureheight}{5.1cm}
%
%
\begin{tikzpicture}[scale=0.55]

\begin{axis}[%
width=\figurewidth,
height=\figureheight,
at={(0.92in,0.642in)},
scale only axis,
point meta min=0.0798866003679426,
point meta max=0.0848439015849966,
axis on top,
xmin=-1.00200400801603,
xmax=1.00200400801603,
xlabel style={font=\color{white!15!black}},
xlabel={$\text{South }\leftarrow\text{ Direction }\rightarrow\text{ North}$},
y dir=reverse,
ymin=-1.00200400801603,
ymax=1.00200400801603,
ylabel style={font=\color{white!15!black}},
ylabel={$\text{East }\leftarrow\text{ Direction }\rightarrow\text{ West}$},
axis background/.style={fill=white},
legend style={legend cell align=left, align=left, draw=white!15!black},
colormap={mymap}{[1pt] rgb(0pt)=(0.2422,0.1504,0.6603); rgb(1pt)=(0.25039,0.164995,0.707614); rgb(2pt)=(0.257771,0.181781,0.751138); rgb(3pt)=(0.264729,0.197757,0.795214); rgb(4pt)=(0.270648,0.214676,0.836371); rgb(5pt)=(0.275114,0.234238,0.870986); rgb(6pt)=(0.2783,0.255871,0.899071); rgb(7pt)=(0.280333,0.278233,0.9221); rgb(8pt)=(0.281338,0.300595,0.941376); rgb(9pt)=(0.281014,0.322757,0.957886); rgb(10pt)=(0.279467,0.344671,0.971676); rgb(11pt)=(0.275971,0.366681,0.982905); rgb(12pt)=(0.269914,0.3892,0.9906); rgb(13pt)=(0.260243,0.412329,0.995157); rgb(14pt)=(0.244033,0.435833,0.998833); rgb(15pt)=(0.220643,0.460257,0.997286); rgb(16pt)=(0.196333,0.484719,0.989152); rgb(17pt)=(0.183405,0.507371,0.979795); rgb(18pt)=(0.178643,0.528857,0.968157); rgb(19pt)=(0.176438,0.549905,0.952019); rgb(20pt)=(0.168743,0.570262,0.935871); rgb(21pt)=(0.154,0.5902,0.9218); rgb(22pt)=(0.146029,0.609119,0.907857); rgb(23pt)=(0.138024,0.627629,0.89729); rgb(24pt)=(0.124814,0.645929,0.888343); rgb(25pt)=(0.111252,0.6635,0.876314); rgb(26pt)=(0.0952095,0.679829,0.859781); rgb(27pt)=(0.0688714,0.694771,0.839357); rgb(28pt)=(0.0296667,0.708167,0.816333); rgb(29pt)=(0.00357143,0.720267,0.7917); rgb(30pt)=(0.00665714,0.731214,0.766014); rgb(31pt)=(0.0433286,0.741095,0.73941); rgb(32pt)=(0.0963952,0.75,0.712038); rgb(33pt)=(0.140771,0.7584,0.684157); rgb(34pt)=(0.1717,0.766962,0.655443); rgb(35pt)=(0.193767,0.775767,0.6251); rgb(36pt)=(0.216086,0.7843,0.5923); rgb(37pt)=(0.246957,0.791795,0.556743); rgb(38pt)=(0.290614,0.79729,0.518829); rgb(39pt)=(0.340643,0.8008,0.478857); rgb(40pt)=(0.3909,0.802871,0.435448); rgb(41pt)=(0.445629,0.802419,0.390919); rgb(42pt)=(0.5044,0.7993,0.348); rgb(43pt)=(0.561562,0.794233,0.304481); rgb(44pt)=(0.617395,0.787619,0.261238); rgb(45pt)=(0.671986,0.779271,0.2227); rgb(46pt)=(0.7242,0.769843,0.191029); rgb(47pt)=(0.773833,0.759805,0.16461); rgb(48pt)=(0.820314,0.749814,0.153529); rgb(49pt)=(0.863433,0.7406,0.159633); rgb(50pt)=(0.903543,0.733029,0.177414); rgb(51pt)=(0.939257,0.728786,0.209957); rgb(52pt)=(0.972757,0.729771,0.239443); rgb(53pt)=(0.995648,0.743371,0.237148); rgb(54pt)=(0.996986,0.765857,0.219943); rgb(55pt)=(0.995205,0.789252,0.202762); rgb(56pt)=(0.9892,0.813567,0.188533); rgb(57pt)=(0.978629,0.838629,0.176557); rgb(58pt)=(0.967648,0.8639,0.16429); rgb(59pt)=(0.96101,0.889019,0.153676); rgb(60pt)=(0.959671,0.913457,0.142257); rgb(61pt)=(0.962795,0.937338,0.12651); rgb(62pt)=(0.969114,0.960629,0.106362); rgb(63pt)=(0.9769,0.9839,0.0805)},
colorbar
]
\addplot [forget plot] graphics [xmin=-1.00200400801603, xmax=1.00200400801603, ymin=-1.00200400801603, ymax=1.00200400801603] {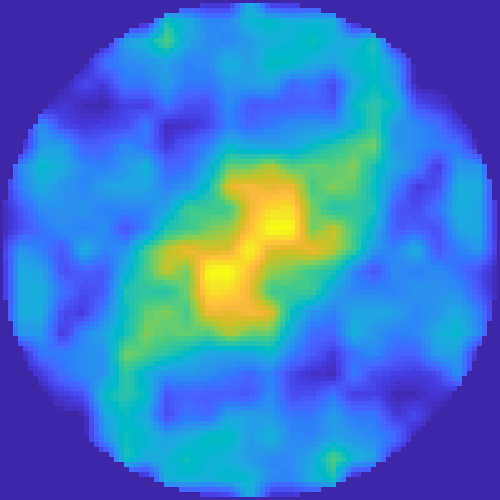};
\end{axis}
\end{tikzpicture}%
	\caption{$64$-bit Data (S1)}
	\label{fig:AARTFAAC:s3:mle:uncert:64bit}
\end{subfigure}%
    ~ 
\begin{subfigure}[t]{0.25\textwidth}
\centering
	\setlength{\figurewidth}{5.1cm}
	\setlength{\figureheight}{5.1cm}
%
%
\begin{tikzpicture}[scale=0.55]

\begin{axis}[%
width=\figurewidth,
height=\figureheight,
at={(0.92in,0.642in)},
scale only axis,
point meta min=0.125661313533783,
point meta max=0.132344577676994,
axis on top,
xmin=-1.00200400801603,
xmax=1.00200400801603,
xlabel style={font=\color{white!15!black}},
xlabel={$\text{South }\leftarrow\text{ Direction }\rightarrow\text{ North}$},
y dir=reverse,
ymin=-1.00200400801603,
ymax=1.00200400801603,
ylabel style={font=\color{white!15!black}},
axis background/.style={fill=white},
legend style={legend cell align=left, align=left, draw=white!15!black},
colormap={mymap}{[1pt] rgb(0pt)=(0.2422,0.1504,0.6603); rgb(1pt)=(0.25039,0.164995,0.707614); rgb(2pt)=(0.257771,0.181781,0.751138); rgb(3pt)=(0.264729,0.197757,0.795214); rgb(4pt)=(0.270648,0.214676,0.836371); rgb(5pt)=(0.275114,0.234238,0.870986); rgb(6pt)=(0.2783,0.255871,0.899071); rgb(7pt)=(0.280333,0.278233,0.9221); rgb(8pt)=(0.281338,0.300595,0.941376); rgb(9pt)=(0.281014,0.322757,0.957886); rgb(10pt)=(0.279467,0.344671,0.971676); rgb(11pt)=(0.275971,0.366681,0.982905); rgb(12pt)=(0.269914,0.3892,0.9906); rgb(13pt)=(0.260243,0.412329,0.995157); rgb(14pt)=(0.244033,0.435833,0.998833); rgb(15pt)=(0.220643,0.460257,0.997286); rgb(16pt)=(0.196333,0.484719,0.989152); rgb(17pt)=(0.183405,0.507371,0.979795); rgb(18pt)=(0.178643,0.528857,0.968157); rgb(19pt)=(0.176438,0.549905,0.952019); rgb(20pt)=(0.168743,0.570262,0.935871); rgb(21pt)=(0.154,0.5902,0.9218); rgb(22pt)=(0.146029,0.609119,0.907857); rgb(23pt)=(0.138024,0.627629,0.89729); rgb(24pt)=(0.124814,0.645929,0.888343); rgb(25pt)=(0.111252,0.6635,0.876314); rgb(26pt)=(0.0952095,0.679829,0.859781); rgb(27pt)=(0.0688714,0.694771,0.839357); rgb(28pt)=(0.0296667,0.708167,0.816333); rgb(29pt)=(0.00357143,0.720267,0.7917); rgb(30pt)=(0.00665714,0.731214,0.766014); rgb(31pt)=(0.0433286,0.741095,0.73941); rgb(32pt)=(0.0963952,0.75,0.712038); rgb(33pt)=(0.140771,0.7584,0.684157); rgb(34pt)=(0.1717,0.766962,0.655443); rgb(35pt)=(0.193767,0.775767,0.6251); rgb(36pt)=(0.216086,0.7843,0.5923); rgb(37pt)=(0.246957,0.791795,0.556743); rgb(38pt)=(0.290614,0.79729,0.518829); rgb(39pt)=(0.340643,0.8008,0.478857); rgb(40pt)=(0.3909,0.802871,0.435448); rgb(41pt)=(0.445629,0.802419,0.390919); rgb(42pt)=(0.5044,0.7993,0.348); rgb(43pt)=(0.561562,0.794233,0.304481); rgb(44pt)=(0.617395,0.787619,0.261238); rgb(45pt)=(0.671986,0.779271,0.2227); rgb(46pt)=(0.7242,0.769843,0.191029); rgb(47pt)=(0.773833,0.759805,0.16461); rgb(48pt)=(0.820314,0.749814,0.153529); rgb(49pt)=(0.863433,0.7406,0.159633); rgb(50pt)=(0.903543,0.733029,0.177414); rgb(51pt)=(0.939257,0.728786,0.209957); rgb(52pt)=(0.972757,0.729771,0.239443); rgb(53pt)=(0.995648,0.743371,0.237148); rgb(54pt)=(0.996986,0.765857,0.219943); rgb(55pt)=(0.995205,0.789252,0.202762); rgb(56pt)=(0.9892,0.813567,0.188533); rgb(57pt)=(0.978629,0.838629,0.176557); rgb(58pt)=(0.967648,0.8639,0.16429); rgb(59pt)=(0.96101,0.889019,0.153676); rgb(60pt)=(0.959671,0.913457,0.142257); rgb(61pt)=(0.962795,0.937338,0.12651); rgb(62pt)=(0.969114,0.960629,0.106362); rgb(63pt)=(0.9769,0.9839,0.0805)},
colorbar
]
\addplot [forget plot] graphics [xmin=-1.00200400801603, xmax=1.00200400801603, ymin=-1.00200400801603, ymax=1.00200400801603] {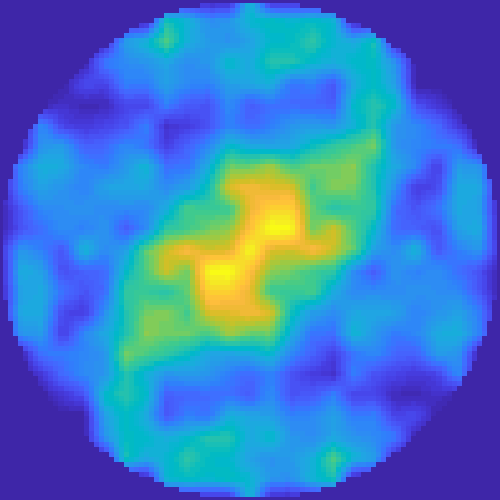};
\end{axis}
\end{tikzpicture}%
	\caption{$1$-bit Data (S1)}
	\label{fig:AARTFAAC:s3:cmle:uncert:1bit}
\end{subfigure}

\begin{subfigure}[t]{0.25\textwidth}
    \centering
	\setlength{\figurewidth}{5.1cm}
	\setlength{\figureheight}{5.1cm}
%
%
\begin{tikzpicture}[scale=0.55]

\begin{axis}[%
width=\figurewidth,
height=\figureheight,
at={(0.92in,0.642in)},
scale only axis,
point meta min=0.0790345383145937,
point meta max=0.0841499897924925,
axis on top,
xmin=-1.00200400801603,
xmax=1.00200400801603,
xlabel style={font=\color{white!15!black}},
xlabel={$\text{South }\leftarrow\text{ Direction }\rightarrow\text{ North}$},
y dir=reverse,
ymin=-1.00200400801603,
ymax=1.00200400801603,
ylabel style={font=\color{white!15!black}},
ylabel={$\text{East }\leftarrow\text{ Direction }\rightarrow\text{ West}$},
axis background/.style={fill=white},
legend style={legend cell align=left, align=left, draw=white!15!black},
colormap={mymap}{[1pt] rgb(0pt)=(0.2422,0.1504,0.6603); rgb(1pt)=(0.25039,0.164995,0.707614); rgb(2pt)=(0.257771,0.181781,0.751138); rgb(3pt)=(0.264729,0.197757,0.795214); rgb(4pt)=(0.270648,0.214676,0.836371); rgb(5pt)=(0.275114,0.234238,0.870986); rgb(6pt)=(0.2783,0.255871,0.899071); rgb(7pt)=(0.280333,0.278233,0.9221); rgb(8pt)=(0.281338,0.300595,0.941376); rgb(9pt)=(0.281014,0.322757,0.957886); rgb(10pt)=(0.279467,0.344671,0.971676); rgb(11pt)=(0.275971,0.366681,0.982905); rgb(12pt)=(0.269914,0.3892,0.9906); rgb(13pt)=(0.260243,0.412329,0.995157); rgb(14pt)=(0.244033,0.435833,0.998833); rgb(15pt)=(0.220643,0.460257,0.997286); rgb(16pt)=(0.196333,0.484719,0.989152); rgb(17pt)=(0.183405,0.507371,0.979795); rgb(18pt)=(0.178643,0.528857,0.968157); rgb(19pt)=(0.176438,0.549905,0.952019); rgb(20pt)=(0.168743,0.570262,0.935871); rgb(21pt)=(0.154,0.5902,0.9218); rgb(22pt)=(0.146029,0.609119,0.907857); rgb(23pt)=(0.138024,0.627629,0.89729); rgb(24pt)=(0.124814,0.645929,0.888343); rgb(25pt)=(0.111252,0.6635,0.876314); rgb(26pt)=(0.0952095,0.679829,0.859781); rgb(27pt)=(0.0688714,0.694771,0.839357); rgb(28pt)=(0.0296667,0.708167,0.816333); rgb(29pt)=(0.00357143,0.720267,0.7917); rgb(30pt)=(0.00665714,0.731214,0.766014); rgb(31pt)=(0.0433286,0.741095,0.73941); rgb(32pt)=(0.0963952,0.75,0.712038); rgb(33pt)=(0.140771,0.7584,0.684157); rgb(34pt)=(0.1717,0.766962,0.655443); rgb(35pt)=(0.193767,0.775767,0.6251); rgb(36pt)=(0.216086,0.7843,0.5923); rgb(37pt)=(0.246957,0.791795,0.556743); rgb(38pt)=(0.290614,0.79729,0.518829); rgb(39pt)=(0.340643,0.8008,0.478857); rgb(40pt)=(0.3909,0.802871,0.435448); rgb(41pt)=(0.445629,0.802419,0.390919); rgb(42pt)=(0.5044,0.7993,0.348); rgb(43pt)=(0.561562,0.794233,0.304481); rgb(44pt)=(0.617395,0.787619,0.261238); rgb(45pt)=(0.671986,0.779271,0.2227); rgb(46pt)=(0.7242,0.769843,0.191029); rgb(47pt)=(0.773833,0.759805,0.16461); rgb(48pt)=(0.820314,0.749814,0.153529); rgb(49pt)=(0.863433,0.7406,0.159633); rgb(50pt)=(0.903543,0.733029,0.177414); rgb(51pt)=(0.939257,0.728786,0.209957); rgb(52pt)=(0.972757,0.729771,0.239443); rgb(53pt)=(0.995648,0.743371,0.237148); rgb(54pt)=(0.996986,0.765857,0.219943); rgb(55pt)=(0.995205,0.789252,0.202762); rgb(56pt)=(0.9892,0.813567,0.188533); rgb(57pt)=(0.978629,0.838629,0.176557); rgb(58pt)=(0.967648,0.8639,0.16429); rgb(59pt)=(0.96101,0.889019,0.153676); rgb(60pt)=(0.959671,0.913457,0.142257); rgb(61pt)=(0.962795,0.937338,0.12651); rgb(62pt)=(0.969114,0.960629,0.106362); rgb(63pt)=(0.9769,0.9839,0.0805)},
colorbar
]
\addplot [forget plot] graphics [xmin=-1.00200400801603, xmax=1.00200400801603, ymin=-1.00200400801603, ymax=1.00200400801603] {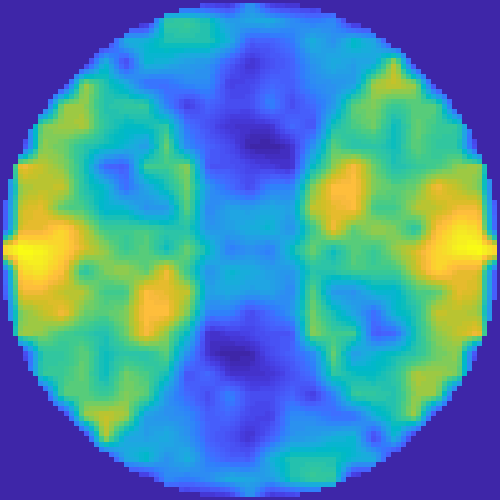};
\end{axis}
\end{tikzpicture}%
	\caption{$64$-bit Data (S2)}
	\label{fig:AARTFAAC:s4:mle:uncert:64bit}
\end{subfigure}%
    ~ 
\begin{subfigure}[t]{0.25\textwidth}
\centering
	\setlength{\figurewidth}{5.1cm}
	\setlength{\figureheight}{5.1cm}
%
%
\begin{tikzpicture}[scale=0.55]

\begin{axis}[%
width=\figurewidth,
height=\figureheight,
at={(0.92in,0.642in)},
scale only axis,
point meta min=0.124435515292639,
point meta max=0.131254335723414,
axis on top,
xmin=-1.00200400801603,
xmax=1.00200400801603,
xlabel style={font=\color{white!15!black}},
xlabel={$\text{South }\leftarrow\text{ Direction }\rightarrow\text{ North}$},
y dir=reverse,
ymin=-1.00200400801603,
ymax=1.00200400801603,
ylabel style={font=\color{white!15!black}},
axis background/.style={fill=white},
legend style={legend cell align=left, align=left, draw=white!15!black},
colormap={mymap}{[1pt] rgb(0pt)=(0.2422,0.1504,0.6603); rgb(1pt)=(0.25039,0.164995,0.707614); rgb(2pt)=(0.257771,0.181781,0.751138); rgb(3pt)=(0.264729,0.197757,0.795214); rgb(4pt)=(0.270648,0.214676,0.836371); rgb(5pt)=(0.275114,0.234238,0.870986); rgb(6pt)=(0.2783,0.255871,0.899071); rgb(7pt)=(0.280333,0.278233,0.9221); rgb(8pt)=(0.281338,0.300595,0.941376); rgb(9pt)=(0.281014,0.322757,0.957886); rgb(10pt)=(0.279467,0.344671,0.971676); rgb(11pt)=(0.275971,0.366681,0.982905); rgb(12pt)=(0.269914,0.3892,0.9906); rgb(13pt)=(0.260243,0.412329,0.995157); rgb(14pt)=(0.244033,0.435833,0.998833); rgb(15pt)=(0.220643,0.460257,0.997286); rgb(16pt)=(0.196333,0.484719,0.989152); rgb(17pt)=(0.183405,0.507371,0.979795); rgb(18pt)=(0.178643,0.528857,0.968157); rgb(19pt)=(0.176438,0.549905,0.952019); rgb(20pt)=(0.168743,0.570262,0.935871); rgb(21pt)=(0.154,0.5902,0.9218); rgb(22pt)=(0.146029,0.609119,0.907857); rgb(23pt)=(0.138024,0.627629,0.89729); rgb(24pt)=(0.124814,0.645929,0.888343); rgb(25pt)=(0.111252,0.6635,0.876314); rgb(26pt)=(0.0952095,0.679829,0.859781); rgb(27pt)=(0.0688714,0.694771,0.839357); rgb(28pt)=(0.0296667,0.708167,0.816333); rgb(29pt)=(0.00357143,0.720267,0.7917); rgb(30pt)=(0.00665714,0.731214,0.766014); rgb(31pt)=(0.0433286,0.741095,0.73941); rgb(32pt)=(0.0963952,0.75,0.712038); rgb(33pt)=(0.140771,0.7584,0.684157); rgb(34pt)=(0.1717,0.766962,0.655443); rgb(35pt)=(0.193767,0.775767,0.6251); rgb(36pt)=(0.216086,0.7843,0.5923); rgb(37pt)=(0.246957,0.791795,0.556743); rgb(38pt)=(0.290614,0.79729,0.518829); rgb(39pt)=(0.340643,0.8008,0.478857); rgb(40pt)=(0.3909,0.802871,0.435448); rgb(41pt)=(0.445629,0.802419,0.390919); rgb(42pt)=(0.5044,0.7993,0.348); rgb(43pt)=(0.561562,0.794233,0.304481); rgb(44pt)=(0.617395,0.787619,0.261238); rgb(45pt)=(0.671986,0.779271,0.2227); rgb(46pt)=(0.7242,0.769843,0.191029); rgb(47pt)=(0.773833,0.759805,0.16461); rgb(48pt)=(0.820314,0.749814,0.153529); rgb(49pt)=(0.863433,0.7406,0.159633); rgb(50pt)=(0.903543,0.733029,0.177414); rgb(51pt)=(0.939257,0.728786,0.209957); rgb(52pt)=(0.972757,0.729771,0.239443); rgb(53pt)=(0.995648,0.743371,0.237148); rgb(54pt)=(0.996986,0.765857,0.219943); rgb(55pt)=(0.995205,0.789252,0.202762); rgb(56pt)=(0.9892,0.813567,0.188533); rgb(57pt)=(0.978629,0.838629,0.176557); rgb(58pt)=(0.967648,0.8639,0.16429); rgb(59pt)=(0.96101,0.889019,0.153676); rgb(60pt)=(0.959671,0.913457,0.142257); rgb(61pt)=(0.962795,0.937338,0.12651); rgb(62pt)=(0.969114,0.960629,0.106362); rgb(63pt)=(0.9769,0.9839,0.0805)},
colorbar
]
\addplot [forget plot] graphics [xmin=-1.00200400801603, xmax=1.00200400801603, ymin=-1.00200400801603, ymax=1.00200400801603] {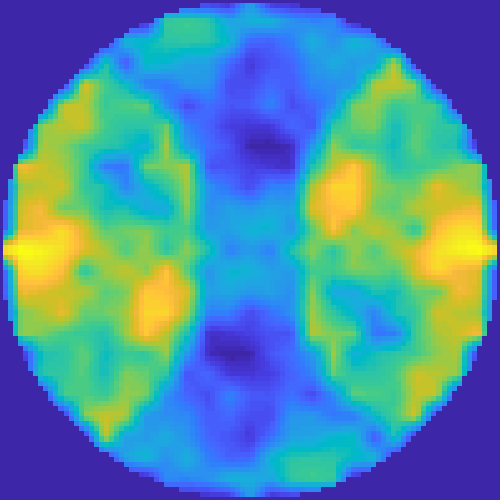};
\end{axis}
\end{tikzpicture}%
	\caption{$1$-bit Data (S2)}
	\label{fig:AARTFAAC:s4:cmle:uncert:1bit}
\end{subfigure}%

\begin{subfigure}[t]{0.25\textwidth}
    \centering
	\setlength{\figurewidth}{5.1cm}
	\setlength{\figureheight}{5.1cm}
%
%
\begin{tikzpicture}[scale=0.55]

\begin{axis}[%
width=\figurewidth,
height=\figureheight,
at={(0.92in,0.642in)},
scale only axis,
point meta min=0.0798901207653091,
point meta max=0.0861564789048747,
axis on top,
xmin=-1.00200400801603,
xmax=1.00200400801603,
xlabel style={font=\color{white!15!black}},
xlabel={$\text{South }\leftarrow\text{ Direction }\rightarrow\text{ North}$},
y dir=reverse,
ymin=-1.00200400801603,
ymax=1.00200400801603,
ylabel style={font=\color{white!15!black}},
ylabel={$\text{East }\leftarrow\text{ Direction }\rightarrow\text{ West}$},
axis background/.style={fill=white},
legend style={legend cell align=left, align=left, draw=white!15!black},
colormap={mymap}{[1pt] rgb(0pt)=(0.2422,0.1504,0.6603); rgb(1pt)=(0.25039,0.164995,0.707614); rgb(2pt)=(0.257771,0.181781,0.751138); rgb(3pt)=(0.264729,0.197757,0.795214); rgb(4pt)=(0.270648,0.214676,0.836371); rgb(5pt)=(0.275114,0.234238,0.870986); rgb(6pt)=(0.2783,0.255871,0.899071); rgb(7pt)=(0.280333,0.278233,0.9221); rgb(8pt)=(0.281338,0.300595,0.941376); rgb(9pt)=(0.281014,0.322757,0.957886); rgb(10pt)=(0.279467,0.344671,0.971676); rgb(11pt)=(0.275971,0.366681,0.982905); rgb(12pt)=(0.269914,0.3892,0.9906); rgb(13pt)=(0.260243,0.412329,0.995157); rgb(14pt)=(0.244033,0.435833,0.998833); rgb(15pt)=(0.220643,0.460257,0.997286); rgb(16pt)=(0.196333,0.484719,0.989152); rgb(17pt)=(0.183405,0.507371,0.979795); rgb(18pt)=(0.178643,0.528857,0.968157); rgb(19pt)=(0.176438,0.549905,0.952019); rgb(20pt)=(0.168743,0.570262,0.935871); rgb(21pt)=(0.154,0.5902,0.9218); rgb(22pt)=(0.146029,0.609119,0.907857); rgb(23pt)=(0.138024,0.627629,0.89729); rgb(24pt)=(0.124814,0.645929,0.888343); rgb(25pt)=(0.111252,0.6635,0.876314); rgb(26pt)=(0.0952095,0.679829,0.859781); rgb(27pt)=(0.0688714,0.694771,0.839357); rgb(28pt)=(0.0296667,0.708167,0.816333); rgb(29pt)=(0.00357143,0.720267,0.7917); rgb(30pt)=(0.00665714,0.731214,0.766014); rgb(31pt)=(0.0433286,0.741095,0.73941); rgb(32pt)=(0.0963952,0.75,0.712038); rgb(33pt)=(0.140771,0.7584,0.684157); rgb(34pt)=(0.1717,0.766962,0.655443); rgb(35pt)=(0.193767,0.775767,0.6251); rgb(36pt)=(0.216086,0.7843,0.5923); rgb(37pt)=(0.246957,0.791795,0.556743); rgb(38pt)=(0.290614,0.79729,0.518829); rgb(39pt)=(0.340643,0.8008,0.478857); rgb(40pt)=(0.3909,0.802871,0.435448); rgb(41pt)=(0.445629,0.802419,0.390919); rgb(42pt)=(0.5044,0.7993,0.348); rgb(43pt)=(0.561562,0.794233,0.304481); rgb(44pt)=(0.617395,0.787619,0.261238); rgb(45pt)=(0.671986,0.779271,0.2227); rgb(46pt)=(0.7242,0.769843,0.191029); rgb(47pt)=(0.773833,0.759805,0.16461); rgb(48pt)=(0.820314,0.749814,0.153529); rgb(49pt)=(0.863433,0.7406,0.159633); rgb(50pt)=(0.903543,0.733029,0.177414); rgb(51pt)=(0.939257,0.728786,0.209957); rgb(52pt)=(0.972757,0.729771,0.239443); rgb(53pt)=(0.995648,0.743371,0.237148); rgb(54pt)=(0.996986,0.765857,0.219943); rgb(55pt)=(0.995205,0.789252,0.202762); rgb(56pt)=(0.9892,0.813567,0.188533); rgb(57pt)=(0.978629,0.838629,0.176557); rgb(58pt)=(0.967648,0.8639,0.16429); rgb(59pt)=(0.96101,0.889019,0.153676); rgb(60pt)=(0.959671,0.913457,0.142257); rgb(61pt)=(0.962795,0.937338,0.12651); rgb(62pt)=(0.969114,0.960629,0.106362); rgb(63pt)=(0.9769,0.9839,0.0805)},
colorbar
]
\addplot [forget plot] graphics [xmin=-1.00200400801603, xmax=1.00200400801603, ymin=-1.00200400801603, ymax=1.00200400801603] {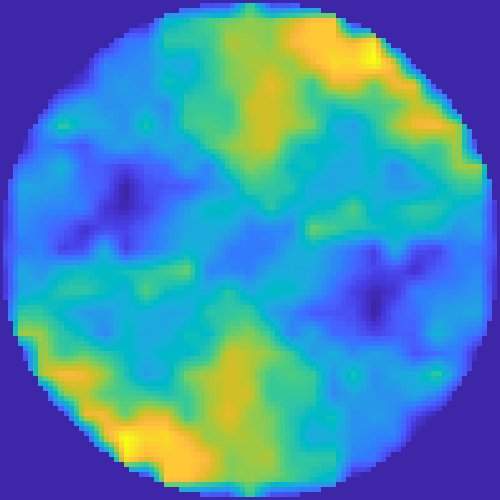};
\end{axis}
\end{tikzpicture}%
	\caption{$64$-bit Data (S3)}
	\label{fig:AARTFAAC:s6:mle:uncert:64bit}
\end{subfigure}%
    ~ 
\begin{subfigure}[t]{0.25\textwidth}
\centering
	\setlength{\figurewidth}{5.1cm}
	\setlength{\figureheight}{5.1cm}
%
%
\begin{tikzpicture}[scale=0.55]

\begin{axis}[%
width=\figurewidth,
height=\figureheight,
at={(0.92in,0.642in)},
scale only axis,
point meta min=0.126126963888099,
point meta max=0.134287736427357,
axis on top,
xmin=-1.00200400801603,
xmax=1.00200400801603,
xlabel style={font=\color{white!15!black}},
xlabel={$\text{South }\leftarrow\text{ Direction }\rightarrow\text{ North}$},
y dir=reverse,
ymin=-1.00200400801603,
ymax=1.00200400801603,
ylabel style={font=\color{white!15!black}},
axis background/.style={fill=white},
legend style={legend cell align=left, align=left, draw=white!15!black},
colormap={mymap}{[1pt] rgb(0pt)=(0.2422,0.1504,0.6603); rgb(1pt)=(0.25039,0.164995,0.707614); rgb(2pt)=(0.257771,0.181781,0.751138); rgb(3pt)=(0.264729,0.197757,0.795214); rgb(4pt)=(0.270648,0.214676,0.836371); rgb(5pt)=(0.275114,0.234238,0.870986); rgb(6pt)=(0.2783,0.255871,0.899071); rgb(7pt)=(0.280333,0.278233,0.9221); rgb(8pt)=(0.281338,0.300595,0.941376); rgb(9pt)=(0.281014,0.322757,0.957886); rgb(10pt)=(0.279467,0.344671,0.971676); rgb(11pt)=(0.275971,0.366681,0.982905); rgb(12pt)=(0.269914,0.3892,0.9906); rgb(13pt)=(0.260243,0.412329,0.995157); rgb(14pt)=(0.244033,0.435833,0.998833); rgb(15pt)=(0.220643,0.460257,0.997286); rgb(16pt)=(0.196333,0.484719,0.989152); rgb(17pt)=(0.183405,0.507371,0.979795); rgb(18pt)=(0.178643,0.528857,0.968157); rgb(19pt)=(0.176438,0.549905,0.952019); rgb(20pt)=(0.168743,0.570262,0.935871); rgb(21pt)=(0.154,0.5902,0.9218); rgb(22pt)=(0.146029,0.609119,0.907857); rgb(23pt)=(0.138024,0.627629,0.89729); rgb(24pt)=(0.124814,0.645929,0.888343); rgb(25pt)=(0.111252,0.6635,0.876314); rgb(26pt)=(0.0952095,0.679829,0.859781); rgb(27pt)=(0.0688714,0.694771,0.839357); rgb(28pt)=(0.0296667,0.708167,0.816333); rgb(29pt)=(0.00357143,0.720267,0.7917); rgb(30pt)=(0.00665714,0.731214,0.766014); rgb(31pt)=(0.0433286,0.741095,0.73941); rgb(32pt)=(0.0963952,0.75,0.712038); rgb(33pt)=(0.140771,0.7584,0.684157); rgb(34pt)=(0.1717,0.766962,0.655443); rgb(35pt)=(0.193767,0.775767,0.6251); rgb(36pt)=(0.216086,0.7843,0.5923); rgb(37pt)=(0.246957,0.791795,0.556743); rgb(38pt)=(0.290614,0.79729,0.518829); rgb(39pt)=(0.340643,0.8008,0.478857); rgb(40pt)=(0.3909,0.802871,0.435448); rgb(41pt)=(0.445629,0.802419,0.390919); rgb(42pt)=(0.5044,0.7993,0.348); rgb(43pt)=(0.561562,0.794233,0.304481); rgb(44pt)=(0.617395,0.787619,0.261238); rgb(45pt)=(0.671986,0.779271,0.2227); rgb(46pt)=(0.7242,0.769843,0.191029); rgb(47pt)=(0.773833,0.759805,0.16461); rgb(48pt)=(0.820314,0.749814,0.153529); rgb(49pt)=(0.863433,0.7406,0.159633); rgb(50pt)=(0.903543,0.733029,0.177414); rgb(51pt)=(0.939257,0.728786,0.209957); rgb(52pt)=(0.972757,0.729771,0.239443); rgb(53pt)=(0.995648,0.743371,0.237148); rgb(54pt)=(0.996986,0.765857,0.219943); rgb(55pt)=(0.995205,0.789252,0.202762); rgb(56pt)=(0.9892,0.813567,0.188533); rgb(57pt)=(0.978629,0.838629,0.176557); rgb(58pt)=(0.967648,0.8639,0.16429); rgb(59pt)=(0.96101,0.889019,0.153676); rgb(60pt)=(0.959671,0.913457,0.142257); rgb(61pt)=(0.962795,0.937338,0.12651); rgb(62pt)=(0.969114,0.960629,0.106362); rgb(63pt)=(0.9769,0.9839,0.0805)},
colorbar
]
\addplot [forget plot] graphics [xmin=-1.00200400801603, xmax=1.00200400801603, ymin=-1.00200400801603, ymax=1.00200400801603] {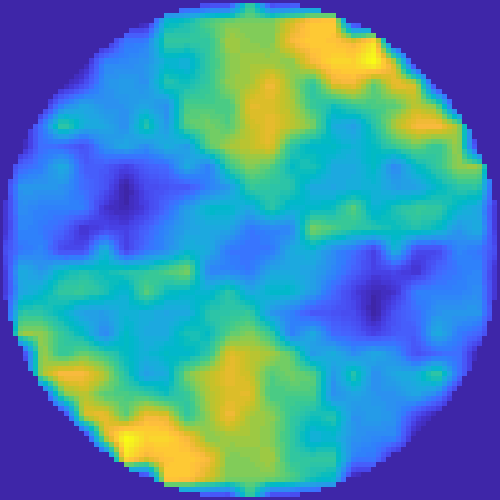};
\end{axis}
\end{tikzpicture}%
	\caption{$1$-bit Data (S3)}
	\label{fig:AARTFAAC:s6:cmle:uncert:1bit}
\end{subfigure}%
\caption{AARTFAAC - Uncertainty ($M=46, D_s=441$)}
\label{fig:LOFAR:aartfaac:indi:uncert}
\end{figure}
Like for the CS302 dataset, the measurements are calibrated such that their sampling precision is $64$-bit. The receive bandwidth is $2B_\text{Y} = \SI{195312.5}{\hertz}$, centered at a frequency of approximately $\SI{62.90}{\mega\hertz}$. The dataset contains $N=\num{195313}$ Nyquist samples per antenna output, corresponding to an observation time of $T=\SI{1}{\second}$. For radio interferometric imaging with the scoring techniques \eqref{update:scoring:rule} and \eqref{scoring:rule:quant}, we proceed as for CS302. Fig. \ref{fig:LOFAR:aartfaac:indi:image} shows the all-sky images obtained when using the measurement data of each telescope station individually while assuming $D_\text{S}=441$ astronomical sources (25 pixels). Comparing the results obtained with the $64$-bit and the $1$-bit radio data, hardly any visual difference can be observed. Fig. \ref{fig:LOFAR:aartfaac:indi:uncert} shows the six corresponding interferometric uncertainty maps which are obtained for a scenario with equal power levels $\theta_{d, \text{qual}}=\SI{-30}{\decibel}, \forall d$. Like for CS302, the standard deviation of the intensity estimates increases by $1.58$ when hard-limiting the data such that the $1$-bit loss can be compensated by increasing the observation interval $T$ by a factor of $2.5$.

\begin{figure}[!t]
\centering
    \begin{subfigure}[t]{0.24\textwidth}
    \centering
	\setlength{\figurewidth}{5.1cm}
	\setlength{\figureheight}{5.1cm}
%
%
\begin{tikzpicture}[scale=0.55]

\begin{axis}[%
width=\figurewidth,
height=\figureheight,
at={(0.92in,0.642in)},
scale only axis,
point meta min=1e-08,
point meta max=0.0019099534261469,
axis on top,
xmin=-1.00200400801603,
xmax=1.00200400801603,
xlabel style={font=\color{white!15!black}},
xlabel={$\text{South }\leftarrow\text{ Direction }\rightarrow\text{ North}$},
y dir=reverse,
ymin=-1.00200400801603,
ymax=1.00200400801603,
ylabel style={font=\color{white!15!black}},
ylabel={$\text{East }\leftarrow\text{ Direction }\rightarrow\text{ West}$},
axis background/.style={fill=white},
legend style={legend cell align=left, align=left, draw=white!15!black},
colormap={mymap}{[1pt] rgb(0pt)=(0.2422,0.1504,0.6603); rgb(1pt)=(0.25039,0.164995,0.707614); rgb(2pt)=(0.257771,0.181781,0.751138); rgb(3pt)=(0.264729,0.197757,0.795214); rgb(4pt)=(0.270648,0.214676,0.836371); rgb(5pt)=(0.275114,0.234238,0.870986); rgb(6pt)=(0.2783,0.255871,0.899071); rgb(7pt)=(0.280333,0.278233,0.9221); rgb(8pt)=(0.281338,0.300595,0.941376); rgb(9pt)=(0.281014,0.322757,0.957886); rgb(10pt)=(0.279467,0.344671,0.971676); rgb(11pt)=(0.275971,0.366681,0.982905); rgb(12pt)=(0.269914,0.3892,0.9906); rgb(13pt)=(0.260243,0.412329,0.995157); rgb(14pt)=(0.244033,0.435833,0.998833); rgb(15pt)=(0.220643,0.460257,0.997286); rgb(16pt)=(0.196333,0.484719,0.989152); rgb(17pt)=(0.183405,0.507371,0.979795); rgb(18pt)=(0.178643,0.528857,0.968157); rgb(19pt)=(0.176438,0.549905,0.952019); rgb(20pt)=(0.168743,0.570262,0.935871); rgb(21pt)=(0.154,0.5902,0.9218); rgb(22pt)=(0.146029,0.609119,0.907857); rgb(23pt)=(0.138024,0.627629,0.89729); rgb(24pt)=(0.124814,0.645929,0.888343); rgb(25pt)=(0.111252,0.6635,0.876314); rgb(26pt)=(0.0952095,0.679829,0.859781); rgb(27pt)=(0.0688714,0.694771,0.839357); rgb(28pt)=(0.0296667,0.708167,0.816333); rgb(29pt)=(0.00357143,0.720267,0.7917); rgb(30pt)=(0.00665714,0.731214,0.766014); rgb(31pt)=(0.0433286,0.741095,0.73941); rgb(32pt)=(0.0963952,0.75,0.712038); rgb(33pt)=(0.140771,0.7584,0.684157); rgb(34pt)=(0.1717,0.766962,0.655443); rgb(35pt)=(0.193767,0.775767,0.6251); rgb(36pt)=(0.216086,0.7843,0.5923); rgb(37pt)=(0.246957,0.791795,0.556743); rgb(38pt)=(0.290614,0.79729,0.518829); rgb(39pt)=(0.340643,0.8008,0.478857); rgb(40pt)=(0.3909,0.802871,0.435448); rgb(41pt)=(0.445629,0.802419,0.390919); rgb(42pt)=(0.5044,0.7993,0.348); rgb(43pt)=(0.561562,0.794233,0.304481); rgb(44pt)=(0.617395,0.787619,0.261238); rgb(45pt)=(0.671986,0.779271,0.2227); rgb(46pt)=(0.7242,0.769843,0.191029); rgb(47pt)=(0.773833,0.759805,0.16461); rgb(48pt)=(0.820314,0.749814,0.153529); rgb(49pt)=(0.863433,0.7406,0.159633); rgb(50pt)=(0.903543,0.733029,0.177414); rgb(51pt)=(0.939257,0.728786,0.209957); rgb(52pt)=(0.972757,0.729771,0.239443); rgb(53pt)=(0.995648,0.743371,0.237148); rgb(54pt)=(0.996986,0.765857,0.219943); rgb(55pt)=(0.995205,0.789252,0.202762); rgb(56pt)=(0.9892,0.813567,0.188533); rgb(57pt)=(0.978629,0.838629,0.176557); rgb(58pt)=(0.967648,0.8639,0.16429); rgb(59pt)=(0.96101,0.889019,0.153676); rgb(60pt)=(0.959671,0.913457,0.142257); rgb(61pt)=(0.962795,0.937338,0.12651); rgb(62pt)=(0.969114,0.960629,0.106362); rgb(63pt)=(0.9769,0.9839,0.0805)},
colorbar
]
\addplot [forget plot] graphics [xmin=-1.00200400801603, xmax=1.00200400801603, ymin=-1.00200400801603, ymax=1.00200400801603] {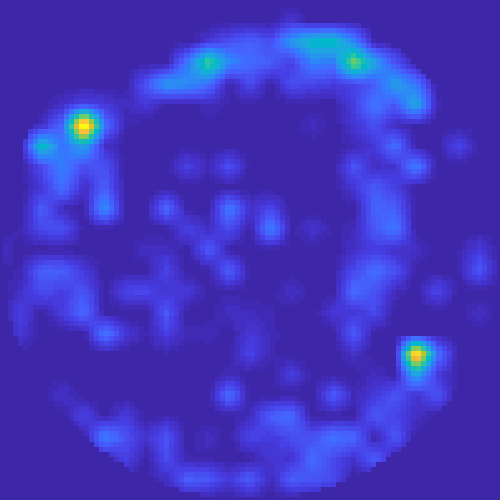};
\end{axis}
\end{tikzpicture}%
	\caption{$64$-bit Data (S1$+$S2)}
	\label{fig:AARTFAAC:s34:mle:image:64bit}
\end{subfigure}%
    ~ 
\begin{subfigure}[t]{0.25\textwidth}
\centering
	\setlength{\figurewidth}{5.1cm}
	\setlength{\figureheight}{5.1cm}
%
%
\begin{tikzpicture}[scale=0.55]

\begin{axis}[%
width=\figurewidth,
height=\figureheight,
at={(0.92in,0.642in)},
scale only axis,
point meta min=1e-08,
point meta max=0.0020236820619554,
axis on top,
xmin=-1.00200400801603,
xmax=1.00200400801603,
xlabel style={font=\color{white!15!black}},
xlabel={$\text{South }\leftarrow\text{ Direction }\rightarrow\text{ North}$},
y dir=reverse,
ymin=-1.00200400801603,
ymax=1.00200400801603,
ylabel style={font=\color{white!15!black}},
axis background/.style={fill=white},
legend style={legend cell align=left, align=left, draw=white!15!black},
colormap={mymap}{[1pt] rgb(0pt)=(0.2422,0.1504,0.6603); rgb(1pt)=(0.25039,0.164995,0.707614); rgb(2pt)=(0.257771,0.181781,0.751138); rgb(3pt)=(0.264729,0.197757,0.795214); rgb(4pt)=(0.270648,0.214676,0.836371); rgb(5pt)=(0.275114,0.234238,0.870986); rgb(6pt)=(0.2783,0.255871,0.899071); rgb(7pt)=(0.280333,0.278233,0.9221); rgb(8pt)=(0.281338,0.300595,0.941376); rgb(9pt)=(0.281014,0.322757,0.957886); rgb(10pt)=(0.279467,0.344671,0.971676); rgb(11pt)=(0.275971,0.366681,0.982905); rgb(12pt)=(0.269914,0.3892,0.9906); rgb(13pt)=(0.260243,0.412329,0.995157); rgb(14pt)=(0.244033,0.435833,0.998833); rgb(15pt)=(0.220643,0.460257,0.997286); rgb(16pt)=(0.196333,0.484719,0.989152); rgb(17pt)=(0.183405,0.507371,0.979795); rgb(18pt)=(0.178643,0.528857,0.968157); rgb(19pt)=(0.176438,0.549905,0.952019); rgb(20pt)=(0.168743,0.570262,0.935871); rgb(21pt)=(0.154,0.5902,0.9218); rgb(22pt)=(0.146029,0.609119,0.907857); rgb(23pt)=(0.138024,0.627629,0.89729); rgb(24pt)=(0.124814,0.645929,0.888343); rgb(25pt)=(0.111252,0.6635,0.876314); rgb(26pt)=(0.0952095,0.679829,0.859781); rgb(27pt)=(0.0688714,0.694771,0.839357); rgb(28pt)=(0.0296667,0.708167,0.816333); rgb(29pt)=(0.00357143,0.720267,0.7917); rgb(30pt)=(0.00665714,0.731214,0.766014); rgb(31pt)=(0.0433286,0.741095,0.73941); rgb(32pt)=(0.0963952,0.75,0.712038); rgb(33pt)=(0.140771,0.7584,0.684157); rgb(34pt)=(0.1717,0.766962,0.655443); rgb(35pt)=(0.193767,0.775767,0.6251); rgb(36pt)=(0.216086,0.7843,0.5923); rgb(37pt)=(0.246957,0.791795,0.556743); rgb(38pt)=(0.290614,0.79729,0.518829); rgb(39pt)=(0.340643,0.8008,0.478857); rgb(40pt)=(0.3909,0.802871,0.435448); rgb(41pt)=(0.445629,0.802419,0.390919); rgb(42pt)=(0.5044,0.7993,0.348); rgb(43pt)=(0.561562,0.794233,0.304481); rgb(44pt)=(0.617395,0.787619,0.261238); rgb(45pt)=(0.671986,0.779271,0.2227); rgb(46pt)=(0.7242,0.769843,0.191029); rgb(47pt)=(0.773833,0.759805,0.16461); rgb(48pt)=(0.820314,0.749814,0.153529); rgb(49pt)=(0.863433,0.7406,0.159633); rgb(50pt)=(0.903543,0.733029,0.177414); rgb(51pt)=(0.939257,0.728786,0.209957); rgb(52pt)=(0.972757,0.729771,0.239443); rgb(53pt)=(0.995648,0.743371,0.237148); rgb(54pt)=(0.996986,0.765857,0.219943); rgb(55pt)=(0.995205,0.789252,0.202762); rgb(56pt)=(0.9892,0.813567,0.188533); rgb(57pt)=(0.978629,0.838629,0.176557); rgb(58pt)=(0.967648,0.8639,0.16429); rgb(59pt)=(0.96101,0.889019,0.153676); rgb(60pt)=(0.959671,0.913457,0.142257); rgb(61pt)=(0.962795,0.937338,0.12651); rgb(62pt)=(0.969114,0.960629,0.106362); rgb(63pt)=(0.9769,0.9839,0.0805)},
colorbar
]
\addplot [forget plot] graphics [xmin=-1.00200400801603, xmax=1.00200400801603, ymin=-1.00200400801603, ymax=1.00200400801603] {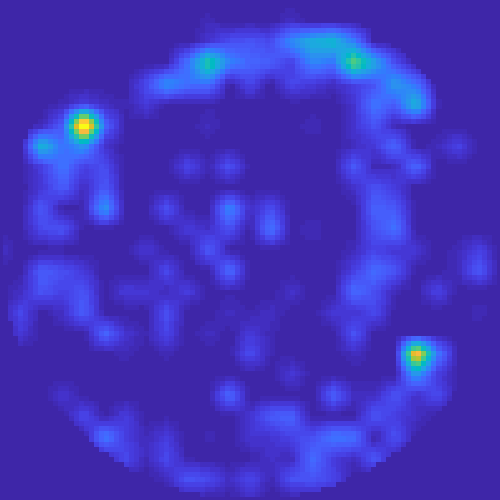};
\end{axis}
\end{tikzpicture}%
	\caption{$1$-bit Data (S1$+$S2)}
	\label{fig:AARTFAAC:s34:cmle:image:1bit}
\end{subfigure}%
\caption{AARTFAAC - All-Sky Image ($M=92, D_\text{S}=441$)}
\label{fig:LOFAR:aartfaac:s34:image}
\vspace{0.5cm}
    \begin{subfigure}[t]{0.24\textwidth}
    \centering
	\setlength{\figurewidth}{5.1cm}
	\setlength{\figureheight}{5.1cm}
%
%
\begin{tikzpicture}[scale=0.55]

\begin{axis}[%
width=\figurewidth,
height=\figureheight,
at={(0.92in,0.642in)},
scale only axis,
point meta min=0.036850664562341,
point meta max=0.0379692194331185,
axis on top,
xmin=-1.00200400801603,
xmax=1.00200400801603,
xlabel style={font=\color{white!15!black}},
xlabel={$\text{South }\leftarrow\text{ Direction }\rightarrow\text{ North}$},
y dir=reverse,
ymin=-1.00200400801603,
ymax=1.00200400801603,
ylabel style={font=\color{white!15!black}},
ylabel={$\text{East }\leftarrow\text{ Direction }\rightarrow\text{ West}$},
axis background/.style={fill=white},
legend style={legend cell align=left, align=left, draw=white!15!black},
colormap={mymap}{[1pt] rgb(0pt)=(0.2422,0.1504,0.6603); rgb(1pt)=(0.25039,0.164995,0.707614); rgb(2pt)=(0.257771,0.181781,0.751138); rgb(3pt)=(0.264729,0.197757,0.795214); rgb(4pt)=(0.270648,0.214676,0.836371); rgb(5pt)=(0.275114,0.234238,0.870986); rgb(6pt)=(0.2783,0.255871,0.899071); rgb(7pt)=(0.280333,0.278233,0.9221); rgb(8pt)=(0.281338,0.300595,0.941376); rgb(9pt)=(0.281014,0.322757,0.957886); rgb(10pt)=(0.279467,0.344671,0.971676); rgb(11pt)=(0.275971,0.366681,0.982905); rgb(12pt)=(0.269914,0.3892,0.9906); rgb(13pt)=(0.260243,0.412329,0.995157); rgb(14pt)=(0.244033,0.435833,0.998833); rgb(15pt)=(0.220643,0.460257,0.997286); rgb(16pt)=(0.196333,0.484719,0.989152); rgb(17pt)=(0.183405,0.507371,0.979795); rgb(18pt)=(0.178643,0.528857,0.968157); rgb(19pt)=(0.176438,0.549905,0.952019); rgb(20pt)=(0.168743,0.570262,0.935871); rgb(21pt)=(0.154,0.5902,0.9218); rgb(22pt)=(0.146029,0.609119,0.907857); rgb(23pt)=(0.138024,0.627629,0.89729); rgb(24pt)=(0.124814,0.645929,0.888343); rgb(25pt)=(0.111252,0.6635,0.876314); rgb(26pt)=(0.0952095,0.679829,0.859781); rgb(27pt)=(0.0688714,0.694771,0.839357); rgb(28pt)=(0.0296667,0.708167,0.816333); rgb(29pt)=(0.00357143,0.720267,0.7917); rgb(30pt)=(0.00665714,0.731214,0.766014); rgb(31pt)=(0.0433286,0.741095,0.73941); rgb(32pt)=(0.0963952,0.75,0.712038); rgb(33pt)=(0.140771,0.7584,0.684157); rgb(34pt)=(0.1717,0.766962,0.655443); rgb(35pt)=(0.193767,0.775767,0.6251); rgb(36pt)=(0.216086,0.7843,0.5923); rgb(37pt)=(0.246957,0.791795,0.556743); rgb(38pt)=(0.290614,0.79729,0.518829); rgb(39pt)=(0.340643,0.8008,0.478857); rgb(40pt)=(0.3909,0.802871,0.435448); rgb(41pt)=(0.445629,0.802419,0.390919); rgb(42pt)=(0.5044,0.7993,0.348); rgb(43pt)=(0.561562,0.794233,0.304481); rgb(44pt)=(0.617395,0.787619,0.261238); rgb(45pt)=(0.671986,0.779271,0.2227); rgb(46pt)=(0.7242,0.769843,0.191029); rgb(47pt)=(0.773833,0.759805,0.16461); rgb(48pt)=(0.820314,0.749814,0.153529); rgb(49pt)=(0.863433,0.7406,0.159633); rgb(50pt)=(0.903543,0.733029,0.177414); rgb(51pt)=(0.939257,0.728786,0.209957); rgb(52pt)=(0.972757,0.729771,0.239443); rgb(53pt)=(0.995648,0.743371,0.237148); rgb(54pt)=(0.996986,0.765857,0.219943); rgb(55pt)=(0.995205,0.789252,0.202762); rgb(56pt)=(0.9892,0.813567,0.188533); rgb(57pt)=(0.978629,0.838629,0.176557); rgb(58pt)=(0.967648,0.8639,0.16429); rgb(59pt)=(0.96101,0.889019,0.153676); rgb(60pt)=(0.959671,0.913457,0.142257); rgb(61pt)=(0.962795,0.937338,0.12651); rgb(62pt)=(0.969114,0.960629,0.106362); rgb(63pt)=(0.9769,0.9839,0.0805)},
colorbar
]
\addplot [forget plot] graphics [xmin=-1.00200400801603, xmax=1.00200400801603, ymin=-1.00200400801603, ymax=1.00200400801603] {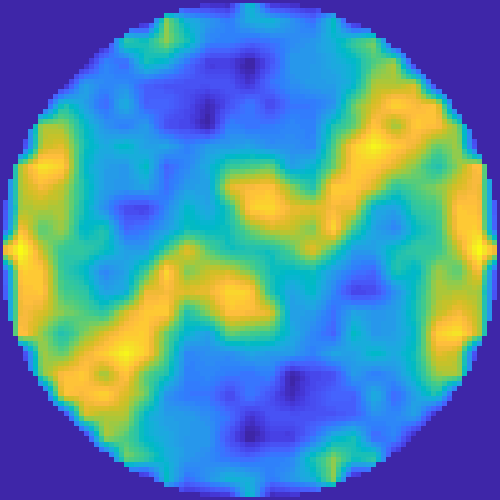};
\end{axis}
\end{tikzpicture}%
	\caption{$64$-bit Data (S1$+$S2)}
	\label{fig:AARTFAAC:s34:mle:uncert:64bit}
\end{subfigure}%
    ~ 
\begin{subfigure}[t]{0.25\textwidth}
\centering
	\setlength{\figurewidth}{5.1cm}
	\setlength{\figureheight}{5.1cm}
%
%
\begin{tikzpicture}[scale=0.55]

\begin{axis}[%
width=\figurewidth,
height=\figureheight,
at={(0.92in,0.642in)},
scale only axis,
point meta min=0.0578894355121394,
point meta max=0.0589650480181006,
axis on top,
xmin=-1.00200400801603,
xmax=1.00200400801603,
xlabel style={font=\color{white!15!black}},
xlabel={$\text{South }\leftarrow\text{ Direction }\rightarrow\text{ North}$},
y dir=reverse,
ymin=-1.00200400801603,
ymax=1.00200400801603,
ylabel style={font=\color{white!15!black}},
axis background/.style={fill=white},
legend style={legend cell align=left, align=left, draw=white!15!black},
colormap={mymap}{[1pt] rgb(0pt)=(0.2422,0.1504,0.6603); rgb(1pt)=(0.25039,0.164995,0.707614); rgb(2pt)=(0.257771,0.181781,0.751138); rgb(3pt)=(0.264729,0.197757,0.795214); rgb(4pt)=(0.270648,0.214676,0.836371); rgb(5pt)=(0.275114,0.234238,0.870986); rgb(6pt)=(0.2783,0.255871,0.899071); rgb(7pt)=(0.280333,0.278233,0.9221); rgb(8pt)=(0.281338,0.300595,0.941376); rgb(9pt)=(0.281014,0.322757,0.957886); rgb(10pt)=(0.279467,0.344671,0.971676); rgb(11pt)=(0.275971,0.366681,0.982905); rgb(12pt)=(0.269914,0.3892,0.9906); rgb(13pt)=(0.260243,0.412329,0.995157); rgb(14pt)=(0.244033,0.435833,0.998833); rgb(15pt)=(0.220643,0.460257,0.997286); rgb(16pt)=(0.196333,0.484719,0.989152); rgb(17pt)=(0.183405,0.507371,0.979795); rgb(18pt)=(0.178643,0.528857,0.968157); rgb(19pt)=(0.176438,0.549905,0.952019); rgb(20pt)=(0.168743,0.570262,0.935871); rgb(21pt)=(0.154,0.5902,0.9218); rgb(22pt)=(0.146029,0.609119,0.907857); rgb(23pt)=(0.138024,0.627629,0.89729); rgb(24pt)=(0.124814,0.645929,0.888343); rgb(25pt)=(0.111252,0.6635,0.876314); rgb(26pt)=(0.0952095,0.679829,0.859781); rgb(27pt)=(0.0688714,0.694771,0.839357); rgb(28pt)=(0.0296667,0.708167,0.816333); rgb(29pt)=(0.00357143,0.720267,0.7917); rgb(30pt)=(0.00665714,0.731214,0.766014); rgb(31pt)=(0.0433286,0.741095,0.73941); rgb(32pt)=(0.0963952,0.75,0.712038); rgb(33pt)=(0.140771,0.7584,0.684157); rgb(34pt)=(0.1717,0.766962,0.655443); rgb(35pt)=(0.193767,0.775767,0.6251); rgb(36pt)=(0.216086,0.7843,0.5923); rgb(37pt)=(0.246957,0.791795,0.556743); rgb(38pt)=(0.290614,0.79729,0.518829); rgb(39pt)=(0.340643,0.8008,0.478857); rgb(40pt)=(0.3909,0.802871,0.435448); rgb(41pt)=(0.445629,0.802419,0.390919); rgb(42pt)=(0.5044,0.7993,0.348); rgb(43pt)=(0.561562,0.794233,0.304481); rgb(44pt)=(0.617395,0.787619,0.261238); rgb(45pt)=(0.671986,0.779271,0.2227); rgb(46pt)=(0.7242,0.769843,0.191029); rgb(47pt)=(0.773833,0.759805,0.16461); rgb(48pt)=(0.820314,0.749814,0.153529); rgb(49pt)=(0.863433,0.7406,0.159633); rgb(50pt)=(0.903543,0.733029,0.177414); rgb(51pt)=(0.939257,0.728786,0.209957); rgb(52pt)=(0.972757,0.729771,0.239443); rgb(53pt)=(0.995648,0.743371,0.237148); rgb(54pt)=(0.996986,0.765857,0.219943); rgb(55pt)=(0.995205,0.789252,0.202762); rgb(56pt)=(0.9892,0.813567,0.188533); rgb(57pt)=(0.978629,0.838629,0.176557); rgb(58pt)=(0.967648,0.8639,0.16429); rgb(59pt)=(0.96101,0.889019,0.153676); rgb(60pt)=(0.959671,0.913457,0.142257); rgb(61pt)=(0.962795,0.937338,0.12651); rgb(62pt)=(0.969114,0.960629,0.106362); rgb(63pt)=(0.9769,0.9839,0.0805)},
colorbar
]
\addplot [forget plot] graphics [xmin=-1.00200400801603, xmax=1.00200400801603, ymin=-1.00200400801603, ymax=1.00200400801603] {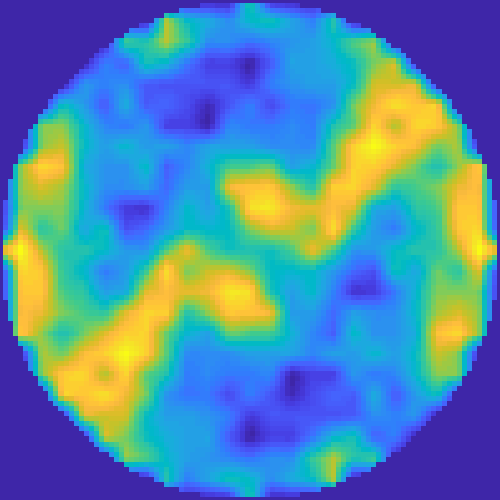};
\end{axis}
\end{tikzpicture}%
	\caption{$1$-bit Data (S1$+$S2)}
	\label{fig:AARTFAAC:s34:cmle:uncert:1bit}
\end{subfigure}%
\caption{AARTFAAC - Uncertainty ($M=92, D_\text{S}=441$)}
\label{fig:AARTFAAC:s34:uncert}
\end{figure}

To analyze the effect of increasing the number of sensors, Fig. \ref{fig:LOFAR:aartfaac:s34:image} depicts the all-sky images when using station S1 and S2 jointly to form an interferometric array with $M=92$ elements which estimates the intensity of the electromagnetic emission from $D_\text{S}=441$ directions.
Fig. \ref{fig:AARTFAAC:s34:uncert} shows the corresponding performance. The binary telescope with $M=92$ antennas outperforms all three $64$-bit stations with $M=46$ array elements in terms of imaging uncertainty, although the measurement data is by a factor $32$ smaller. These results confirm that reducing the A/D resolution to $1$-bit and doubling the number of receivers is an effective method to obtain high-performance low-complexity sensing architectures \cite{SteinFauss19}.

\begin{figure}[!b]
\centering
    \begin{subfigure}[t]{0.24\textwidth}
    \centering
	\setlength{\figurewidth}{5.1cm}
	\setlength{\figureheight}{5.1cm}
%
%
\begin{tikzpicture}[scale=0.55]

\begin{axis}[%
width=\figurewidth,
height=\figureheight,
at={(0.92in,0.642in)},
scale only axis,
point meta min=1e-08,
point meta max=0.00182288795759687,
axis on top,
xmin=-1.00200400801603,
xmax=1.00200400801603,
xlabel style={font=\color{white!15!black}},
xlabel={$\text{South }\leftarrow\text{ Direction }\rightarrow\text{ North}$},
y dir=reverse,
ymin=-1.00200400801603,
ymax=1.00200400801603,
ylabel style={font=\color{white!15!black}},
ylabel={$\text{East }\leftarrow\text{ Direction }\rightarrow\text{ West}$},
axis background/.style={fill=white},
legend style={legend cell align=left, align=left, draw=white!15!black},
colormap={mymap}{[1pt] rgb(0pt)=(0.2422,0.1504,0.6603); rgb(1pt)=(0.25039,0.164995,0.707614); rgb(2pt)=(0.257771,0.181781,0.751138); rgb(3pt)=(0.264729,0.197757,0.795214); rgb(4pt)=(0.270648,0.214676,0.836371); rgb(5pt)=(0.275114,0.234238,0.870986); rgb(6pt)=(0.2783,0.255871,0.899071); rgb(7pt)=(0.280333,0.278233,0.9221); rgb(8pt)=(0.281338,0.300595,0.941376); rgb(9pt)=(0.281014,0.322757,0.957886); rgb(10pt)=(0.279467,0.344671,0.971676); rgb(11pt)=(0.275971,0.366681,0.982905); rgb(12pt)=(0.269914,0.3892,0.9906); rgb(13pt)=(0.260243,0.412329,0.995157); rgb(14pt)=(0.244033,0.435833,0.998833); rgb(15pt)=(0.220643,0.460257,0.997286); rgb(16pt)=(0.196333,0.484719,0.989152); rgb(17pt)=(0.183405,0.507371,0.979795); rgb(18pt)=(0.178643,0.528857,0.968157); rgb(19pt)=(0.176438,0.549905,0.952019); rgb(20pt)=(0.168743,0.570262,0.935871); rgb(21pt)=(0.154,0.5902,0.9218); rgb(22pt)=(0.146029,0.609119,0.907857); rgb(23pt)=(0.138024,0.627629,0.89729); rgb(24pt)=(0.124814,0.645929,0.888343); rgb(25pt)=(0.111252,0.6635,0.876314); rgb(26pt)=(0.0952095,0.679829,0.859781); rgb(27pt)=(0.0688714,0.694771,0.839357); rgb(28pt)=(0.0296667,0.708167,0.816333); rgb(29pt)=(0.00357143,0.720267,0.7917); rgb(30pt)=(0.00665714,0.731214,0.766014); rgb(31pt)=(0.0433286,0.741095,0.73941); rgb(32pt)=(0.0963952,0.75,0.712038); rgb(33pt)=(0.140771,0.7584,0.684157); rgb(34pt)=(0.1717,0.766962,0.655443); rgb(35pt)=(0.193767,0.775767,0.6251); rgb(36pt)=(0.216086,0.7843,0.5923); rgb(37pt)=(0.246957,0.791795,0.556743); rgb(38pt)=(0.290614,0.79729,0.518829); rgb(39pt)=(0.340643,0.8008,0.478857); rgb(40pt)=(0.3909,0.802871,0.435448); rgb(41pt)=(0.445629,0.802419,0.390919); rgb(42pt)=(0.5044,0.7993,0.348); rgb(43pt)=(0.561562,0.794233,0.304481); rgb(44pt)=(0.617395,0.787619,0.261238); rgb(45pt)=(0.671986,0.779271,0.2227); rgb(46pt)=(0.7242,0.769843,0.191029); rgb(47pt)=(0.773833,0.759805,0.16461); rgb(48pt)=(0.820314,0.749814,0.153529); rgb(49pt)=(0.863433,0.7406,0.159633); rgb(50pt)=(0.903543,0.733029,0.177414); rgb(51pt)=(0.939257,0.728786,0.209957); rgb(52pt)=(0.972757,0.729771,0.239443); rgb(53pt)=(0.995648,0.743371,0.237148); rgb(54pt)=(0.996986,0.765857,0.219943); rgb(55pt)=(0.995205,0.789252,0.202762); rgb(56pt)=(0.9892,0.813567,0.188533); rgb(57pt)=(0.978629,0.838629,0.176557); rgb(58pt)=(0.967648,0.8639,0.16429); rgb(59pt)=(0.96101,0.889019,0.153676); rgb(60pt)=(0.959671,0.913457,0.142257); rgb(61pt)=(0.962795,0.937338,0.12651); rgb(62pt)=(0.969114,0.960629,0.106362); rgb(63pt)=(0.9769,0.9839,0.0805)},
colorbar
]
\addplot [forget plot] graphics [xmin=-1.00200400801603, xmax=1.00200400801603, ymin=-1.00200400801603, ymax=1.00200400801603] {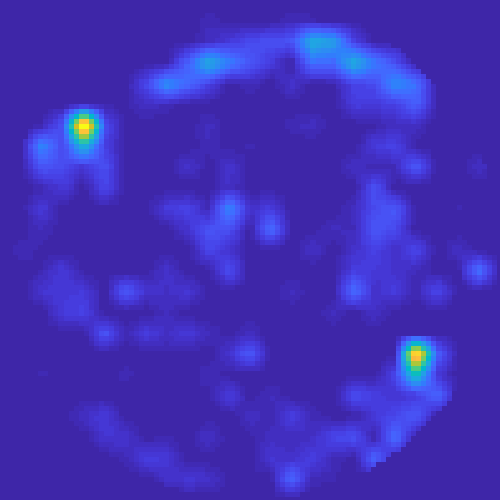};
\end{axis}
\end{tikzpicture}%
	\caption{$64$-bit Data (S1$+$S2$+$S3)}
	\label{fig:AARTFAAC:s346:mle:image:64bit}
\end{subfigure}%
    ~ 
\begin{subfigure}[t]{0.25\textwidth}
\centering
	\setlength{\figurewidth}{5.1cm}
	\setlength{\figureheight}{5.1cm}
%
%
\begin{tikzpicture}[scale=0.55]

\begin{axis}[%
width=\figurewidth,
height=\figureheight,
at={(0.92in,0.642in)},
scale only axis,
point meta min=1e-08,
point meta max=0.0019068208621445,
axis on top,
xmin=-1.00200400801603,
xmax=1.00200400801603,
xlabel style={font=\color{white!15!black}},
xlabel={$\text{South }\leftarrow\text{ Direction }\rightarrow\text{ North}$},
y dir=reverse,
ymin=-1.00200400801603,
ymax=1.00200400801603,
ylabel style={font=\color{white!15!black}},
axis background/.style={fill=white},
legend style={legend cell align=left, align=left, draw=white!15!black},
colormap={mymap}{[1pt] rgb(0pt)=(0.2422,0.1504,0.6603); rgb(1pt)=(0.25039,0.164995,0.707614); rgb(2pt)=(0.257771,0.181781,0.751138); rgb(3pt)=(0.264729,0.197757,0.795214); rgb(4pt)=(0.270648,0.214676,0.836371); rgb(5pt)=(0.275114,0.234238,0.870986); rgb(6pt)=(0.2783,0.255871,0.899071); rgb(7pt)=(0.280333,0.278233,0.9221); rgb(8pt)=(0.281338,0.300595,0.941376); rgb(9pt)=(0.281014,0.322757,0.957886); rgb(10pt)=(0.279467,0.344671,0.971676); rgb(11pt)=(0.275971,0.366681,0.982905); rgb(12pt)=(0.269914,0.3892,0.9906); rgb(13pt)=(0.260243,0.412329,0.995157); rgb(14pt)=(0.244033,0.435833,0.998833); rgb(15pt)=(0.220643,0.460257,0.997286); rgb(16pt)=(0.196333,0.484719,0.989152); rgb(17pt)=(0.183405,0.507371,0.979795); rgb(18pt)=(0.178643,0.528857,0.968157); rgb(19pt)=(0.176438,0.549905,0.952019); rgb(20pt)=(0.168743,0.570262,0.935871); rgb(21pt)=(0.154,0.5902,0.9218); rgb(22pt)=(0.146029,0.609119,0.907857); rgb(23pt)=(0.138024,0.627629,0.89729); rgb(24pt)=(0.124814,0.645929,0.888343); rgb(25pt)=(0.111252,0.6635,0.876314); rgb(26pt)=(0.0952095,0.679829,0.859781); rgb(27pt)=(0.0688714,0.694771,0.839357); rgb(28pt)=(0.0296667,0.708167,0.816333); rgb(29pt)=(0.00357143,0.720267,0.7917); rgb(30pt)=(0.00665714,0.731214,0.766014); rgb(31pt)=(0.0433286,0.741095,0.73941); rgb(32pt)=(0.0963952,0.75,0.712038); rgb(33pt)=(0.140771,0.7584,0.684157); rgb(34pt)=(0.1717,0.766962,0.655443); rgb(35pt)=(0.193767,0.775767,0.6251); rgb(36pt)=(0.216086,0.7843,0.5923); rgb(37pt)=(0.246957,0.791795,0.556743); rgb(38pt)=(0.290614,0.79729,0.518829); rgb(39pt)=(0.340643,0.8008,0.478857); rgb(40pt)=(0.3909,0.802871,0.435448); rgb(41pt)=(0.445629,0.802419,0.390919); rgb(42pt)=(0.5044,0.7993,0.348); rgb(43pt)=(0.561562,0.794233,0.304481); rgb(44pt)=(0.617395,0.787619,0.261238); rgb(45pt)=(0.671986,0.779271,0.2227); rgb(46pt)=(0.7242,0.769843,0.191029); rgb(47pt)=(0.773833,0.759805,0.16461); rgb(48pt)=(0.820314,0.749814,0.153529); rgb(49pt)=(0.863433,0.7406,0.159633); rgb(50pt)=(0.903543,0.733029,0.177414); rgb(51pt)=(0.939257,0.728786,0.209957); rgb(52pt)=(0.972757,0.729771,0.239443); rgb(53pt)=(0.995648,0.743371,0.237148); rgb(54pt)=(0.996986,0.765857,0.219943); rgb(55pt)=(0.995205,0.789252,0.202762); rgb(56pt)=(0.9892,0.813567,0.188533); rgb(57pt)=(0.978629,0.838629,0.176557); rgb(58pt)=(0.967648,0.8639,0.16429); rgb(59pt)=(0.96101,0.889019,0.153676); rgb(60pt)=(0.959671,0.913457,0.142257); rgb(61pt)=(0.962795,0.937338,0.12651); rgb(62pt)=(0.969114,0.960629,0.106362); rgb(63pt)=(0.9769,0.9839,0.0805)},
colorbar
]
\addplot [forget plot] graphics [xmin=-1.00200400801603, xmax=1.00200400801603, ymin=-1.00200400801603, ymax=1.00200400801603] {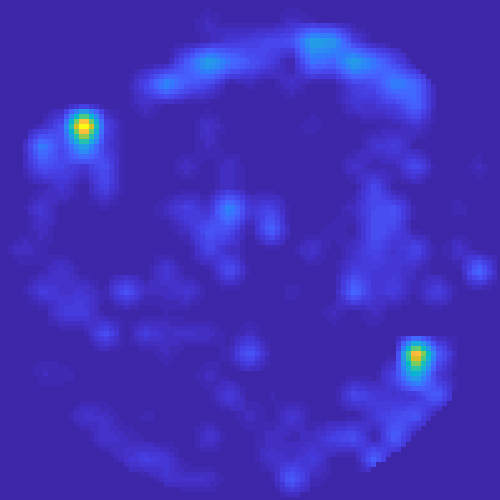};
\end{axis}
\end{tikzpicture}%
	\caption{$1$-bit Data (S1$+$S2$+$S3)}
	\label{fig:AARTFAAC:s346:cmle:image:1bit}
\end{subfigure}%
\caption{AARTFAAC - All-Sky Image ($M=138, D_\text{S}=441$)}
\label{fig:LOFAR:aartfaac:s346:image}
\vspace{0.5cm}
    \begin{subfigure}[t]{0.24\textwidth}
    \centering
	\setlength{\figurewidth}{5.1cm}
	\setlength{\figureheight}{5.1cm}
%
%
\begin{tikzpicture}[scale=0.55]

\begin{axis}[%
width=\figurewidth,
height=\figureheight,
at={(0.92in,0.642in)},
scale only axis,
point meta min=0.0253150710038305,
point meta max=0.0260384865089471,
axis on top,
xmin=-1.00200400801603,
xmax=1.00200400801603,
xlabel style={font=\color{white!15!black}},
xlabel={$\text{South }\leftarrow\text{ Direction }\rightarrow\text{ North}$},
y dir=reverse,
ymin=-1.00200400801603,
ymax=1.00200400801603,
ylabel style={font=\color{white!15!black}},
ylabel={$\text{East }\leftarrow\text{ Direction }\rightarrow\text{ West}$},
axis background/.style={fill=white},
legend style={legend cell align=left, align=left, draw=white!15!black},
colormap={mymap}{[1pt] rgb(0pt)=(0.2422,0.1504,0.6603); rgb(1pt)=(0.25039,0.164995,0.707614); rgb(2pt)=(0.257771,0.181781,0.751138); rgb(3pt)=(0.264729,0.197757,0.795214); rgb(4pt)=(0.270648,0.214676,0.836371); rgb(5pt)=(0.275114,0.234238,0.870986); rgb(6pt)=(0.2783,0.255871,0.899071); rgb(7pt)=(0.280333,0.278233,0.9221); rgb(8pt)=(0.281338,0.300595,0.941376); rgb(9pt)=(0.281014,0.322757,0.957886); rgb(10pt)=(0.279467,0.344671,0.971676); rgb(11pt)=(0.275971,0.366681,0.982905); rgb(12pt)=(0.269914,0.3892,0.9906); rgb(13pt)=(0.260243,0.412329,0.995157); rgb(14pt)=(0.244033,0.435833,0.998833); rgb(15pt)=(0.220643,0.460257,0.997286); rgb(16pt)=(0.196333,0.484719,0.989152); rgb(17pt)=(0.183405,0.507371,0.979795); rgb(18pt)=(0.178643,0.528857,0.968157); rgb(19pt)=(0.176438,0.549905,0.952019); rgb(20pt)=(0.168743,0.570262,0.935871); rgb(21pt)=(0.154,0.5902,0.9218); rgb(22pt)=(0.146029,0.609119,0.907857); rgb(23pt)=(0.138024,0.627629,0.89729); rgb(24pt)=(0.124814,0.645929,0.888343); rgb(25pt)=(0.111252,0.6635,0.876314); rgb(26pt)=(0.0952095,0.679829,0.859781); rgb(27pt)=(0.0688714,0.694771,0.839357); rgb(28pt)=(0.0296667,0.708167,0.816333); rgb(29pt)=(0.00357143,0.720267,0.7917); rgb(30pt)=(0.00665714,0.731214,0.766014); rgb(31pt)=(0.0433286,0.741095,0.73941); rgb(32pt)=(0.0963952,0.75,0.712038); rgb(33pt)=(0.140771,0.7584,0.684157); rgb(34pt)=(0.1717,0.766962,0.655443); rgb(35pt)=(0.193767,0.775767,0.6251); rgb(36pt)=(0.216086,0.7843,0.5923); rgb(37pt)=(0.246957,0.791795,0.556743); rgb(38pt)=(0.290614,0.79729,0.518829); rgb(39pt)=(0.340643,0.8008,0.478857); rgb(40pt)=(0.3909,0.802871,0.435448); rgb(41pt)=(0.445629,0.802419,0.390919); rgb(42pt)=(0.5044,0.7993,0.348); rgb(43pt)=(0.561562,0.794233,0.304481); rgb(44pt)=(0.617395,0.787619,0.261238); rgb(45pt)=(0.671986,0.779271,0.2227); rgb(46pt)=(0.7242,0.769843,0.191029); rgb(47pt)=(0.773833,0.759805,0.16461); rgb(48pt)=(0.820314,0.749814,0.153529); rgb(49pt)=(0.863433,0.7406,0.159633); rgb(50pt)=(0.903543,0.733029,0.177414); rgb(51pt)=(0.939257,0.728786,0.209957); rgb(52pt)=(0.972757,0.729771,0.239443); rgb(53pt)=(0.995648,0.743371,0.237148); rgb(54pt)=(0.996986,0.765857,0.219943); rgb(55pt)=(0.995205,0.789252,0.202762); rgb(56pt)=(0.9892,0.813567,0.188533); rgb(57pt)=(0.978629,0.838629,0.176557); rgb(58pt)=(0.967648,0.8639,0.16429); rgb(59pt)=(0.96101,0.889019,0.153676); rgb(60pt)=(0.959671,0.913457,0.142257); rgb(61pt)=(0.962795,0.937338,0.12651); rgb(62pt)=(0.969114,0.960629,0.106362); rgb(63pt)=(0.9769,0.9839,0.0805)},
colorbar
]
\addplot [forget plot] graphics [xmin=-1.00200400801603, xmax=1.00200400801603, ymin=-1.00200400801603, ymax=1.00200400801603] {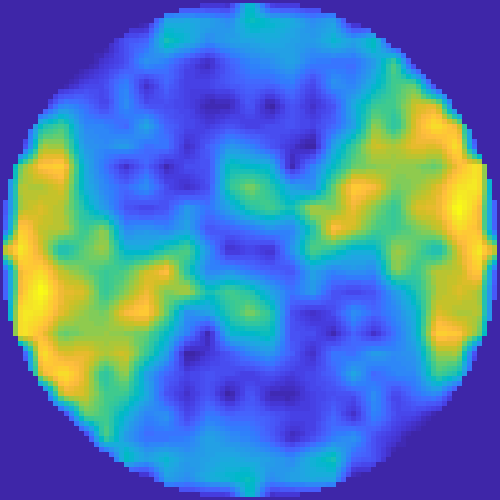};
\end{axis}
\end{tikzpicture}%
	\caption{$64$-bit Data (S1$+$S2$+$S3)}
	\label{fig:AARTFAAC:s346:mle:uncert:64bit}
\end{subfigure}%
    ~ 
\begin{subfigure}[t]{0.25\textwidth}
\centering
	\setlength{\figurewidth}{5.1cm}
	\setlength{\figureheight}{5.1cm}
%
%
\begin{tikzpicture}[scale=0.55]

\begin{axis}[%
width=\figurewidth,
height=\figureheight,
at={(0.92in,0.642in)},
scale only axis,
point meta min=0.0392322965082606,
point meta max=0.0399754776134471,
axis on top,
xmin=-1.00200400801603,
xmax=1.00200400801603,
xlabel style={font=\color{white!15!black}},
xlabel={$\text{South }\leftarrow\text{ Direction }\rightarrow\text{ North}$},
y dir=reverse,
ymin=-1.00200400801603,
ymax=1.00200400801603,
ylabel style={font=\color{white!15!black}},
axis background/.style={fill=white},
legend style={legend cell align=left, align=left, draw=white!15!black},
colormap={mymap}{[1pt] rgb(0pt)=(0.2422,0.1504,0.6603); rgb(1pt)=(0.25039,0.164995,0.707614); rgb(2pt)=(0.257771,0.181781,0.751138); rgb(3pt)=(0.264729,0.197757,0.795214); rgb(4pt)=(0.270648,0.214676,0.836371); rgb(5pt)=(0.275114,0.234238,0.870986); rgb(6pt)=(0.2783,0.255871,0.899071); rgb(7pt)=(0.280333,0.278233,0.9221); rgb(8pt)=(0.281338,0.300595,0.941376); rgb(9pt)=(0.281014,0.322757,0.957886); rgb(10pt)=(0.279467,0.344671,0.971676); rgb(11pt)=(0.275971,0.366681,0.982905); rgb(12pt)=(0.269914,0.3892,0.9906); rgb(13pt)=(0.260243,0.412329,0.995157); rgb(14pt)=(0.244033,0.435833,0.998833); rgb(15pt)=(0.220643,0.460257,0.997286); rgb(16pt)=(0.196333,0.484719,0.989152); rgb(17pt)=(0.183405,0.507371,0.979795); rgb(18pt)=(0.178643,0.528857,0.968157); rgb(19pt)=(0.176438,0.549905,0.952019); rgb(20pt)=(0.168743,0.570262,0.935871); rgb(21pt)=(0.154,0.5902,0.9218); rgb(22pt)=(0.146029,0.609119,0.907857); rgb(23pt)=(0.138024,0.627629,0.89729); rgb(24pt)=(0.124814,0.645929,0.888343); rgb(25pt)=(0.111252,0.6635,0.876314); rgb(26pt)=(0.0952095,0.679829,0.859781); rgb(27pt)=(0.0688714,0.694771,0.839357); rgb(28pt)=(0.0296667,0.708167,0.816333); rgb(29pt)=(0.00357143,0.720267,0.7917); rgb(30pt)=(0.00665714,0.731214,0.766014); rgb(31pt)=(0.0433286,0.741095,0.73941); rgb(32pt)=(0.0963952,0.75,0.712038); rgb(33pt)=(0.140771,0.7584,0.684157); rgb(34pt)=(0.1717,0.766962,0.655443); rgb(35pt)=(0.193767,0.775767,0.6251); rgb(36pt)=(0.216086,0.7843,0.5923); rgb(37pt)=(0.246957,0.791795,0.556743); rgb(38pt)=(0.290614,0.79729,0.518829); rgb(39pt)=(0.340643,0.8008,0.478857); rgb(40pt)=(0.3909,0.802871,0.435448); rgb(41pt)=(0.445629,0.802419,0.390919); rgb(42pt)=(0.5044,0.7993,0.348); rgb(43pt)=(0.561562,0.794233,0.304481); rgb(44pt)=(0.617395,0.787619,0.261238); rgb(45pt)=(0.671986,0.779271,0.2227); rgb(46pt)=(0.7242,0.769843,0.191029); rgb(47pt)=(0.773833,0.759805,0.16461); rgb(48pt)=(0.820314,0.749814,0.153529); rgb(49pt)=(0.863433,0.7406,0.159633); rgb(50pt)=(0.903543,0.733029,0.177414); rgb(51pt)=(0.939257,0.728786,0.209957); rgb(52pt)=(0.972757,0.729771,0.239443); rgb(53pt)=(0.995648,0.743371,0.237148); rgb(54pt)=(0.996986,0.765857,0.219943); rgb(55pt)=(0.995205,0.789252,0.202762); rgb(56pt)=(0.9892,0.813567,0.188533); rgb(57pt)=(0.978629,0.838629,0.176557); rgb(58pt)=(0.967648,0.8639,0.16429); rgb(59pt)=(0.96101,0.889019,0.153676); rgb(60pt)=(0.959671,0.913457,0.142257); rgb(61pt)=(0.962795,0.937338,0.12651); rgb(62pt)=(0.969114,0.960629,0.106362); rgb(63pt)=(0.9769,0.9839,0.0805)},
colorbar
]
\addplot [forget plot] graphics [xmin=-1.00200400801603, xmax=1.00200400801603, ymin=-1.00200400801603, ymax=1.00200400801603] {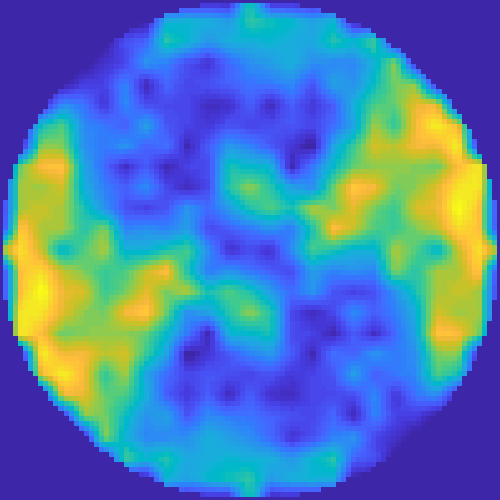};
\end{axis}
\end{tikzpicture}%
	\caption{$1$-bit Data (S1$+$S2$+$S3)}
	\label{fig:AARTFAAC:s346:cmle:uncert:1bit}
\end{subfigure}%
\caption{AARTFAAC - Uncertainty ($M=138, D_\text{S}=441$)}
\label{fig:AARTFAAC:s346:uncert}
\end{figure}

\begin{figure}
\centering
    \begin{subfigure}[t]{0.24\textwidth}
    \centering
	\setlength{\figurewidth}{5.1cm}
	\setlength{\figureheight}{5.1cm}
%
%
\begin{tikzpicture}[scale=0.55]

\begin{axis}[%
width=\figurewidth,
height=\figureheight,
at={(0.92in,0.642in)},
scale only axis,
point meta min=1e-08,
point meta max=0.00307767432197427,
axis on top,
xmin=-1.00200400801603,
xmax=1.00200400801603,
xlabel style={font=\color{white!15!black}},
xlabel={$\text{South }\leftarrow\text{ Direction }\rightarrow\text{ North}$},
y dir=reverse,
ymin=-1.00200400801603,
ymax=1.00200400801603,
ylabel style={font=\color{white!15!black}},
ylabel={$\text{East }\leftarrow\text{ Direction }\rightarrow\text{ West}$},
axis background/.style={fill=white},
legend style={legend cell align=left, align=left, draw=white!15!black},
colormap={mymap}{[1pt] rgb(0pt)=(0.2422,0.1504,0.6603); rgb(1pt)=(0.25039,0.164995,0.707614); rgb(2pt)=(0.257771,0.181781,0.751138); rgb(3pt)=(0.264729,0.197757,0.795214); rgb(4pt)=(0.270648,0.214676,0.836371); rgb(5pt)=(0.275114,0.234238,0.870986); rgb(6pt)=(0.2783,0.255871,0.899071); rgb(7pt)=(0.280333,0.278233,0.9221); rgb(8pt)=(0.281338,0.300595,0.941376); rgb(9pt)=(0.281014,0.322757,0.957886); rgb(10pt)=(0.279467,0.344671,0.971676); rgb(11pt)=(0.275971,0.366681,0.982905); rgb(12pt)=(0.269914,0.3892,0.9906); rgb(13pt)=(0.260243,0.412329,0.995157); rgb(14pt)=(0.244033,0.435833,0.998833); rgb(15pt)=(0.220643,0.460257,0.997286); rgb(16pt)=(0.196333,0.484719,0.989152); rgb(17pt)=(0.183405,0.507371,0.979795); rgb(18pt)=(0.178643,0.528857,0.968157); rgb(19pt)=(0.176438,0.549905,0.952019); rgb(20pt)=(0.168743,0.570262,0.935871); rgb(21pt)=(0.154,0.5902,0.9218); rgb(22pt)=(0.146029,0.609119,0.907857); rgb(23pt)=(0.138024,0.627629,0.89729); rgb(24pt)=(0.124814,0.645929,0.888343); rgb(25pt)=(0.111252,0.6635,0.876314); rgb(26pt)=(0.0952095,0.679829,0.859781); rgb(27pt)=(0.0688714,0.694771,0.839357); rgb(28pt)=(0.0296667,0.708167,0.816333); rgb(29pt)=(0.00357143,0.720267,0.7917); rgb(30pt)=(0.00665714,0.731214,0.766014); rgb(31pt)=(0.0433286,0.741095,0.73941); rgb(32pt)=(0.0963952,0.75,0.712038); rgb(33pt)=(0.140771,0.7584,0.684157); rgb(34pt)=(0.1717,0.766962,0.655443); rgb(35pt)=(0.193767,0.775767,0.6251); rgb(36pt)=(0.216086,0.7843,0.5923); rgb(37pt)=(0.246957,0.791795,0.556743); rgb(38pt)=(0.290614,0.79729,0.518829); rgb(39pt)=(0.340643,0.8008,0.478857); rgb(40pt)=(0.3909,0.802871,0.435448); rgb(41pt)=(0.445629,0.802419,0.390919); rgb(42pt)=(0.5044,0.7993,0.348); rgb(43pt)=(0.561562,0.794233,0.304481); rgb(44pt)=(0.617395,0.787619,0.261238); rgb(45pt)=(0.671986,0.779271,0.2227); rgb(46pt)=(0.7242,0.769843,0.191029); rgb(47pt)=(0.773833,0.759805,0.16461); rgb(48pt)=(0.820314,0.749814,0.153529); rgb(49pt)=(0.863433,0.7406,0.159633); rgb(50pt)=(0.903543,0.733029,0.177414); rgb(51pt)=(0.939257,0.728786,0.209957); rgb(52pt)=(0.972757,0.729771,0.239443); rgb(53pt)=(0.995648,0.743371,0.237148); rgb(54pt)=(0.996986,0.765857,0.219943); rgb(55pt)=(0.995205,0.789252,0.202762); rgb(56pt)=(0.9892,0.813567,0.188533); rgb(57pt)=(0.978629,0.838629,0.176557); rgb(58pt)=(0.967648,0.8639,0.16429); rgb(59pt)=(0.96101,0.889019,0.153676); rgb(60pt)=(0.959671,0.913457,0.142257); rgb(61pt)=(0.962795,0.937338,0.12651); rgb(62pt)=(0.969114,0.960629,0.106362); rgb(63pt)=(0.9769,0.9839,0.0805)},
colorbar
]
\addplot [forget plot] graphics [xmin=-1.00200400801603, xmax=1.00200400801603, ymin=-1.00200400801603, ymax=1.00200400801603] {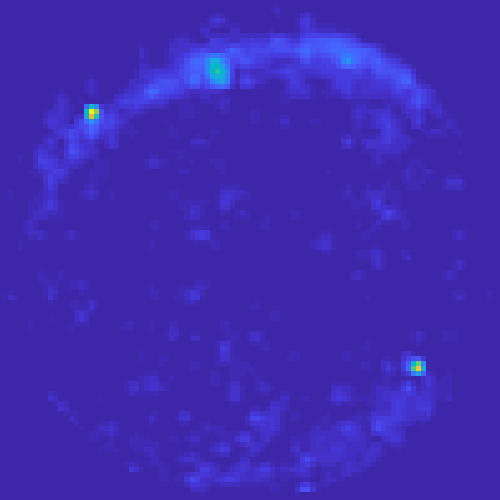};
\end{axis}
\end{tikzpicture}%
	\caption{$64$-bit Data (S1$+$S2$+$S3)}
	\label{fig:AARTFAAC:s346:mle:image:p50:64bit}
\end{subfigure}%
    ~ 
\begin{subfigure}[t]{0.25\textwidth}
\centering
	\setlength{\figurewidth}{5.1cm}
	\setlength{\figureheight}{5.1cm}
%
%
\begin{tikzpicture}[scale=0.55]

\begin{axis}[%
width=\figurewidth,
height=\figureheight,
at={(0.92in,0.642in)},
scale only axis,
point meta min=1e-08,
point meta max=0.00317475801750691,
axis on top,
xmin=-1.00200400801603,
xmax=1.00200400801603,
xlabel style={font=\color{white!15!black}},
xlabel={$\text{South }\leftarrow\text{ Direction }\rightarrow\text{ North}$},
y dir=reverse,
ymin=-1.00200400801603,
ymax=1.00200400801603,
ylabel style={font=\color{white!15!black}},
axis background/.style={fill=white},
legend style={legend cell align=left, align=left, draw=white!15!black},
colormap={mymap}{[1pt] rgb(0pt)=(0.2422,0.1504,0.6603); rgb(1pt)=(0.25039,0.164995,0.707614); rgb(2pt)=(0.257771,0.181781,0.751138); rgb(3pt)=(0.264729,0.197757,0.795214); rgb(4pt)=(0.270648,0.214676,0.836371); rgb(5pt)=(0.275114,0.234238,0.870986); rgb(6pt)=(0.2783,0.255871,0.899071); rgb(7pt)=(0.280333,0.278233,0.9221); rgb(8pt)=(0.281338,0.300595,0.941376); rgb(9pt)=(0.281014,0.322757,0.957886); rgb(10pt)=(0.279467,0.344671,0.971676); rgb(11pt)=(0.275971,0.366681,0.982905); rgb(12pt)=(0.269914,0.3892,0.9906); rgb(13pt)=(0.260243,0.412329,0.995157); rgb(14pt)=(0.244033,0.435833,0.998833); rgb(15pt)=(0.220643,0.460257,0.997286); rgb(16pt)=(0.196333,0.484719,0.989152); rgb(17pt)=(0.183405,0.507371,0.979795); rgb(18pt)=(0.178643,0.528857,0.968157); rgb(19pt)=(0.176438,0.549905,0.952019); rgb(20pt)=(0.168743,0.570262,0.935871); rgb(21pt)=(0.154,0.5902,0.9218); rgb(22pt)=(0.146029,0.609119,0.907857); rgb(23pt)=(0.138024,0.627629,0.89729); rgb(24pt)=(0.124814,0.645929,0.888343); rgb(25pt)=(0.111252,0.6635,0.876314); rgb(26pt)=(0.0952095,0.679829,0.859781); rgb(27pt)=(0.0688714,0.694771,0.839357); rgb(28pt)=(0.0296667,0.708167,0.816333); rgb(29pt)=(0.00357143,0.720267,0.7917); rgb(30pt)=(0.00665714,0.731214,0.766014); rgb(31pt)=(0.0433286,0.741095,0.73941); rgb(32pt)=(0.0963952,0.75,0.712038); rgb(33pt)=(0.140771,0.7584,0.684157); rgb(34pt)=(0.1717,0.766962,0.655443); rgb(35pt)=(0.193767,0.775767,0.6251); rgb(36pt)=(0.216086,0.7843,0.5923); rgb(37pt)=(0.246957,0.791795,0.556743); rgb(38pt)=(0.290614,0.79729,0.518829); rgb(39pt)=(0.340643,0.8008,0.478857); rgb(40pt)=(0.3909,0.802871,0.435448); rgb(41pt)=(0.445629,0.802419,0.390919); rgb(42pt)=(0.5044,0.7993,0.348); rgb(43pt)=(0.561562,0.794233,0.304481); rgb(44pt)=(0.617395,0.787619,0.261238); rgb(45pt)=(0.671986,0.779271,0.2227); rgb(46pt)=(0.7242,0.769843,0.191029); rgb(47pt)=(0.773833,0.759805,0.16461); rgb(48pt)=(0.820314,0.749814,0.153529); rgb(49pt)=(0.863433,0.7406,0.159633); rgb(50pt)=(0.903543,0.733029,0.177414); rgb(51pt)=(0.939257,0.728786,0.209957); rgb(52pt)=(0.972757,0.729771,0.239443); rgb(53pt)=(0.995648,0.743371,0.237148); rgb(54pt)=(0.996986,0.765857,0.219943); rgb(55pt)=(0.995205,0.789252,0.202762); rgb(56pt)=(0.9892,0.813567,0.188533); rgb(57pt)=(0.978629,0.838629,0.176557); rgb(58pt)=(0.967648,0.8639,0.16429); rgb(59pt)=(0.96101,0.889019,0.153676); rgb(60pt)=(0.959671,0.913457,0.142257); rgb(61pt)=(0.962795,0.937338,0.12651); rgb(62pt)=(0.969114,0.960629,0.106362); rgb(63pt)=(0.9769,0.9839,0.0805)},
colorbar
]
\addplot [forget plot] graphics [xmin=-1.00200400801603, xmax=1.00200400801603, ymin=-1.00200400801603, ymax=1.00200400801603] {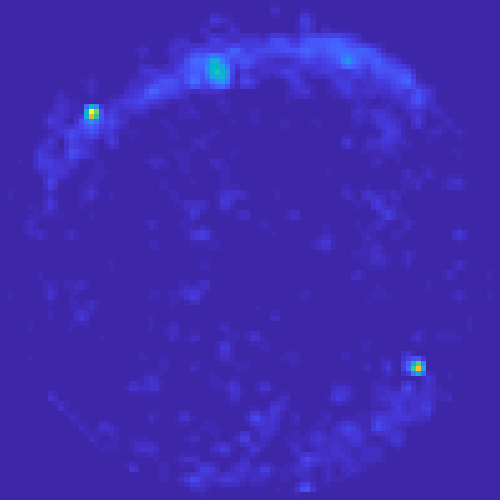};
\end{axis}
\end{tikzpicture}%
	\caption{$1$-bit Data (S1$+$S2$+$S3)}
	\label{fig:AARTFAAC:s346:cmle:image:p50:1bit}
\end{subfigure}%
\caption{AARTFAAC - All-Sky Image ($M=138, D_\text{S}=1876$)}
\label{fig:LOFAR:aartfaac:s346:image:p50}
\vspace{0.5cm}
    \begin{subfigure}[t]{0.24\textwidth}
    \centering
	\setlength{\figurewidth}{5.1cm}
	\setlength{\figureheight}{5.1cm}
%
%
\begin{tikzpicture}[scale=0.55]

\begin{axis}[%
width=\figurewidth,
height=\figureheight,
at={(0.92in,0.642in)},
scale only axis,
point meta min=0.0495539319166857,
point meta max=0.0519221667706263,
axis on top,
xmin=-1.00200400801603,
xmax=1.00200400801603,
xlabel style={font=\color{white!15!black}},
xlabel={$\text{South }\leftarrow\text{ Direction }\rightarrow\text{ North}$},
y dir=reverse,
ymin=-1.00200400801603,
ymax=1.00200400801603,
ylabel style={font=\color{white!15!black}},
ylabel={$\text{East }\leftarrow\text{ Direction }\rightarrow\text{ West}$},
axis background/.style={fill=white},
legend style={legend cell align=left, align=left, draw=white!15!black},
colormap={mymap}{[1pt] rgb(0pt)=(0.2422,0.1504,0.6603); rgb(1pt)=(0.25039,0.164995,0.707614); rgb(2pt)=(0.257771,0.181781,0.751138); rgb(3pt)=(0.264729,0.197757,0.795214); rgb(4pt)=(0.270648,0.214676,0.836371); rgb(5pt)=(0.275114,0.234238,0.870986); rgb(6pt)=(0.2783,0.255871,0.899071); rgb(7pt)=(0.280333,0.278233,0.9221); rgb(8pt)=(0.281338,0.300595,0.941376); rgb(9pt)=(0.281014,0.322757,0.957886); rgb(10pt)=(0.279467,0.344671,0.971676); rgb(11pt)=(0.275971,0.366681,0.982905); rgb(12pt)=(0.269914,0.3892,0.9906); rgb(13pt)=(0.260243,0.412329,0.995157); rgb(14pt)=(0.244033,0.435833,0.998833); rgb(15pt)=(0.220643,0.460257,0.997286); rgb(16pt)=(0.196333,0.484719,0.989152); rgb(17pt)=(0.183405,0.507371,0.979795); rgb(18pt)=(0.178643,0.528857,0.968157); rgb(19pt)=(0.176438,0.549905,0.952019); rgb(20pt)=(0.168743,0.570262,0.935871); rgb(21pt)=(0.154,0.5902,0.9218); rgb(22pt)=(0.146029,0.609119,0.907857); rgb(23pt)=(0.138024,0.627629,0.89729); rgb(24pt)=(0.124814,0.645929,0.888343); rgb(25pt)=(0.111252,0.6635,0.876314); rgb(26pt)=(0.0952095,0.679829,0.859781); rgb(27pt)=(0.0688714,0.694771,0.839357); rgb(28pt)=(0.0296667,0.708167,0.816333); rgb(29pt)=(0.00357143,0.720267,0.7917); rgb(30pt)=(0.00665714,0.731214,0.766014); rgb(31pt)=(0.0433286,0.741095,0.73941); rgb(32pt)=(0.0963952,0.75,0.712038); rgb(33pt)=(0.140771,0.7584,0.684157); rgb(34pt)=(0.1717,0.766962,0.655443); rgb(35pt)=(0.193767,0.775767,0.6251); rgb(36pt)=(0.216086,0.7843,0.5923); rgb(37pt)=(0.246957,0.791795,0.556743); rgb(38pt)=(0.290614,0.79729,0.518829); rgb(39pt)=(0.340643,0.8008,0.478857); rgb(40pt)=(0.3909,0.802871,0.435448); rgb(41pt)=(0.445629,0.802419,0.390919); rgb(42pt)=(0.5044,0.7993,0.348); rgb(43pt)=(0.561562,0.794233,0.304481); rgb(44pt)=(0.617395,0.787619,0.261238); rgb(45pt)=(0.671986,0.779271,0.2227); rgb(46pt)=(0.7242,0.769843,0.191029); rgb(47pt)=(0.773833,0.759805,0.16461); rgb(48pt)=(0.820314,0.749814,0.153529); rgb(49pt)=(0.863433,0.7406,0.159633); rgb(50pt)=(0.903543,0.733029,0.177414); rgb(51pt)=(0.939257,0.728786,0.209957); rgb(52pt)=(0.972757,0.729771,0.239443); rgb(53pt)=(0.995648,0.743371,0.237148); rgb(54pt)=(0.996986,0.765857,0.219943); rgb(55pt)=(0.995205,0.789252,0.202762); rgb(56pt)=(0.9892,0.813567,0.188533); rgb(57pt)=(0.978629,0.838629,0.176557); rgb(58pt)=(0.967648,0.8639,0.16429); rgb(59pt)=(0.96101,0.889019,0.153676); rgb(60pt)=(0.959671,0.913457,0.142257); rgb(61pt)=(0.962795,0.937338,0.12651); rgb(62pt)=(0.969114,0.960629,0.106362); rgb(63pt)=(0.9769,0.9839,0.0805)},
colorbar
]
\addplot [forget plot] graphics [xmin=-1.00200400801603, xmax=1.00200400801603, ymin=-1.00200400801603, ymax=1.00200400801603] {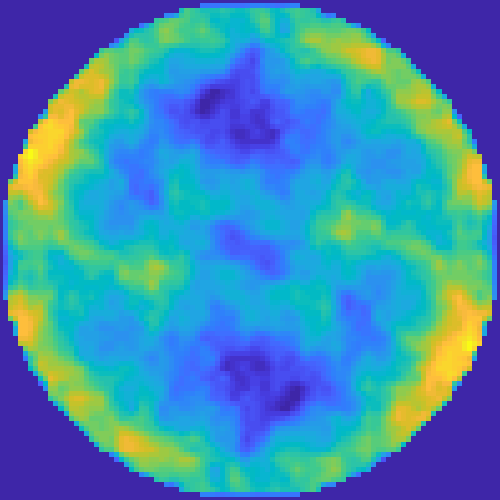};
\end{axis}
\end{tikzpicture}%
	\caption{$64$-bit Data (S1$+$S2$+$S3)}
	\label{fig:AARTFAAC:s346:mle:uncert:p50:64bit}
\end{subfigure}%
    ~ 
\begin{subfigure}[t]{0.25\textwidth}
\centering
	\setlength{\figurewidth}{5.1cm}
	\setlength{\figureheight}{5.1cm}
%
%
\begin{tikzpicture}[scale=0.55]

\begin{axis}[%
width=\figurewidth,
height=\figureheight,
at={(0.92in,0.642in)},
scale only axis,
point meta min=0.078518507756812,
point meta max=0.0809022460918489,
axis on top,
xmin=-1.00200400801603,
xmax=1.00200400801603,
xlabel style={font=\color{white!15!black}},
xlabel={$\text{South }\leftarrow\text{ Direction }\rightarrow\text{ North}$},
y dir=reverse,
ymin=-1.00200400801603,
ymax=1.00200400801603,
ylabel style={font=\color{white!15!black}},
axis background/.style={fill=white},
legend style={legend cell align=left, align=left, draw=white!15!black},
colormap={mymap}{[1pt] rgb(0pt)=(0.2422,0.1504,0.6603); rgb(1pt)=(0.25039,0.164995,0.707614); rgb(2pt)=(0.257771,0.181781,0.751138); rgb(3pt)=(0.264729,0.197757,0.795214); rgb(4pt)=(0.270648,0.214676,0.836371); rgb(5pt)=(0.275114,0.234238,0.870986); rgb(6pt)=(0.2783,0.255871,0.899071); rgb(7pt)=(0.280333,0.278233,0.9221); rgb(8pt)=(0.281338,0.300595,0.941376); rgb(9pt)=(0.281014,0.322757,0.957886); rgb(10pt)=(0.279467,0.344671,0.971676); rgb(11pt)=(0.275971,0.366681,0.982905); rgb(12pt)=(0.269914,0.3892,0.9906); rgb(13pt)=(0.260243,0.412329,0.995157); rgb(14pt)=(0.244033,0.435833,0.998833); rgb(15pt)=(0.220643,0.460257,0.997286); rgb(16pt)=(0.196333,0.484719,0.989152); rgb(17pt)=(0.183405,0.507371,0.979795); rgb(18pt)=(0.178643,0.528857,0.968157); rgb(19pt)=(0.176438,0.549905,0.952019); rgb(20pt)=(0.168743,0.570262,0.935871); rgb(21pt)=(0.154,0.5902,0.9218); rgb(22pt)=(0.146029,0.609119,0.907857); rgb(23pt)=(0.138024,0.627629,0.89729); rgb(24pt)=(0.124814,0.645929,0.888343); rgb(25pt)=(0.111252,0.6635,0.876314); rgb(26pt)=(0.0952095,0.679829,0.859781); rgb(27pt)=(0.0688714,0.694771,0.839357); rgb(28pt)=(0.0296667,0.708167,0.816333); rgb(29pt)=(0.00357143,0.720267,0.7917); rgb(30pt)=(0.00665714,0.731214,0.766014); rgb(31pt)=(0.0433286,0.741095,0.73941); rgb(32pt)=(0.0963952,0.75,0.712038); rgb(33pt)=(0.140771,0.7584,0.684157); rgb(34pt)=(0.1717,0.766962,0.655443); rgb(35pt)=(0.193767,0.775767,0.6251); rgb(36pt)=(0.216086,0.7843,0.5923); rgb(37pt)=(0.246957,0.791795,0.556743); rgb(38pt)=(0.290614,0.79729,0.518829); rgb(39pt)=(0.340643,0.8008,0.478857); rgb(40pt)=(0.3909,0.802871,0.435448); rgb(41pt)=(0.445629,0.802419,0.390919); rgb(42pt)=(0.5044,0.7993,0.348); rgb(43pt)=(0.561562,0.794233,0.304481); rgb(44pt)=(0.617395,0.787619,0.261238); rgb(45pt)=(0.671986,0.779271,0.2227); rgb(46pt)=(0.7242,0.769843,0.191029); rgb(47pt)=(0.773833,0.759805,0.16461); rgb(48pt)=(0.820314,0.749814,0.153529); rgb(49pt)=(0.863433,0.7406,0.159633); rgb(50pt)=(0.903543,0.733029,0.177414); rgb(51pt)=(0.939257,0.728786,0.209957); rgb(52pt)=(0.972757,0.729771,0.239443); rgb(53pt)=(0.995648,0.743371,0.237148); rgb(54pt)=(0.996986,0.765857,0.219943); rgb(55pt)=(0.995205,0.789252,0.202762); rgb(56pt)=(0.9892,0.813567,0.188533); rgb(57pt)=(0.978629,0.838629,0.176557); rgb(58pt)=(0.967648,0.8639,0.16429); rgb(59pt)=(0.96101,0.889019,0.153676); rgb(60pt)=(0.959671,0.913457,0.142257); rgb(61pt)=(0.962795,0.937338,0.12651); rgb(62pt)=(0.969114,0.960629,0.106362); rgb(63pt)=(0.9769,0.9839,0.0805)},
colorbar
]
\addplot [forget plot] graphics [xmin=-1.00200400801603, xmax=1.00200400801603, ymin=-1.00200400801603, ymax=1.00200400801603] {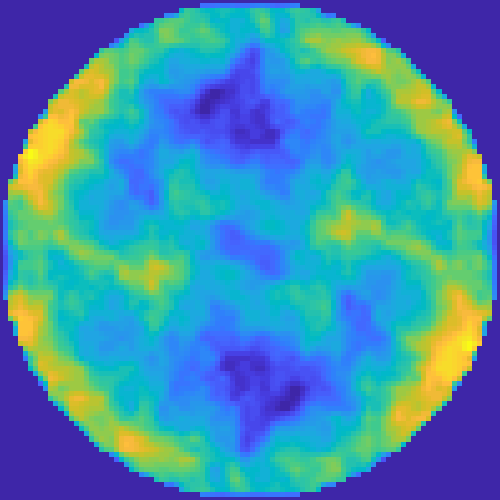};
\end{axis}
\end{tikzpicture}%
	\caption{$1$-bit Data (S1$+$S2$+$S3)}
	\label{fig:AARTFAAC:s346:cmle:uncert:p50:1bit}
\end{subfigure}%
\caption{AARTFAAC - Uncertainty ($M=138, D_\text{S}=1876$)}
\label{fig:AARTFAAC:s346:uncert:p50}
\end{figure}

Fig. \ref{fig:LOFAR:aartfaac:s346:image} shows the all-sky images when using the radio measurements of all three LOFAR stations jointly ($M=138$) to resolve $D_\text{S}=441$ source power levels, while Fig. \ref{fig:AARTFAAC:s346:uncert} shows the resulting interferometric uncertainty. The additional radio antennas further push the sensitivity of the telescope systems with high-resolution and low-resolution A/D conversion.
With $M=138$ sensing elements, it is possible to increase the number of sources to $D_\text{S}=1876$ ($50$ pixels). The interferometric algorithms then provide a higher spatial resolution on the all-sky images, which is demonstrated in Fig. \ref{fig:LOFAR:aartfaac:s346:image:p50}.
Analyzing Fig. \ref{fig:AARTFAAC:s346:uncert:p50} in comparison to Fig. \ref{fig:AARTFAAC:s346:uncert}, it can be seen that the interferometric sensitivity decreases when increasing the spatial imaging resolution. However, the algorithm reconstructing $D_\text{S}=1876$ intensity values from the measurements of $M=138$ binary antennas provides a higher spatial resolution at a lower interferometric uncertainty than the three systems using $M=46$ antennas with $64$-bit A/D resolution to estimate $D_\text{S}=441$ power levels. The binary telescope with $M=138$ array elements achieves this while processing a data volume which is more than $21$ times smaller than with $M=46$ sensors featuring $64$-bit outputs.
\section{Discussion}
Complementing the discourse, certain aspects related to the different views on image reconstruction in radio astronomy, likelihood-oriented processing, radio interferometry with quantized signals, and storage-based telescopes are discussed.

\subsection{Deterministic and Probabilistic Perspectives}
To classify the presented methods and results, we try to connect them with other imaging approaches in radio astronomy:
\subsubsection{Fourier Analysis}
Classical image reconstruction approaches in radio astronomy \cite{Hoegbom74,Clark80,Schwab84,Cornwell08,Cornwell08_W} are based on the van Cittert-Zernike theorem \cite{Cittert34,Zernike38} and assume that the electromagnetic power distribution can be mapped onto the celestial sphere. Geometric arguments and two-dimensional Fourier analysis is used to establish a deterministic relationship between the intensity values on the sphere and the correlations (visibilities) between signals of antenna pairs (baseline) at the receiver. This analytic relationship in connection with sampling theorems is used to formulate the reconstruction of the radio image as a deconvolution problem. Since these methods are formulated from a deterministic viewpoint, they require sufficiently long integration time $T$ and do not make use of the stochastic properties of the measurement system. 

A first step towards a probabilistic perspective in the context of deconvolution methods is the use of the maximum entropy principle \cite{Gull78, Cornwell85,Narayan86}. Out of all images which can be brought into agreement with the measured correlations, the one which maximizes a certain entropy measure is selected. The random nature of the empirical correlations and the intensity values is considered element-wise such that the statistical dependencies between these variables is not exploited during the reconstruction. The entropy measure itself can be seen as regularizing probabilistic prior knowledge \eqref{prior:phy}. Methods from compressed sensing \cite{Bobin08,Wiaux09} are based on the idea that the image can be linearly decomposed in a sparse random basis. This approach can be seen as a probabilistic variant of the Fourier analysis which features a deterministic basis. Additional probabilistic traits are notable here when interpreting the norm which determines the degree of sparseness as prior knowledge \eqref{prior:phy}. 
\subsubsection{Array Signal Processing}
An important contribution, which detaches the discussion from the deterministic viewpoint of the Fourier analysis and opens up a receiver-centric probabilistic view onto radio astronomical imaging, is the modeling of the noisy empirical correlations at the receiver and the summary of all correlation values in linear matrix-vector equations \cite{Leshem00}. Such the rich set of array processing and linear algebra methods can be used for imaging in radio astronomy \cite{Leshem01,VanDerTol07,BenDavid08,Wijnholds08,Levanda10,Veen13,Levanda13,Sardarabadi16,Naghibzadeh18}. Within this framework, methods such as covariance matching and weighted least squares now offer the possibility of exploiting statistical dependencies between the empirical correlations during the image reconstruction.
\subsubsection{Hardware-aware Statistical Signal Processing}
Within such a probabilistic understanding of astronomy this work goes one step further. It models the radio signals as function of the intensity parameters at the front-end behind the antennas and considers hardware effects during the transition from analog sensor signals to digital measurement data. For the precise probabilistic modeling of the digital measurement data, it deviates from the idealistic assumption of a Gaussian distribution and abstracts the data model in the exponential family. This class of distributions covers a broad range of sensor data models relevant in practice and exhibits favorable mathematical properties. For interferometric imaging, this work relies purely on the information embedded in the sensor measurements and makes no prior assumptions \eqref{prior:phy} about the structure of the radio image. By means of parametric statistical methods, which are known to meet certain optimality criteria, the discussion on algorithmic approaches is restricted. Committing to these techniques, the quality of the array system can be quantitatively determined and optimized based on estimation-theoretic information measures. The results obtained with LOFAR measurements imply that, for the task of interferometric imaging, the sensor hardware and data communication infrastructure of current radio telescopes are over-specified. 
\subsection{Challenges in Likelihood-oriented Data Processing}
Besides their numerous advantages, the use of optimal statistical algorithms also raises issues. For example, the scoring method \eqref{update:scoring:rule} used here to solve the maximum-likelihood optimization problem \eqref{definition:mle} only guarantees a fast convergence to one of the roots of the score \eqref{ideal:score}. If this function has multiple roots, the question arises whether the solution found is the global maximum of the likelihood \cite{Small00}. It is known that within the exponential family, under a smooth parameterization of the statistical weights, the score function has a single root. These conditions are fulfilled for the model with high A/D resolution \eqref{multivariate:gauss:covariance:parameter} and linear covariance parameters \eqref{definition:received:covariance}. To the best of our knowledge, the effect of signal quantization on the score is an open problem. The results obtained here with hard-limited sensor data and the behavior of the scoring method during our tests indicate that quantization of the measurements has no significant effect on the structure of the score function \eqref{ideal:score}.

An obstacle for optimal processing methods such as the MLE \eqref{definition:mle} is computational complexity. In the literature, this is frequently found as an argument to resort to suboptimal methods. At first glance one is apt to agree. For example, the update term \eqref{scoring:update:quant} shows that for exponential family distributions the covariance matrix of the sufficient statistics \eqref{definition:covariance:transformed} must be inverted. The multivariate Gaussian model with $2M$ variables \eqref{multivariate:gauss:covariance:parameter} exhibits $C=2M^2+M$ of these statistics, such that the inversion has to support a matrix size of $38226 \times 38226$ when considering $M=138$ antennas. With a second look it can be seen that matrices like \eqref{definition:covariance:transformed} have structure. That is why the update term \eqref{scoring:update:substitution} can be computed with inverting a $276 \times 276$ matrix. When calculating the image in Fig. \ref{fig:AARTFAAC:s346:mle:image:64bit}, one iteration of Fisher scoring \eqref{update:scoring:rule} on a simple computer (\SI{1.6}{\giga\hertz}, \SI{4}{\giga\byte} RAM, MATLAB) takes $\SI{1.5}{\second}$. Certainly too much time for a real-time application on a mobile device. But for radio astronomy, which has no time requirements after the correlation stage \eqref{definition:empirical:cov:y} and for which supercomputers with computing power in the range of PFLOPS are available today, definitely feasible. Additionally, the computational cost of the MLE vanishes constantly at the rate of Moore's law. Note that in return for the computational effort the MLE algorithm is consistent and efficient, two important criteria for high-performance sensor data processing in scientific applications.
\subsection{Van Vleck Corrections vs. Multivariate Binary MLEs}
The idea of using quantized data for radio astronomy is not new, see, e.g., \cite[pp. 254]{BookThompson}. The standard approach is to apply the Van Vleck corrections, see, e.g., \cite{Johnson13}. There, through the inverse of the arc-sine law \eqref{arcsin:law}, one transforms the elements of the quantized empirical covariance \eqref{empirical:auxiliary:statistics:quant} independently from each other with the goal to obtain the unquantized empirical covariance \eqref{definition:empirical:cov:y}. Then one works under the assumption of an ideal $\infty$-bit digital signal acquisition and, therefore, with the formulas of the Gaussian framework \eqref{multivariate:gauss:covariance:parameter}. An issue in this correction technique is that the arc-sine law \eqref{arcsin:law} is an analytic relationship connecting the two asymptotic covariances \eqref{covariance:unquantized} and \eqref{covariance:quantized}. With a finite number of samples $N$, the elements of the two empirical covariances \eqref{definition:empirical:cov:y} and \eqref{empirical:auxiliary:statistics:quant}, deviate randomly from their asymptotic versions. Further, the elements in \eqref{definition:empirical:cov:y} and \eqref{empirical:auxiliary:statistics:quant} have specific statistical dependencies. Therefore, applying element-wise corrections on the random variables \eqref{empirical:auxiliary:statistics:quant} introduces nonlinear random effects which are challenging to characterize. As such, the Van Vleck corrections cause an additional loss of information and potential bias. Determining the precise impact of the Van Vleck corrections on the final interferometric imaging solution is, to the best of our knowledge, an open problem. The technique employed here addresses the task of processing the $1$-bit sensor data differently. Aiming for consistent and efficient likelihood-based statistical processing by solving the multivariate binary MLE, the presented approach exploits the complete covariance structure \eqref{covariance:unquantized:reduced} before the hard-limiter when estimating the intensity values jointly. A computationally tractable solution is obtained by iteratively updating an auxiliary score function \eqref{conservative:score} and information matrix \eqref{pessimistic:fisher:matrix} of the multivariate binary data model at the hard-limiter output \eqref{hard:limiter}. Using all empirical elements in \eqref{empirical:auxiliary:statistics:quant} jointly while taking into account their dependencies characterized by \eqref{definition:covariance:transformed}, this enables finding an estimate for the intensities in the linear covariance structure \eqref{covariance:unquantized:reduced}, which attains the analytically tractable performance \eqref{definition:error:mle:quant}. As a result, the effect of coarsely quantizing the radio measurements onto the interferometric imaging solution can be determined precisely.
\subsection{Storage-based Radio Telescope Systems}
Memory-based radio telescope systems are already being discussed today in the context of LOFAR. In \cite{Gunst18} the authors provide a detailed overview of the scientific advantages under a memory-based telescope architecture and estimate that the approach will be feasible for LOFAR in $10$ years. The design concept discussed here is different. The proposed radio telescope architecture radically follows the idea of shifting analog circuit complexity into the digital domain to exploit its exponential technology progress to a full extend. The power and hardware resources freed through this extreme layout paradigm are invested in a higher number of radio sensors, leading to improved measurement sensitivity and spatial resolution. A memory-based telescope design is, therefore, not the motivation for our discussion but a natural consequence. Setting the amplitude resolution of the signal acquisition at the antennas to $1$-bit would reduce the sensor data volume of the LOFAR telescope by one order of magnitude and make a memory-based design as outlined by \cite{Gunst18} possible today.
\section{Conclusion}
We have proposed a progressive all-digital radio telescope architecture based on probabilistic binary sensing and statistical data processing. The resource-efficient signal acquisition and the small resulting data volume of \emph{The Massive Binary Radio Lenses} allow using significantly more distributed radio sensors which increases the sensitivity of astronomical surveys and their spatial resolution. The fundamental principles of the envisioned system have been explained, and the basic building blocks defined. We have delivered an overview of the scientific, technical, and institutional advantages. To provide a first convincing proof of concept, we have performed hardware-aware probabilistic modeling of sensor outputs in an exemplary binary radio telescope system and derived a consistent interferometric image reconstruction method along the principle of maximum-likelihood. Further, we have provided an analytical measure for the interferometric performance achieved with the derived method in binary radio telescope systems. Visual comparison of all-sky images obtained with large-scale array measurements from LOFAR and the provided interferometric uncertainty analysis demonstrate that massive binary sensing has the potential to form one of the key technologies for future all-digital radio telescope systems.
\section{Acknowledgement}
The provision of the calibrated LOFAR measurements by Stefan J. Wijnholds from Netherlands Institute for Radio Astronomy (ASTRON) and his advice in handling the telescope data are gratefully acknowledged. Several technical discussions on radio telescope systems with him and his comments on the manuscript have helped to improve the presentation. Several academic discussions on radio astronomy and array processing with, as well as comments on the manuscript by Alle-Jan van der Veen from Delft University of Technology (TU Delft) have fostered the presented research and are gratefully acknowledged. The numerous comments by the fifteen anonymous reviewers of the RADIANT research proposal were helpful when improving the concepts outlined in this manuscript and are, therefore, gratefully acknowledged. The author would also like to thank Albert-Jan Boonstra and Mark Ruiter from Netherlands Institute for Radio Astronomy (ASTRON) for useful technical discussions and personal encouragement when preparing this technology study in August 2017. For the courage and confidence to support this research administratively and financially, the author is indebted to the Deutsche Forschungsgemeinschaft (DFG, German Research Foundation).

\end{document}